\documentclass[%
 reprint,
nofootinbib,
 amsmath,amssymb,
 aps,
]{revtex4-2}

\usepackage{graphicx}
\usepackage{dcolumn}
\usepackage{bm}

\begin{document}

\preprint{APS/123-QED}

\title{Paraxial diffusion-field retrieval}

\author{David M.~Paganin}
 \affiliation{School of Physics and Astronomy, Monash University, Clayton, Victoria, 3800, Australia}
 
\author{Daniele Pelliccia}
\affiliation{Instruments \& Data Tools Pty Ltd, Victoria 3178, Australia}

\author{Kaye S.~Morgan}
 \affiliation{School of Physics and Astronomy, Monash University, Clayton, Victoria, 3800, Australia~}

\date{\today}

\begin{abstract}
Unresolved spatially-random microstructure, in an illuminated sample, can lead to position-dependent blur when an image of that sample is formed.  For a small propagation distance, between the exit surface of the sample and the entrance surface of a position-sensitive detector, the paraxial approximation implies that the blurring influence of the sample may be modeled using an anomalous-diffusion field.  This diffusion field may have a scalar or tensor character, depending on whether the random microstructure has an autocorrelation function that is rotationally isotropic or anisotropic, respectively. Partial differential equations are written down and then solved, in a closed-form manner, for several variants of the inverse problem of diffusion-field retrieval given suitable intensity images.  Both uniform-illumination and structured-illumination schemes are considered. Links are made, between the recovered diffusion field and certain statistical properties of the unresolved microstructure.  The developed theory---which may be viewed as a crudely parallel form of small-angle scattering under the Guinier approximation---is applicable to a range of paraxial radiation and matter fields, such as visible light, x rays, neutrons, and electrons. 
\end{abstract}

\maketitle


\section{Introduction}

Pollen is a thermometer. According to this key result from the theory \cite{EinsteinBrownianMotion} of Brownian motion \cite{BrownOnBrownianMotion}, one initially-localized pollen cluster, placed on a smooth still water surface, has a width that diffuses in time at a rate proportional to the square root of the water temperature \cite{EinsteinBrownianMotion}.  A set of pollen clusters, placed at different locations on the water surface, will therefore diffuse at different rates if the water temperature depends on position. Even though the pollen-buffeting water molecules are neither spatially nor temporally resolved, the influence of this unresolved spatiotemporal substructure may be both visualized and quantified, by watching each individual pollen cluster spread with time.  Upon measuring the pollen-cluster diffusion coefficient $D(x,y,t)$, as a function of transverse position $(x,y)$ and time $t$, the corresponding temperature distribution $T(x,y,t)$ may be inferred.  

Turn from Brownian motion to imaging physics.  Replace water-kicked pollen grains with paraxial light rays---or x rays, electrons, neutrons, etc.---that are buffeted by spatially-random unresolved microstructures \cite{Pagot2003,Yashiro2011,Lynch2011,Modregger2012,Yashiro2015} as they traverse an imaged sample. A narrow pencil \cite{BornWolf} of initially-parallel rays will diffuse as it traverses the spatially-random medium \cite{ChernovBook1960}, with the associated position-dependent diffusion coefficient being a measure of the statistical properties of the traversed medium.  The more the pencil is diffusively broadened, the stronger the spatial fluctuations in the traversed material must be.  

\begin{figure}[ht!]
\centering
\includegraphics[width=0.99\columnwidth]{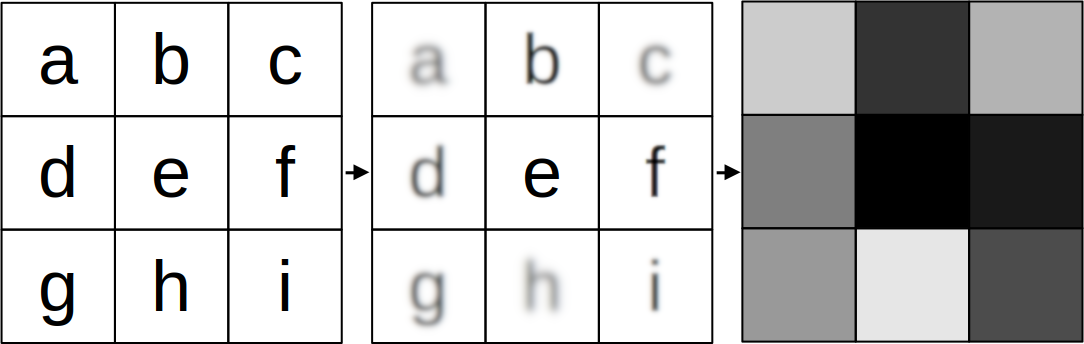}
\caption{The intensity distribution in the left panel is blurred, in a position-dependent manner, to give the image in the middle panel. The position-dependent degree of blurring, namely the ``diffusion field'', is indicated in the right panel.  ``Diffusion-field retrieval'' seeks to reconstruct the rightmost panel, given the images in the left and middle panels. }
\label{fig:DiffusionField}
\end{figure}

To visualize the imaging experiment we have in mind, consider Fig.~\ref{fig:DiffusionField}.  A thin sample is illuminated with paraxial radiation or matter waves, via an imaging system with sufficiently low degree of coherence \cite{Zernike1938} that the effects of refraction and interference may be ignored.  Suppose the intensity distribution, at the exit surface of the sample, is given by the left panel of Fig.~\ref{fig:DiffusionField}.  Upon free-space propagation---or, equivalently, defocus---the exit-surface image will blur in a position-dependent manner, as shown in the middle panel of Fig.~\ref{fig:DiffusionField}.  We consider this position-dependent blur to be associated with spatially unresolved microstructure (internal roughness) in the sample, as described in the previous paragraph.  By eye, we see that some letters are blurred more than others, while one letter is not blurred at all.  Qualitatively, we expect the associated position-dependent diffusion coefficient---namely the ``diffusion field''---to look as shown in the right panel of Fig.~\ref{fig:DiffusionField}.  This diffusion field is zero for the letter that is not blurred at all, with a large positive value for the letters that are most blurred.  As its name implies, ``diffusion-field retrieval'' seeks to solve the inverse problem \cite{Sabatier2000} of quantitatively recovering the diffusion field in the right panel of Fig.~\ref{fig:DiffusionField}, given intensity measurements such as those in the middle and left panels.  

The present paper studies the problem of diffusion-field retrieval for paraxial radiation and matter fields such as visible light, x rays, neutrons, and electrons.  It draws on several areas of research. The links, between these areas and our core theme of diffusion-field retrieval, will be clarified in later sections of the paper. By way of introduction, however, we list these key ingredients now:
\begin{itemize}
    \item Deterministic methods of phase retrieval, based on the transport-of-intensity equation \cite{Teague1983,TIE==LongReviewArticle2020}, relate longitudinal intensity transport to the phase of a propagating paraxial coherent beam. In a similar vein, the present paper relates longitudinal intensity transport to the diffusion field---associated with a propagating paraxial beam having a sufficiently low degree of coherence that the effects of refraction and Fresnel diffraction may be neglected---and seeks to solve the resulting partial differential equations in a closed-form (non iterative) manner. In analogy with the concept of phase retrieval, we speak of our-approach as ``diffusion-field retrieval''.  This may be thought of as a special case of the Fokker--Planck \cite{Risken1989} extension to the transport-of-intensity equation \cite{MorganPaganin2019,PaganinMorgan2019}, with the diffuse-flow term included and the coherent-flow term taken to be negligible. 
    \item A growing body of relatively recent literature, on diffuse-scatter dark-field imaging using x rays \cite{ MorrisonBrowne1992, SuzukiUchida1995, ando2002, Pagot2003, rigon2003, levine2004, Pfeiffer2008df, wen2008, pfeiffer2009, Bech2010, Yashiro2010, Yashiro2011, Lynch2011, Modregger2012, endrizzi2014,  berujon2012b, zanette2014} and neutrons \cite{Strobl2008, grunzweig2008, Strobl2014}, directly inspires the present work.  These papers study the visibility reduction in a transmitted beam, associated with spatially-random unresolved microstructure within an imaged sample. Closely related work, on decoherence and extinction in the context of propagation-based phase contrast and analyzer-crystal phase contrast \cite{Nesterets2008}, also directly influences the present work. Note, our paper considers a simpler problem than the references that have just been cited, since we ignore the often-important influence of coherence-enabled effects such as sample refraction and Fresnel diffraction. 
    \item  Small-angle neutron and x-ray scattering, from spatially-random microstructure that is not directly spatially resolved, is used to obtain statistical information about such microstructure from far-field diffraction patterns obtained using a localized illumination probe \cite{GlatterKratky1982,SiviaScatteringTheoryBook2011}.  Diffusion-field retrieval may be viewed as a crudely parallel form of small-angle scattering, corresponding to the Guinier regime \cite{GuinierFournetBook} associated with the central portion of the small-angle-scatter cone. 
    \item Ray diffusion is associated with propagation through spatially-random media, such as starlight through the turbulent atmosphere \cite{ChernovBook1960}.  Models for this process enable a recovered diffusion field to be related back to the statistical properties of the sample that resulted in such a diffusion field. 
    \item In the language of transmission electron microscopy (TEM) together with x-ray or neutron scattering, our scalar diffusion field $D(x,y)$ is akin to a position-dependent Debye-Waller factor \cite{CowleyBook}.  In our case, this blurring factor is associated with static random microstructure, rather than  the typical TEM case where it is associated with dynamic thermal fluctuations in atomic positions.  In this sense, our position-dependent blur is more directly analogous to a typical x-ray or neutron case, where Debye-Waller factors may be used to model scattering by surface roughness in reflectivity measurements, as in \citet{Sinha1988}.  Along similar lines, our ensemble averaging---over a set of realizations of the spatially-random microstructure---bears some resemblance to similar averages that are taken in the frozen-phonon model \cite{FrozenPhononModel}, as used in many TEM contexts \cite{KirklandBook}. Analogous ensemble averages occur throughout the theory of physical optics \cite{Pedersen1976,MandelWolf,Nesterets2008,GoodmanSpeckleBook}. 
    \item  Ray and wave propagation through turbid media \cite{TurbidMediumLightScatter,DiffuseLightTransportTutorial2008,DiffuseOpticalTomographyReview2014} is closely related to the refractive and diffractive sample-microstructure models, respectively, that are considered in this paper.  However, we can work with a much simpler model for intensity transport through a spatially-random medium, due to the strongly simplifying---and strongly restrictive---assumption of paraxiality \cite{Khelashvili2006}.
    \item Finally, the theory of anomalous diffusion is of direct relevance \cite{MetzlerKlafter2000,EvangelistaLenziBook2018}.  The position-dependent point-spread function, associated with the diffusion field immediately downstream of an illuminated sample, has a width $\sigma_{\Delta}$ that scales linearly with the distance $\Delta$ between the exit surface of the sample and the entrance surface of a position-sensitive intensity detector.  Since $\sigma_{\Delta}$ does not scale with the square root of the evolution parameter $\Delta$, as would be the case for normal diffusion \cite{EinsteinBrownianMotion,CrankBook}, our forward problem is a case of anomalous diffusion.  
\end{itemize}

The remainder of the paper is structured as follows.   Section~\ref{sec:II} develops a simple model, for the forward problem of how a sample's diffusion field influences the paraxial intensity transport, downstream of the sample. This  finite-difference anomalous-diffusion expression considers the diffusion-inducing spatially-random microstructure to have a correlation function that is rotationally isotropic at each point in the sample, but may vary as a function of position within the sample (over length scales that are large compared to the correlation length of the random microstructure). This allows the diffusion field to be described by a scalar function of transverse position.  Two complementary special cases are delineated, corresponding to (i) a ``large microstructure'' refractive case where {\em all} incident rays diffuse as they pass through the sample, and (ii) a ``small microstructure'' weakly-diffractive case where only a minute fraction of the incident illumination is diffusely scattered by the sample's unresolved random microstructure.  Section~\ref{sec:III} considers the corresponding inverse problem of scalar diffusion-field retrieval, both in the absence of any beam-shaping element for the sample illumination, and then in the presence of such a structured but otherwise unspecified beam-shaping mask.  Section~\ref{sec:IV} generalizes the preceding work, by considering a tensorial version of diffusion-field retrieval.  This case corresponds to rotationally-anisotropic spatially-unresolved sample microstructure, which causes the corresponding position-dependent small-angle-scatter fans to have an elliptical rather than a rotationally-symmetric transverse profile.  The associated position-dependent diffusion field, in this case, is a second-rank symmetric tensor at each point on the exit surface of the sample.   Some broader implications and extensions of diffusion-field retrieval, in both its scalar and tensorial forms, are discussed in Sec.~\ref{sec:V}.  Concluding remarks comprise Sec.~\ref{sec:VI}.      

\section{Paraxial diffusion-field retrieval: the forward problem}\label{sec:II}

\begin{figure}[ht!]
\centering
\includegraphics[width=0.9\columnwidth]{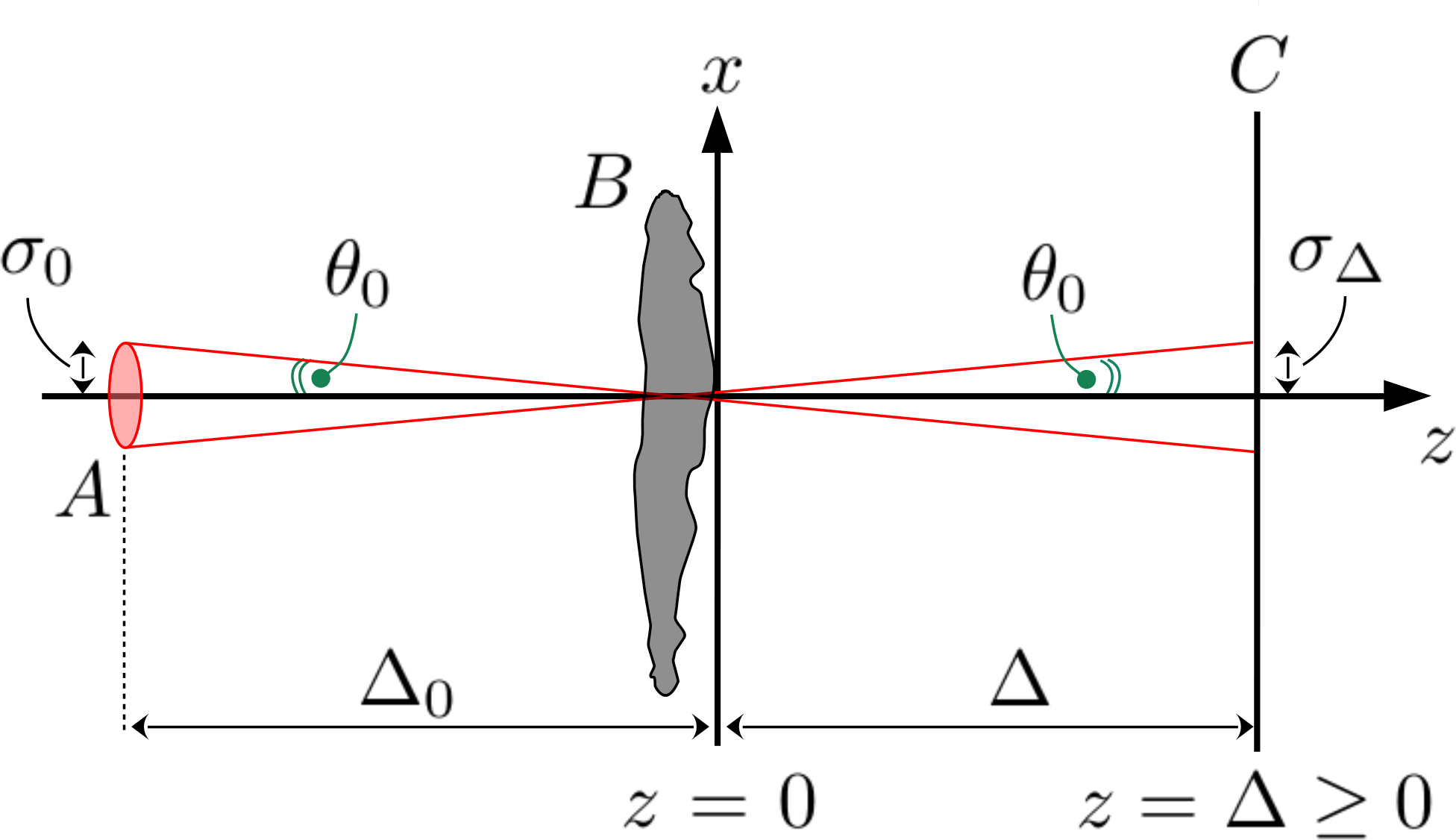}
\caption{Paraxial imaging, using an extended chaotic source $A$ with a sufficiently low degree of coherence, gives a blurred image of an optically thin sample $B$ over the detection plane $C$.}
\label{Fig:SimpleImagingWithBlur}
\end{figure}

Consider the paraxial imaging scenario in Fig.~\ref{Fig:SimpleImagingWithBlur}.   Here, an extended chaotic source $A$ with half-width $\sigma_0$ illuminates a noncrystalline thin sample $B$.\footnote{(i) An ``extended chaotic source'' consists of independent radiators, with emitters at distinct points on such a source having no phase correlation with respect to one another \cite{LoudonBook}.   (ii) ``Thin sample'' means that the projection approximation \cite{Paganin2006} holds.  Stated differently, the transverse energy flow within the volume of the sample may be neglected, by assumption, in calculating the properties of the field at the exit surface of the sample.  Throughout the paper, we take ``thin sample'' to mean ``a sample for which the projection approximation holds''.}  Only one transverse dimension is shown, for simplicity. The distance between the chaotic source and the sample entrance surface is assumed to be sufficiently small, such that there is negligible correlation between the complex disturbance at spatially distinct points over this entrance surface.  Stated more precisely, the complex degree of coherence \cite{BornWolf,MandelWolf,ThinWolfBook} (and the closely-related mutual intensity) are both taken to be delta-correlated, over the sample entrance surface.\footnote{These assumptions imply that the imaging system is linear at the level of intensity, with the output intensity distribution being equal to the smearing of the input intensity distribution with a point-spread function that depends on transverse position in general. When the further assumption of shift invariance is applicable, this relation between input and output intensity reduces to the convolution integral given in Eq.~(\ref{eq:SimpleConvolutionBlur}). For further detail, see p.~320 of \citet{GoodmanStatisticalOpticsBook}, pp~368-369 of \citet{SalehTeichBook}, and pp~530-531 of \citet{BarrettMyersBook}.} For such an imaging system, the intensity distribution---over the detection plane $C$---is blurred in comparison to the intensity distribution over the exit-surface plane $z=0$. 

By assumption, both the illuminating and transmitted imaging quanta (e.g., photons, electrons, neutrons etc.)~are paraxial with respect to the optical axis.  Narrow-energy-band (``narrowband'') illumination is assumed throughout, unless otherwise stated.  Also, again unless otherwise stated, all of our discussions will pertain to both scalar radiation (such as visible light and hard x rays, neglecting the influence of polarization) and scalar  matter waves (such as electrons and neutrons, neglecting the influence of spin). 

Let $x$ denote the transverse coordinate perpendicular to the optical axis $z$, with the thin-sample nominally-planar exit surface corresponding to $z=0$.  The blurred intensity distribution $I(x,z=\Delta \ge 0)$, over the 
plane $C$, may be approximated by the convolution integral \cite{DeblurByDefocus}
\begin{equation}
\label{eq:SimpleConvolutionBlur}
  I(x,z=\Delta \ge 0)=\int_{-\infty}^{\infty} I_0(x')K(x-x',
  \Delta) 
  \, \mathrm{d}x'.
\end{equation}
Here, 
\begin{equation}
\label{eq:DefinitionForInfocusIntensity} 
 I_0(x) \equiv I(x,z= 0)
\end{equation}
denotes the ``in focus'' intensity distribution over the exit surface $z=0$, with the blurring kernel $K(x,\Delta)$ corresponding to the point spread function associated with the magnified (or demagnified) image of the source distribution over the detector plane $C$.  We assume this blurring kernel to be normalized via
\begin{equation}
\label{eq:BlurringKernelNormalised}
 \int_{-\infty}^{\infty} K(x,\Delta)\, \mathrm{d}x = 1,
\end{equation}
and to have vanishing first moment\footnote{This assumption implicitly considers the refractive effects of the sample to be negligible.  Stated differently, we are here assuming that the influence of propagation-based phase contrast \cite{KleinOpat1976,Snigirev,Paganin2006} (also termed out-of-focus contrast \cite{Bremmer1952,CowleyBook}) may be neglected. This assumption, which can break down when the source size is sufficiently small \cite{WilkinsFish}, will be dropped in a companion paper that is currently in preparation. Cf.~Sec.~\ref{Sec:Discussion--FokkerPlanckExtension}. \label{footnote:WeInitialNelgectRefractionOfBlurCone}}
\begin{equation}
\label{eq:VanishingFirstMomentofK}
 \int_{-\infty}^{\infty} x \, K(x,\Delta)\, \mathrm{d}x = 0.
\end{equation}
%

Assume that $K(x,\Delta)$ is twice differentiable with respect to $x$, and has a standard deviation  
\begin{equation}
\label{eq:Variance_of_K}
\sigma_{\Delta}=\sqrt{\int_{-\infty}^{\infty} x^2 \, K(x,\Delta)\, \mathrm{d}x}
\end{equation}
that is small compared to the characteristic transverse length scale over which $I_0(x)$ varies appreciably.  Make the change of variables 
\begin{equation}
\label{eq:SimpleChangeOfVariables}
    X=x-x'
\end{equation}
in Eq.~(\ref{eq:SimpleConvolutionBlur}), Taylor expand $I_0(x-X)$ to second order about $x$, and then use Eqs.~(\ref{eq:BlurringKernelNormalised})-(\ref{eq:Variance_of_K}).  Hence \cite{GureyevLocalDeconvolution2003} 
\begin{equation}
\label{eq:SimpleBlurIn1D}
I(x,z=\Delta \ge 0) \approx \left(1+\frac{1}{2}\sigma_{\Delta}^2 \frac{\mathrm{d}^2}{\mathrm{d} x^2} \right)I(x,z=0).
\end{equation}

Generalizing to two transverse dimensions, corresponding to Cartesian coordinates $(x,y)$ in planes perpendicular to the optical axis $z$, we may then write \cite{Unsharp0,DeblurByDefocus}
\begin{equation}
\label{eq:2DformForDiffusiveBlurOnly}
I(x,y,z=\Delta \ge 0) \approx \left(1+\tfrac{1}{2}\sigma_{\Delta}^2\nabla_{\perp}^2\right)I(x,y,z=0).
\end{equation}
Here, $\nabla_{\perp}^2$ denotes the transverse Laplacian operator
\begin{equation}
\label{eq:TransverseLaplacianDefinition}
\nabla_{\perp}^2 = \frac{\partial^2}{\partial x^2} + \frac{\partial^2}{\partial y^2}.
\end{equation}

In retrospect, we could have written down Eq.~(\ref{eq:2DformForDiffusiveBlurOnly}) on inspection, given that it is a simple forward-finite-difference form of the diffusion equation.  Dimensional analysis shows $\sigma_{\Delta}$ to play the role of a characteristic transverse length scale, which must be on the order of the blur width associated with our paraxial imaging system, since this is the only relevant length scale that can be meaningfully associated with the blurring kernel $K$ that was used to derive Eq.~(\ref{eq:2DformForDiffusiveBlurOnly}). 

It is useful to hold in mind this conceptual connection between diffusion and imaging-system blur, throughout much of the remainder of this paper.  However, it is equally important to note the following key difference.  In conventional diffusion the half-width $\sigma$ of the blurring kernel scales as the square root of an evolution parameter such as time $t$ \cite{EinsteinBrownianMotion}.  In our case, the relevant evolution parameter is the sample-to-detector propagation distance $\Delta \ge 0$, but $\sigma_{\Delta}$ must scale linearly with the parameter $\Delta$, rather than with the square root of this parameter.  In the language of diffusion physics, this is an example of anomalous diffusion.  More precisely, we may write \cite{EvangelistaLenziBook2018}
\begin{equation}
\label{eq:AnomalousDiffusion1}
\sigma_\Delta \propto \Delta^{0.5 + a},
\end{equation}
where
\begin{equation}
\label{eq:AnomalousDiffusion2}
a \ge -\tfrac{1}{2}
\end{equation}
is an anomalous diffusion coefficient.  If this anomalous diffusion coefficient vanishes, normal (i.e., non-anomalous) diffusion holds. In our case, we are in the anomalous-diffusion regime where 
\begin{equation}
\label{eq:BallisticDiffusion}
a=\tfrac{1}{2}, 
\end{equation}
which is known as ballistic superdiffusion \cite{MetzlerKlafter2000}. 

Let us now return to the main thread of our argument. Given the source-to-sample distance $\Delta_0$ and the sample-to-detector distance $\Delta$, similar triangles shows the detector-plane blurring half-width $\sigma_{\Delta}$ to obey
\begin{equation}
\label{eq:SimilarTriangles}
\frac{\sigma_0}{\Delta_0}=\frac{\sigma_{\Delta}}{\Delta}.
\end{equation}  
Equation~(\ref{eq:SimilarTriangles}) allows Eq.~(\ref{eq:2DformForDiffusiveBlurOnly}) to be written as
\begin{equation}
\label{eq:2DVersionOfSimpleBlur}
I(x,y,z=\Delta \ge 0) \approx (1+\Delta^2 D \nabla_{\perp}^2)I(x,y,z=0).
\end{equation}
Above, we use the dimensionless diffusion coefficient \cite{Bech2010}
\begin{equation}
\label{eq:BlurConeAngleSimpleForm}
D \equiv \frac{\sigma_0^2}{2\Delta_0^2} \approx \tfrac{1}{2}\theta_0^2, \quad \theta_0 \ll 1,
\end{equation}
which induces the blurring of the intensity field, as it propagates from the sample exit surface $z=0$ to the detection plane $C$, in Fig.~\ref{Fig:SimpleImagingWithBlur}. Our dimensionless diffusion coefficient is proportional to the square of the half-angle
\begin{equation}
 \theta_0=\tan^{-1}\frac{\sigma_0}{\Delta_0}\approx \frac{\sigma_0}{\Delta_0} 
\end{equation}
associated with the diffusive ray cone, which emerges from each point over the plane $z=0$. 

\begin{figure}[ht!]
\centering
\includegraphics[width=0.99\columnwidth]{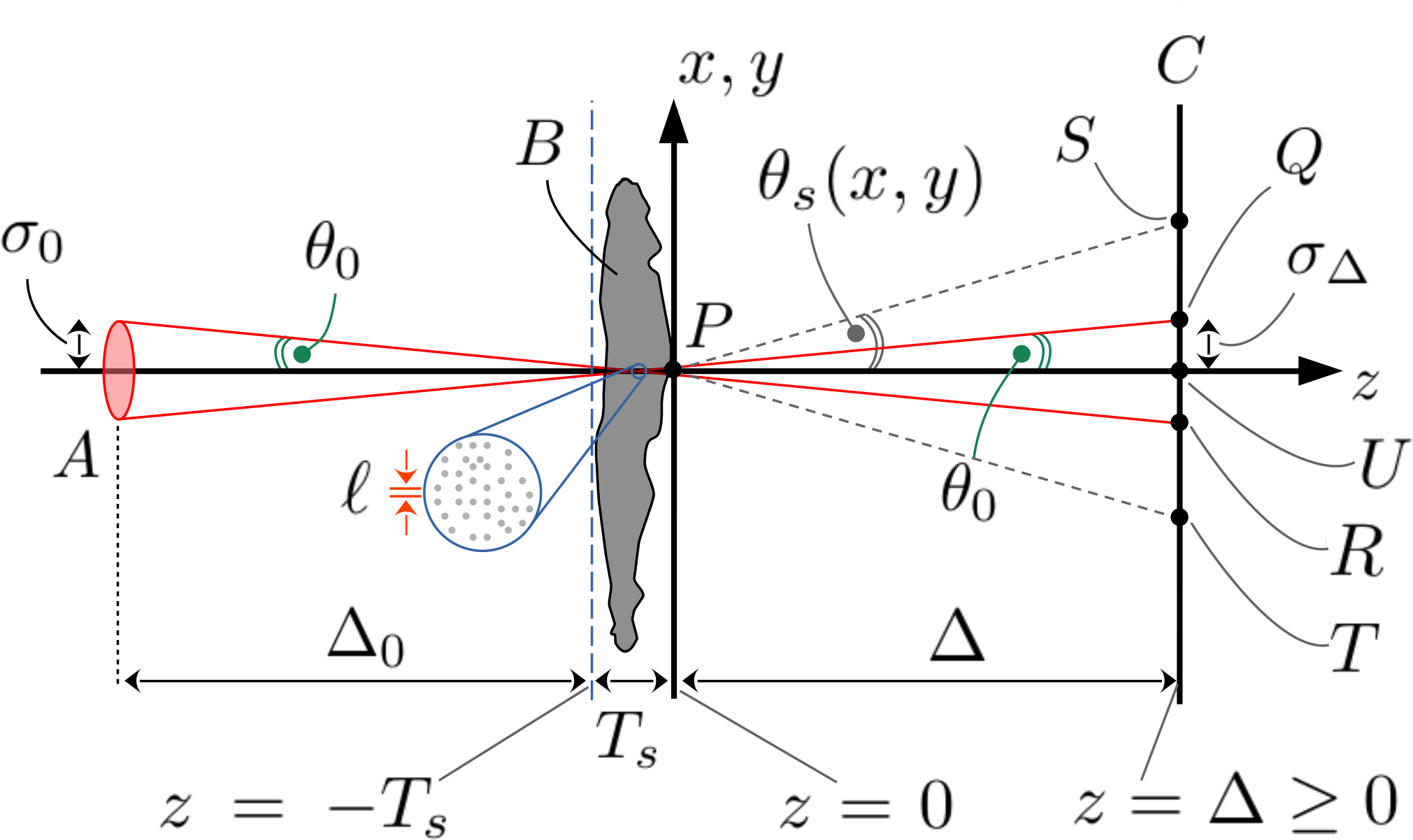}
\caption{Generalization of previous figure, to include the additional blurring effects of the unresolved random sample microstructure, as indicated in the inset.}
\label{Fig:SimpleImagingWithSourceBlurAndSampleBlur}
\end{figure}

The image blur, associated with the extended chaotic source, may be augmented by a contribution that is associated with unresolved sample microstructure \cite{Pagot2003,Yashiro2011,Lynch2011,Modregger2012,Yashiro2015}.  By assumption, this unresolved sample microstructure is taken to be locally spatially random and amorphous, since the effects of partially-ordered or fully-ordered crystalline microstructure lies outside the scope of our paper.   Figure~\ref{Fig:SimpleImagingWithSourceBlurAndSampleBlur} illustrates both the source-blur and microstructure-blur effects, each of which have blur cones with different apex half-angles.  In this diagram, the blur cone $PQR$ associated with source-size effects (in the absence of sample microstructure-induced blur) has been supplemented by the blur cone $PST$ associated with sample microstructure (in the absence of source-size blur).  We use $\theta_0$ and $\theta_s(x,y)$ to denote the respective apex half-angles of these two cones.  The latter half-angle depends on transverse position $(x,y)$ since the amorphous spatially-random sample microstructure (as represented by the inset near the bottom left of the figure) will in general vary in its statistical character, as we move to different transverse locations in the sample.  

To proceed further, introduce the Fresnel number \cite{SalehTeichBook}
\begin{equation}
N_{\textrm{F}}(x,y)=\frac{[\ell(x,y)]^2}{\lambda T_s}
\end{equation}
associated with the passage of the illuminating radiation or matter wave field, from the nominal entrance surface $z=-T_s$ of the thin sample, to the nominal exit surface $z=0$.  Here, $\ell(x,y)$ is the characteristic transverse length scale over which the spatially-random amorphous microstructure fluctuates appreciably, and $\lambda$ is the average wavelength of the narrowband radiation or matter wave field.  By assumption, the fluctuations in sample structure are assumed to have no preferred direction, although we note that this assumption is relaxed in Sec.~\ref{sec:IV}.       

The Fresnel number $N_{\textrm{F}}(x,y)$ enables us to distinguish between two qualitatively different indicative cases that will be considered throughout much of the remainder of the paper. These two cases, which are not comprehensive, indicate two important regimes that may be encountered in practice. Both employ the assumption
\begin{equation}
\label{eq:Ell1MuchBiggerThanTheWavelength}
\ell(x,y) \gg \lambda
\end{equation}
as a necessary condition for their validity. Case \#1 corresponds to the additional condition where 
\begin{equation}
\label{eq:CaseOneFresnelNumberCondition}
N_{\textrm{F}}(x,y) \gg 1 \quad \textrm{(Case~\#1)}.
\end{equation}
Case \#2 corresponds to a complementary regime where 
\begin{equation}
\label{eq:CaseTwoFresnelNumberCondition}
N_{\textrm{F}}(x,y) \ll 1 \quad \textrm{(Case~\#2)}.
\end{equation}
We do not consider the more complicated intermediate regime where $N_\textrm{F}(x,y)$ is on the order of unity.  

Case \#1 corresponds to a geometrical-optics condition for ray propagation in a spatially-random medium.\footnote{See e.g.~the use of Eq.~(\ref{eq:CaseOneFresnelNumberCondition}) on p.~12 of \citet{ChernovBook1960}.} Here, the unresolved microstructure is considered to be sufficiently slowly varying in space that the condition in Eq.~(\ref{eq:CaseOneFresnelNumberCondition}) holds.  Under this model, every incident ray will perform a continuous random walk as it traverses the medium.  This phenomenon of ray diffusion, so called because the average transverse position of an incident ray will diffuse as it refractively traverses the random medium, leads to a cone of rays at the exit surface of the sample whose apex has the half-angle \cite{Pagot2003}
\begin{equation}
\label{eq:TotalBlurAngleCase1}
    \theta(x,y)=\sqrt{\theta_0^2+[\theta_s(x,y)]^2}\quad \textrm{(Case~\#1)}.
\end{equation}
In writing down the preceding expression, we have made the approximation that the following two blur-cone half-angles may be added in quadrature: (i) the half-angle $\theta_0$ due to the extended chaotic source in the absence of additional blurring effects due to sample microstructure, and (ii) the half-angle $\theta_s(x,y)$ due to the sample microstructure in the absence of source-size blur.  This approximation will hold, for example, if our two contributions to image blur may be well approximated by a Gaussian function. Finally, we remark that there is a single diffusive cone at the exit surface of each point of the thin sample, in Case \#1.  

Case \#2 complements the refractive model of the previous case, by instead considering a diffractive regime for scalar wave propagation in a spatially-random medium.\footnote{See e.g.~the use of Eq.~(\ref{eq:CaseTwoFresnelNumberCondition}) on p.~45 of \citet{ChernovBook1960}.  Cf.~comments regarding diffractive versus refractive models for scattering by a random medium, made in the context of x-ray small-angle scattering, in \citet{Beeman1947} and \citet{Dragsdorf1956}.}  Here, the unresolved microstructure is considered to be sufficiently rapidly varying in space that the condition in Eq.~(\ref{eq:CaseTwoFresnelNumberCondition}) holds.  If, in addition, we may also assume the variance of the spatially-random fluctuations in refractive index to be sufficiently small, the first Born approximation may be employed \cite{Messiah,CowleyBook,Paganin2006}. This approximation assumes the fraction $F$, of the incident field that is scattered by the sample microstructure, to be much smaller than unity.  Since this fraction will in general depend upon the transverse location of the illuminated sample, we write the condition on the scattering fraction as
\begin{equation}
\label{eq:F-much-less-than-one-for-case-2}
    0 \le F(x,y) \ll 1 \quad \textrm{(Case~\#2)}.
\end{equation}
Under the first Born approximation, and in contrast to the scenario considered in the previous paragraph, there will be two coaxial diffusive cones at each point on the exit surface of the sample. The inner cone, which will have a half-angle of $\theta_0$, corresponds to the influence of source-size blur on all photons (or electrons, neutrons, muons etc.) that pass through the sample without being scattered by the microstructure.  The outer cone, which corresponds to single scattering under the first Born approximation, will have a half-angle $\sqrt{\theta_0^2+[\theta_s(x,y)]^2}$ that corresponds to the diffuse small-angle scatter associated with the unresolved microstructure.  Under the Guinier approximation, which we henceforth adopt, $\theta_s(x,y)$ may be considered to be a function solely of the microstructure correlation length \cite{SiviaScatteringTheoryBook2011}. Adding the two blur-cone angles in quadrature, and including the influence of the scatter-fraction $F$, the effective blur-width half-angle increases from its source-blur value of $\theta_0$ according to (cf.~Ref.~\cite{Beeman1947}) 
\begin{eqnarray}
\label{eq:TotalBlurAngleCase2}
\theta_0 \longrightarrow \theta(x,y)=\sqrt{\theta_0^2+F(x,y)[\theta_s(x,y)]^2}, \quad\quad\quad\quad\quad \\ \nonumber \quad\quad\quad\quad\textrm{where}\quad  0 \le F(x,y) \ll 1 \quad \textrm{(Case~\#2)}. 
\end{eqnarray}
We reiterate that $\theta(x,y)$, in Eq.~(\ref{eq:TotalBlurAngleCase2}), is an effective blur angle associated with the combined influence of two coaxial diffusive cones.

For much of the remainder of this paper, we can treat Cases \#1 and \#2 simultaneously.  We do this by employing the upper line of Eq.~(\ref{eq:TotalBlurAngleCase2}), on the understanding that setting $F(x,y)=1$ corresponds to Case \#1, with $F(x,y)\ll 1$ for Case \#2.  With this understanding, we may now substitute the upper line of Eq.~(\ref{eq:TotalBlurAngleCase2}) into Eq.~(\ref{eq:BlurConeAngleSimpleForm}), to give the following position-dependent dimensionless diffusion coefficient that includes the influence of both source-size blur and sample-microstructure blur (cf.~Ref.~\cite{PaganinPelliccia2020}): 
\begin{equation}
\label{eq:PositionDependentScalarDiffusionCoefficient}
    D(x,y) = \tfrac{1}{2}\left\{\theta_0^2+F(x,y)[\theta_s(x,y)]^2\right\}.
\end{equation}
Also, since the dimensionless diffusion coefficient now depends on transverse position, Eq.~(\ref{eq:2DVersionOfSimpleBlur}) becomes 
\begin{align}
\nonumber I(&x,y,z=\Delta \ge 0) \\  
&=I(x,y,z=0) +\Delta^2\nabla_{\perp}^2[D(x,y) I(x,y,z=0)].
\label{eq:2DVersionOfSimpleBlurWithGeneralisation}
\end{align}

\begin{figure}[ht!]
\centering
\includegraphics[width=0.75\columnwidth]{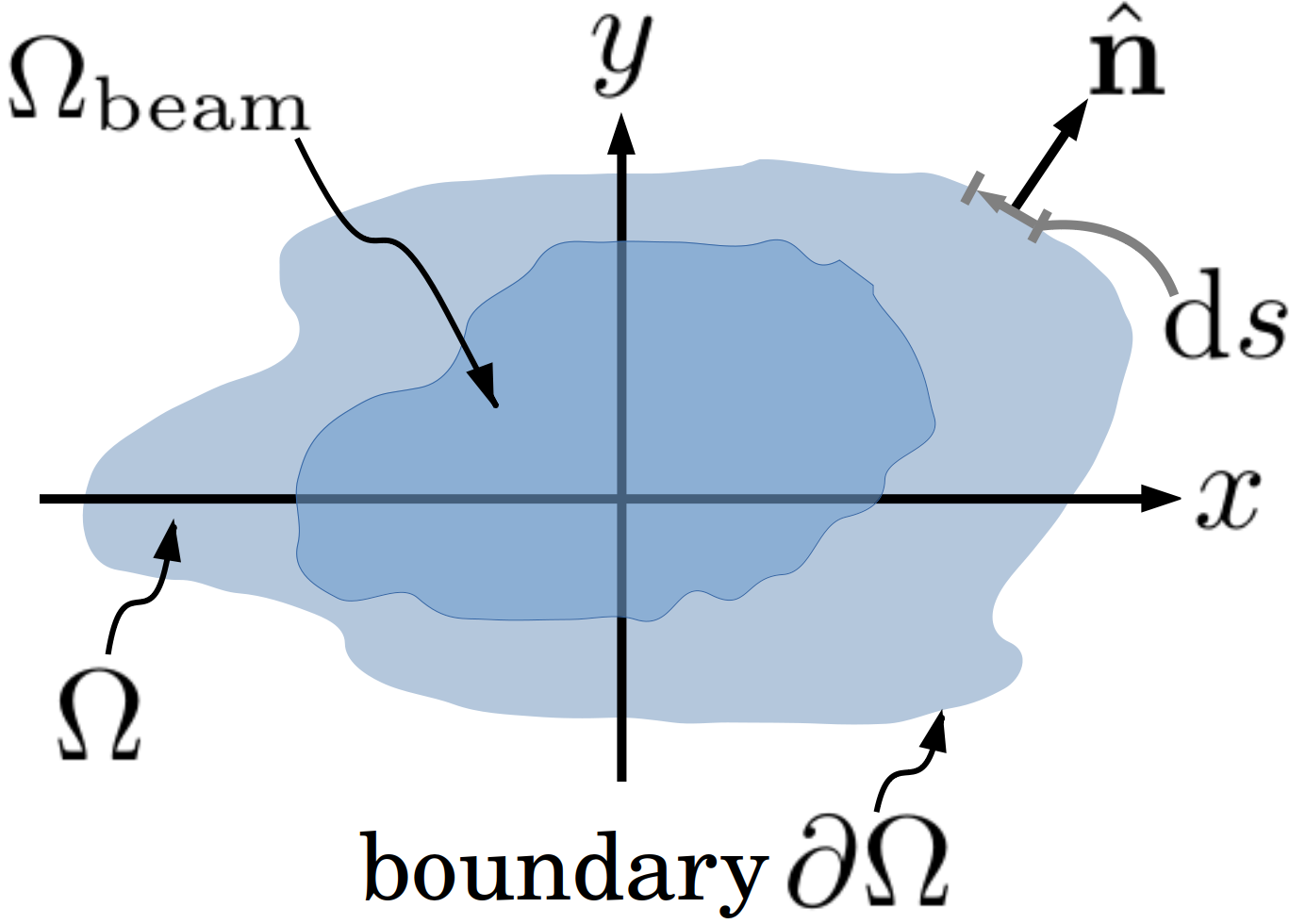}
\caption{Notation for a region $\Omega$, with smooth boundary $\partial\Omega$, in the plane $z=0$ that is perpendicular to the optical axis $z$.  The darker region $\Omega_{\textrm{beam}}$ denotes the area where the beam-like intensity distribution $I(x,y,z=0)$ is non-negligible.}
\label{Fig:EnergyConservation}
\end{figure}

The finite-difference approximation, to the position-dependent anomalous diffusion process as given by Eq.~(\ref{eq:2DVersionOfSimpleBlurWithGeneralisation}), conserves the integrated intensity over planes of constant $z$ that are perpendicular to the optical axis. To see this, consider a finite-energy paraxial beam-like intensity distribution $I(x,y,z=0)$ which vanishes more rapidly than $1/r_{\perp}$ at large transverse radii 
\begin{equation}
\label{eq:TransverseRadius}
r_{\perp}=\sqrt{x^2+y^2}.
\end{equation}
This is illustrated in Fig.~\ref{Fig:EnergyConservation}.  Here, the two-dimensional darker region $\Omega_{\textrm{beam}}$ indicates where in the plane $z=0$ the beam-like intensity distribution $I(x,y,z=0)$ is non-negligible, with this region being entirely contained within the simply-connected smooth-boundary region $\Omega$.  Now integrate Eq.~(\ref{eq:2DVersionOfSimpleBlurWithGeneralisation}) over the region $\Omega$, and apply the Gauss divergence theorem, to give
\begin{align}
\nonumber \iint_{\Omega}&I(x,y,z=\Delta \ge 0) \, \mathrm{d}x\,\mathrm{d}y \\ \nonumber &= \iint_{\Omega}I(x,y,z= 0) \, \mathrm{d}x\,\mathrm{d}y \\ &+ \Delta^2 \oint_{\partial\Omega}\hat{\mathbf{n}}\cdot \{\nabla_{\perp} [D(x,y) I(x,y,z=0)]\}\,\mathrm{d}s.
\label{eq:EnergyConservationIntermediateResult}
\end{align}
Here, $\hat{\mathbf{n}}$ is the outward-pointing unit normal to the smooth boundary $\partial\Omega$ of $\Omega$, $\mathrm{d}s$ is a differential element of arc length along this boundary, and $\nabla_{\perp}$ is the gradient operator in the $(x,y)$ plane.  If $\Omega$ is now taken to be sufficiently large, the final line of the above equation will become vanishingly small, because $I(x,y,z=0)$ vanishes more rapidly than $1/r_{\perp}$ at large $r_{\perp}$. This demonstrates the claim in the first sentence of this paragraph. We have also seen that the term in braces, in the final line of Eq.~(\ref{eq:EnergyConservationIntermediateResult}), is a conserved flux \cite{Press} (Noether current \cite{GoldsteinBook,MandlShawBook}) over the region $\Omega$, associated with our anomalous diffusive flow. 

The argument of the previous paragraph will hold true for any choice of $\Delta \ge 0$, but if the sample-to-detector propagation distance $\Delta$ is taken to be too large, Eq.~(\ref{eq:2DVersionOfSimpleBlurWithGeneralisation}) will cease to give a good approximation for the propagated intensity $I(x,y,z=\Delta)$. To estimate the maximum value of $\Delta$ for which Eq.~(\ref{eq:2DVersionOfSimpleBlurWithGeneralisation}) will be acceptably accurate, we may use dimensional analysis to approximate the position-dependent blurring width as 
\begin{equation}
\label{eq:PositionDependentBlurWidthEstimate}
w(x,y)\approx \Delta\sqrt{D(x,y)}.
\end{equation}
Let the maximum blurring width $w(x,y)$, implied by the above expression, be denoted by $w_{\textrm{max}}$.  Hence
\begin{equation}
\label{eq:MaximalPositionDependentBlurWidthEstimate}
w_{\textrm{max}}\approx \Delta_{\textrm{max}}\sqrt{D_{\textrm{max}}} \approx R,
\end{equation}
where $D_{\textrm{max}}$ is the maximum value of $D(x,y)$, $R$ is the spatial resolution of the imaging system used to measure $I(x,y,z=\Delta)$, and $\Delta_{\textrm{max}}$ is the maximum permissible propagation distance.  Hence
\begin{equation}
\label{eq:MaximumPropagationDistance}
\Delta \le \Delta_{\textrm{max}}\approx\frac{R}{\sqrt{D_{\textrm{max}}}}.
\end{equation}
If this condition is violated, higher-order transverse derivative terms are required, to augment the term proportional to the transverse Laplacian in Eq.~(\ref{eq:2DVersionOfSimpleBlurWithGeneralisation}).  In this case, which we do not consider here, the resulting hierarchy of diffusion terms will closely resemble the construction employed in the diffusive part of the Kramers-Moyal equation \cite{Risken1989,MorganPaganin2019,PaganinMorgan2019}. The coefficients, of these diffusion terms, are related to the moments of the small-angle-scatter functions at each transverse location \cite{DeblurByDefocus,Modregger2012,modregger2017,modregger2018,PaganinMorgan2019}.

\section{Paraxial diffusion-field retrieval: the inverse problem}\label{sec:III}

Equation~(\ref{eq:2DVersionOfSimpleBlurWithGeneralisation}) is our underpinning forward model.  The solution to this equation, in which we use measured intensity data to compute the diffusion field $D(x,y)$, is what we term the ``first inverse problem'' ${\mathcal{I}}_1$.  We may also speak of this as ``diffusion-field retrieval''. If ${\mathcal{I}}_1$ is solved, this enables the ``second inverse problem'' ${\mathcal{I}}_2$ to be considered, namely the question of relating $D(x,y)$ to the position-dependent correlation length $\ell(x,y)$ of the spatially-random microstructure in the thin sample \cite{Morgan2022Preprint} (cf.~Refs.~\cite{Yashiro2010,prade2016short}).  Schematically, we denote these consecutive inverse problems as
\begin{equation}
\label{eq:SchematicChainOfTwoInverseProblemsForScalarD}
 I \stackrel{{\mathcal{I}}_1}{\longrightarrow} D \stackrel{{\mathcal{I}}_2}{\longrightarrow} \ell.
\end{equation}

\subsection{First inverse problem: General remarks}

Assume the thin sample in Fig.~\ref{Fig:SimpleImagingWithSourceBlurAndSampleBlur} to be paraxially illuminated by an approximately spatially-uniform intensity distribution, of intensity $I_u$, from a narrowband source.  Assume the sample to be well contained within the field of view $\Omega_{\textrm{FOV}}$ of the image that is taken in the plane $z=\Delta$, so that $\theta_s(x,y)$ in Eq.~(\ref{eq:PositionDependentScalarDiffusionCoefficient}) can be taken to vanish along the boundary $\partial\Omega_{\textrm{FOV}}$ of this field of view.  Let the sample-to-detector propagation distance $\Delta$ be smaller than the maximum permissible value given by Eq.~(\ref{eq:MaximumPropagationDistance}).\footnote{Given the noise that will inevitably be present in all experimental intensity measurements, it would often be useful to take $\Delta$ to be on the same order as its maximum permissible value of $\Delta_{\textrm{max}}$.}  Suppose, also, that the unpropagated exit-surface intensity distribution $I(x,y,z=0)$ has been measured.  We also assume that $I(x,y,z=0)$ never vanishes anywhere over the field of view, so that (i) there are no zeroes in the intensity field that illuminates the entrance surface of the sample, and (ii) there are no points on the sample that completely absorb the incident illumination.

To solve our first inverse problem $\mathcal{I}_1$, subject to the assumptions outlined in the previous paragraph, we  introduce the known data function
\begin{equation}
\label{eq:DataFunction}    \rho(x,y)=\frac{I(x,y,z=0)-I(x,y,z=\Delta\ge 0)}{\Delta^2}
\end{equation}
and the auxiliary function
\begin{equation}
\label{eq:AuxiliaryFunctionDefinition}
    u(x,y)=D(x,y) I(x,y,z=0).
\end{equation}
Hence Eq.~(\ref{eq:2DVersionOfSimpleBlurWithGeneralisation}) leads to the boundary value problem  
\begin{equation}
\label{eq:BVPforScalarDiffusionFieldRetrieval}
\begin{cases}
  -\nabla_{\perp}^2u(x,y) = \rho(x,y),   \\
  [u(x,y)]_{\partial\Omega_{\textrm{FOV}}} = \tfrac{1}{2}\theta_0^2 I_u,
\end{cases}
\end{equation}
where $[u(x,y)]_{\partial\Omega_{\textrm{FOV}}}$ denotes the values $u(x,y)$ takes over the boundary of the field of view. Note that, in writing the final line of Eq.~(\ref{eq:BVPforScalarDiffusionFieldRetrieval}), use has been made of the previously stated assumption that $\theta_s(x,y)$ vanishes along the boundary of the field of view.  Hence Eq.~(\ref{eq:PositionDependentScalarDiffusionCoefficient}) reduces to $D$ having the constant value of $\tfrac{1}{2}\theta_0^2$ along this boundary.  Moreover, the value of the intensity at the boundary equals the intensity of the uniform illumination $I_u$, since the sample is well contained within the field of view.\footnote{If the assumption of uniform illumination intensity $I_u$ is dropped, and replaced by the weaker assumption of nonuniform illumination intensity $I_n(x,y)$ that does not vanish at any point within the field of view, the lower line of Eq.~(\ref{eq:BVPforScalarDiffusionFieldRetrieval}) will correspond to a nonconstant but otherwise known Dirichlet boundary conditions $[u(x,y)]_{\partial\Omega_{\textrm{FOV}}} = \tfrac{1}{2}\theta_0^2 [I_n(x,y)]_{\partial\Omega_{\textrm{FOV}}}$.}

Equation~(\ref{eq:BVPforScalarDiffusionFieldRetrieval}) is a Poisson equation with constant Dirichlet boundary conditions.  Since this is one of the most commonly encountered partial differential equations in mathematical physics, many methods are available for its solution.  In analytical studies, for example, a standard Green-function solution is available. For the numerical analysis of experimental data, there are again many approaches available, including full multigrid methods \cite{Gureyev_1999}, finite element methods \cite{StraussPDEbook}, relaxation methods \cite{Press}, and methods that employ orthogonal function expansions such as Zernike polynomials \cite{GureyevZernikePaper1995} or Fourier harmonics \cite{Gureyev1997}. If any one of the available methods is used to solve for the auxiliary function $u(x,y)$, then Eq.~(\ref{eq:AuxiliaryFunctionDefinition}) tells us the required diffusion field $D(x,y)$ is given by $u(x,y)/I(x,y,z=0)$. Moreover, since $I(x,y,z=0)$ is assumed to be strictly positive over the field of view, there are no division-by-zero instabilities.   

As a formal symbolic summary, we write 
\begin{equation}
\label{eq:SymbolicSummaryInverseProblem1}
    D(x,y)=\frac{\hat{\mathcal{L}}^{-1}\rho(x,y)}{I(x,y,z=0)}, \quad\textrm{where}\quad\hat{\mathcal{L}}=-\nabla_{\perp}^2.
\end{equation}
Here, $\hat{\mathcal{L}}^{-1}$ denotes the inverse of the negative Laplacian operator $\hat{\mathcal{L}}$, as computed using any of the previously listed methods, consistent with the stated boundary conditions for the auxiliary function.

\subsection{First inverse problem: Fourier-series solution}\label{sec:FourierSeriesSolutionForPoissonEquation}

As an indicative example of the solution to our first inverse problem $\mathcal{I}_1$, which in our view is likely to be of use when analyzing pixelated data that are experimentally obtained using digital cameras, let us consider a simple Fourier-series approach. 

We write the solution to Eq.~(\ref{eq:BVPforScalarDiffusionFieldRetrieval}) in the form
\begin{equation}
\label{eq:PoissonEqnWithZeroDirichletBoundaryConditions}
u(x,y)=v(x,y)+\tfrac{1}{2}\theta_0^2I_u.
\end{equation}
Here, $v(x,y)$ denotes the solution to the Poisson equation for zero Dirichlet boundary conditions, so that
\begin{equation}
\label{eq:BoundaryConditionForV}
\begin{cases}
  -\nabla_{\perp}^2v(x,y) = \rho(x,y),   \\
  [v(x,y)]_{\partial\Omega_{\textrm{FOV}}} =0.
\end{cases}
\end{equation}

Assume the data function $\rho(x,y)$ to have been constructed, via Eq.~(\ref{eq:DataFunction}), based on measurements of the intensity in the planes $z=0$ and $z=\Delta$.  Assume these intensity measurements to cover a rectangular field of view $\Omega_{\textrm{FOV}}$, via a digital camera with $(J+1) \times (L+1)$ equally sized square pixels.  The physical width of each pixel, in units of distance, is $\mathcal{W}$. The center of each pixel in the digital camera  corresponds to the coordinate lattice
\begin{equation}
\label{eq:LatticeOfPixelCoordinates}
(x_j,y_l)=(j\mathcal{W},l\mathcal{W}),\quad \textrm{where }~\begin{cases}
  j=0,1,\cdots,J,   \\
  l=0,1,\cdots,L.
\end{cases}
\end{equation}
The data function is estimated over all of the interior points of this lattice, with the zero boundary conditions being associated with the boundary points of the lattice. 

We now follow a standard approach to solving the Poisson equation with zero boundary conditions.\footnote{See e.g.~pp~1024, 1027, and 1054-1055 of \citet{Press}, upon which Eqs.~(\ref{eq:FiniteDifferencePoissonEquaton})-(\ref{eq:SineTransformSolution}) of the main text are modeled. } Let
\begin{equation}
\label{eq:ShorthandForFunctionsOnLattice}
    v_{j,l}\equiv v(x_j,y_l) \quad \textrm{and}\quad \rho_{j,l}\equiv \rho(x_j,y_l),
\end{equation}
hence the second-order finite-difference form of the first line of Eq.~(\ref{eq:BoundaryConditionForV}) may be written as
\begin{equation}
\label{eq:FiniteDifferencePoissonEquaton}
    v_{j+1,l}+v_{j-1,l}+v_{j,l+1}+v_{j,l-1}-4v_{j,l}
    = -\mathcal{W}^2\rho_{j,l}.
\end{equation}
The required solution is  expressed as a Fourier sine series
\begin{equation}
\label{eq:SineTransformForward}
    v_{j,l}=\frac{4}{JL}
    \sum_{m=1}^{J-1}
    \sum_{n=1}^{L-1}
    {\breve{v}}_{m,n}
    \sin\frac{\pi j m}{J}
    \sin\frac{\pi l n}{L},
\end{equation}
where  $\breve{v}$ is the Fourier sine transform of $v$.  The form of the above equation guarantees that $v_{j,l}$ vanishes at the boundary points where $j=0,J$ or $l=0,L$, as required by our zero Dirichlet boundary conditions in Eq.~(\ref{eq:BoundaryConditionForV}).  Similarly, our data function may be written as
\begin{equation}
\label{eq:SineTransformForward2}
    \rho_{j,l}=\frac{4}{JL}
    \sum_{m=1}^{J-1}
    \sum_{n=1}^{L-1}
    {\breve{\rho}}_{m,n}
    \sin\frac{\pi j m}{J}
    \sin\frac{\pi l n}{L}.
\end{equation}

Substitute Eqs.~(\ref{eq:SineTransformForward}) and (\ref{eq:SineTransformForward2}) into Eq.~(\ref{eq:FiniteDifferencePoissonEquaton}), to give
\begin{equation}
\label{eq:SineTransformSolution}
    \breve{v}_{m,n}=
    \frac{\tfrac{1}{2}\mathcal{W}^2{\breve{\rho}}_{m,n}}{2
    -\cos\frac{\pi m}{J}
    -\cos\frac{\pi n}{L}
    }.
\end{equation}
Applying an inverse sine transform to $\breve{v}_{m,n}$ yields $v_{j,l}$, which may then be substituted into Eq.~(\ref{eq:PoissonEqnWithZeroDirichletBoundaryConditions}) to give
\begin{equation}
u(x_j,y_l)=v(x_j,y_l)+\tfrac{1}{2}\theta_0^2I_u.
\end{equation}
Finally, we employ Eq.~(\ref{eq:AuxiliaryFunctionDefinition}) to transform our auxiliary function $u$ into the required diffusion field.  Hence
\begin{equation}
D(x_j,y_l)=\frac{u(x_j,y_l)}{I(x_j,y_l,z=0)}.
\end{equation}
The recovered diffusion field $D$ is unique, since (i) the Poisson equation for specified zero Dirichlet boundary conditions, namely Eq.~(\ref{eq:BoundaryConditionForV}), is uniquely soluble for $v$ (see e.g.~pp~37-38 of Ref.~\cite{JacksonElectrodynamicsBook} or pp~149-150 of Ref.~\cite{StraussPDEbook}) and (ii) the chain of simple algebraic equations, that leads from $v$ to $u$ and thence to $D$, yields a one-to-one correspondence between $v$ and $D$.

Our unique-recovery Fourier-series solution, to the first inverse problem $\mathcal{I}_1$, may be written in the compact form
\begin{equation}
\label{eq:InverseProblem1==FormalSolution}
D_{j,l}=\frac{
\hat{\mathcal{S'}}_{m\rightarrow j}^{n\rightarrow l}\left[\Phi_{m,n}\left(\hat{\mathcal{S}}_{j\rightarrow m}^{l\rightarrow n}\rho_{j,l}\right)\right] + \tfrac{1}{2}\theta_0^2I_u
}
{
I_{j,l}(z=0)
}.
\end{equation}
Here, ${\hat{\mathcal{S}}}_{j\rightarrow m}^{l\rightarrow n}$ denotes the forward Fourier sine transform using the convention implied by  Eq.~(\ref{eq:SineTransformForward}), $\hat{\mathcal{S'}}_{m\rightarrow j}^{n\rightarrow l}$ is the corresponding inverse sine transform, $\Phi_{m,n}$ is the filter
\begin{equation}
\label{eq:FilterForFirstInverseProblem}
    \Phi_{m,n}=
    \frac{\tfrac{1}{2}\mathcal{W}^2}{2
    -\cos\frac{\pi m}{J}
    -\cos\frac{\pi n}{L}
    },
\end{equation}
and we have introduced the abbreviation
\begin{equation}
\label{eq:AbbreviationForInfocusIntensityMap}
I_{j,l}(z=0) \equiv I(x_j,y_l,z=0).    
\end{equation}

The diffusion-field retrieval in Eqs.~(\ref{eq:InverseProblem1==FormalSolution})-(\ref{eq:FilterForFirstInverseProblem}) is equivalent to the following sequence of steps:
\begin{enumerate}
\item Use a square-pixel digital camera to record the intensity maps $I(x_j,y_l,z=0)$ and $I(x_j,y_l,z=\Delta\ge 0)$, corresponding to infocus and defocused images in Fig.~\ref{Fig:SimpleImagingWithSourceBlurAndSampleBlur}, respectively.  Here, $(x_j,y_l)$ corresponds to the lattice of pixel coordinates in Eq.~(\ref{eq:LatticeOfPixelCoordinates}).
\item Use these two intensity measurements to form the data function $\rho(x_j,y_l)$ in Eq.~(\ref{eq:DataFunction}).
\item Take the Fourier sine transform of the data function, e.g.~using the Fast Sine Transform\footnote{See e.g.~pp~620-623 of \citet{Press}.}. 
\item The resulting array, which is indexed by the pair of integers $(m,n)$, is then multiplied in a pointwise manner, by the filter $\Phi_{m,n}$ in Eq.~(\ref{eq:FilterForFirstInverseProblem}).
\item Take the inverse Fourier sine transform of the resulting array.
\item Add the constant $\tfrac{1}{2}\theta_0^2 I_u$ to every element of the resulting array.  Here, $\theta_0$ is the half-angle that the extended chaotic source subtends at the entrance surface of the thin sample (see Fig.~\ref{Fig:SimpleImagingWithSourceBlurAndSampleBlur}), and $I_u$ is the uniform intensity with which, by assumption, the sample is illuminated.
\item The resulting array, which is indexed by the integers $(j,l)$, is then divided by the zero-defocus intensity map $I_{j,l}(z=0)$ (see Eq.~(\ref{eq:AbbreviationForInfocusIntensityMap})), in a pointwise manner.  This yields the required diffusion field
\begin{equation}
D_{j,l}\equiv D(x_j,y_l),    
\end{equation}
over the same pixelated grid that was used to obtain the measured intensity maps in Step 1.
\end{enumerate}

\subsection{First inverse problem: Alternative formulations}
\label{sec:AlternativeFormulationsForDiffusionField}

Some further flexibility, regarding the inverse problem of determining the diffusion field from suitable intensity measurements, arises upon realizing that the representation of the diffusion field is not unique.  In particular, the diffusion term $\nabla_{\perp}^2 [D(x,y)  I(x,y,z=0)]$ in Eq.~(\ref{eq:2DVersionOfSimpleBlurWithGeneralisation}) may be used to define a different representation $\widetilde{D}(x,y)$ for the diffusion field, which obeys
\begin{align}
\nonumber  \nabla_{\perp}^2[&D(x,y) \, I(x,y,z=0)] \\ &\equiv \nabla_{\perp}\cdot [\widetilde{D}(x,y) \nabla_{\perp} I(x,y,z=0)].
\label{eq:ChangeOfRepresentationForDiffusionField--1}
\end{align}
With this representation for the diffusion field, Eq.~(\ref{eq:2DVersionOfSimpleBlurWithGeneralisation}) may be written in the alternative form (cf.~Eq.~(\ref{eq:TensorDiffusionFieldRetrieval8}))
\begin{align}
\nonumber I(&x,y,z=\Delta \ge 0) =I(x,y,z=0) \\  
 &+\Delta^2\nabla_{\perp}\cdot [\widetilde{D}(x,y) \nabla_{\perp} I(x,y,z=0)].
\label{eq:2DVersionOfSimpleBlurWithGeneralisation==ALTERNATIVE-FORM}
\end{align}

Suppose we choose to work with Eq.~(\ref{eq:2DVersionOfSimpleBlurWithGeneralisation==ALTERNATIVE-FORM}), with intensity measurements $I(x,y,z=\Delta \ge 0)$ and $I(x,y,z=0)$ being used to recover the representation of the diffusion field that is given by $\widetilde{D}(x,y)$. If required, we may make the following uniquely specified change of representation
\begin{equation}
\label{eq:MappingBetweenRepresentationsForD}
 \widetilde{D}(x,y)\longrightarrow D(x,y),
\end{equation}
as follows.  Given the known function $\widetilde{D}(x,y)$, which has been recovered using any suitable solution to Eq.~(\ref{eq:2DVersionOfSimpleBlurWithGeneralisation==ALTERNATIVE-FORM}), compute the right side of Eq.~(\ref{eq:ChangeOfRepresentationForDiffusionField--1}).  The resulting Poisson equation for $D(x,y) I(x,y,z=0)$ (together with the associated constant Dirichlet boundary conditions) is mathematically identical to Eqs.~(\ref{eq:AuxiliaryFunctionDefinition}) and (\ref{eq:BVPforScalarDiffusionFieldRetrieval}), if the data function $\rho$ is replaced by $-\nabla_{\perp}\cdot [\widetilde{D}(x,y) \nabla_{\perp} I(x,y,z=0)]$.  Therefore, in the notation of Eq.~(\ref{eq:SymbolicSummaryInverseProblem1}), the change of representation in Eq.~(\ref{eq:MappingBetweenRepresentationsForD}) has the explicit form
\begin{equation}
\label{eq:SymbolicSummaryConvertingBetweenRepresentations}
    D(x,y)=-\frac{\hat{\mathcal{L}}^{-1}\left\{\nabla_{\perp}\cdot \left[\widetilde{D}(x,y) \nabla_{\perp} I(x,y,z=0)\right]\right\}}{I(x,y,z=0)}.
\end{equation}
This change of representation is unique because the underpinning solution to the Poisson-equation boundary value problem, represented by $\hat{\mathcal{L}}^{-1}$, has a unique solution for the specified Dirichlet boundary conditions.

The ``$D$ representation'' and ``$\widetilde{D}$ representation'' become equivalent when the diffusion field is slowly varying with respect to transverse position.  Under this approximation, the diffusion field commutes with spatial derivative operators to a reasonable degree of accuracy, hence both Eqs.~(\ref{eq:2DVersionOfSimpleBlurWithGeneralisation}) and (\ref{eq:2DVersionOfSimpleBlurWithGeneralisation==ALTERNATIVE-FORM}) reduce to
\begin{align}
\label{eq:2DVersionOfSimpleBlurWithGeneralisation==SLOWLY-VARYING-D}
I(&x,y,z=\Delta \ge 0) \approx I(x,y,z=0)
\\ \nonumber &+\Delta^2D(x,y)\nabla_{\perp}^2I(x,y,z=0) \quad\textrm{(}D\textrm{~slowly varying)}.
\end{align}

In formulations for the inverse problem of diffusion-field retrieval, the ``$D$ representation'' in Eq.~(\ref{eq:2DVersionOfSimpleBlurWithGeneralisation}) leads to second-order partial differential equations for the diffusion field.  Conversely, the ``$\widetilde{D}$ representation'' in Eq.~(\ref{eq:2DVersionOfSimpleBlurWithGeneralisation==ALTERNATIVE-FORM}) leads to partial differential equations that are of first order with respect to spatial derivatives of the diffusion field. While the two representations for the diffusion field are  physically and mathematically equivalent, they may be numerically inequivalent from the perspective of computational speed or stability with respect to noise in the input intensity data.  Moreover, for the purposes of visual inspection or automated evaluation of a recovered diffusion field, the different representations $D(x,y)$ and $\widetilde{D}(x,y)$ may have different qualitative utility, e.g., regarding the ability of either human or machine vision to isolate and visualize morphological features of interest.  Finally, if the diffusion field is a sufficiently slowly varying function of position, our partial differential equations reduce to the algebraic equation in Eq.~(\ref{eq:2DVersionOfSimpleBlurWithGeneralisation==SLOWLY-VARYING-D}), when the intensity measurements are considered as the known functions and the diffusion field is taken to be unknown.

\subsection{First inverse problem: Further examples, employing one or more masks}

Below we give four further examples of diffusion-field retrieval, all of which employ one or more beam-shaping masks.  We use the term ``mask'' to indicate a thin transmissive object whose degree of transmission depends on the transverse coordinates.  This enables controllably-structured incident illumination.  Such a structured-illumination approach is motivated by the use of masks to visualize refractive-index variations, in contexts such as visible-light imaging with periodic-pattern structured illumination \cite{Massig1,Perciante, Massig2}, x-ray single-grid imaging \cite{wen2010,morgan2011quantitative}, and x-ray speckle tracking using spatially-random illumination \cite{berujon2012,Morgan2012}.  While the x-ray techniques in the previous sentence have now been extended to include the simultaneous presence of both refractive and diffusive effects (see e.g.~the review article by \citet{zdora2018}, together with references therein), we here consider the simpler case where refractive effects are absent (cf.~Sec.~\ref{Sec:Discussion--FokkerPlanckExtension}).  In this sense, the following examples may be viewed as diffusion-only variants of speckle tracking and single-grid imaging.     

\begin{figure}[ht!]
\centering
\includegraphics[width=0.99\columnwidth]{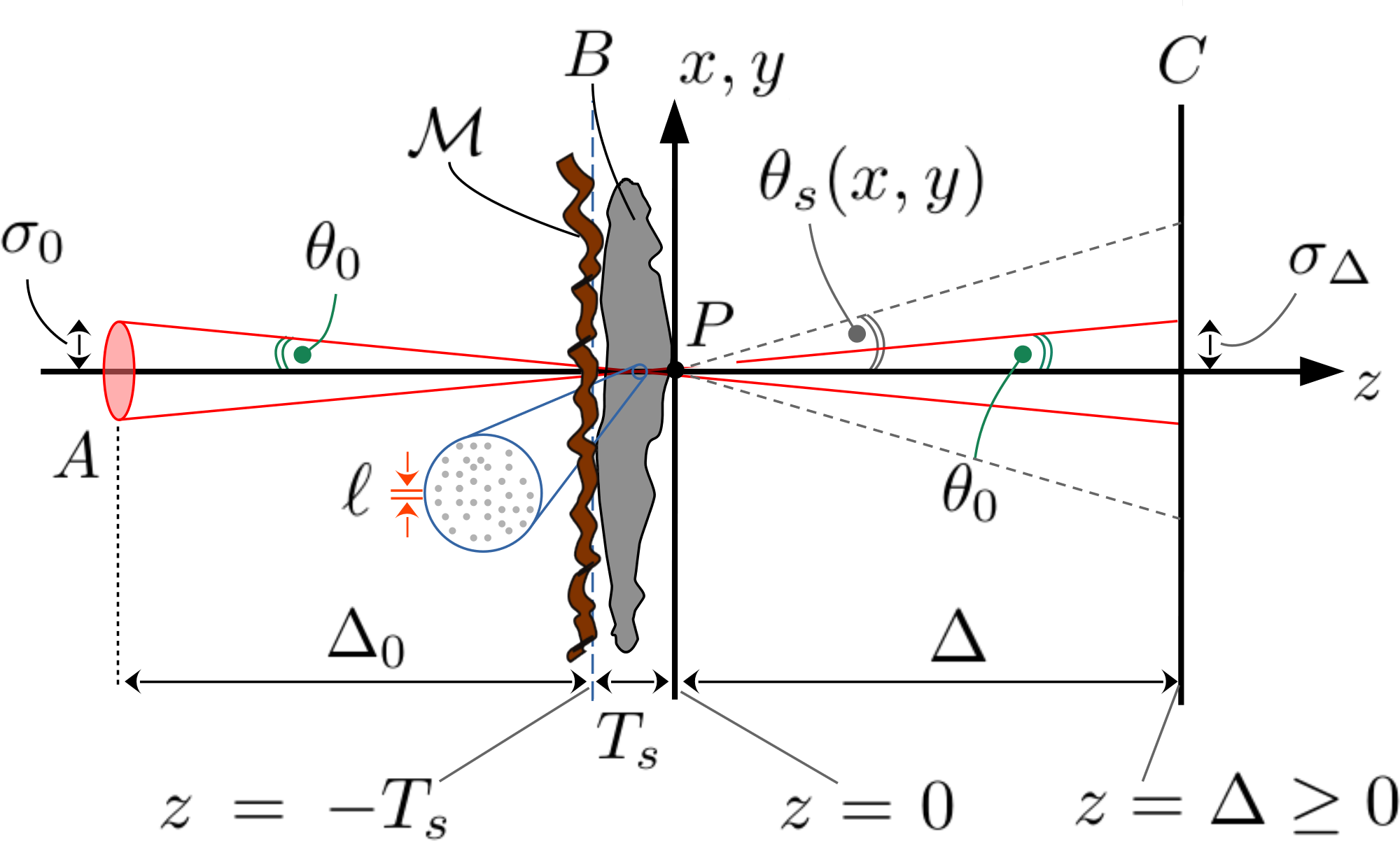}
\caption{Modification of Fig.~3, to include structured illumination using a mask $\mathcal{M}$.  Note, the mask $\mathcal{M}$ may also be placed at the exit surface $z=0$ of the thin sample.}
\label{Fig:SimpleImagingWithSourceBlurAndSampleBlurAndMask}
\end{figure}

Consider Fig.~\ref{Fig:SimpleImagingWithSourceBlurAndSampleBlurAndMask}.  This differs from the preceding situation we have considered, as there is now a mask $\mathcal{M}$ with intensity transmission function $M(x,y)$, located in the plane $z=-T_s$ corresponding to the nominally-planar entrance surface of the thin sample.  Note, the subsequent analysis is unchanged if we instead place the mask in the plane $z=0$ at the exit surface of this sample.  The mask transmission function, which lies between zero and unity at each transverse location $(x,y)$, may correspond to 
\begin{itemize}
\item a spatially-periodic structure such as a grid \cite{Perciante,Massig2,wen2010,morgan2011quantitative}, 
\item a spatially-random structure such as a speckle field \cite{berujon2012,Morgan2012,zdora2017}, a random fractal \cite{SethnaBook,Kingston2022arXiv}, or a binary random mask, or 
\item precisely engineered masks such as modified uniformly redundant arrays \cite{Gottesman1989newFamily}. 
\end{itemize}

The key purpose of the mask is to introduce a spatially rapidly varying transmission function, which amplifies the measured signal for the purposes of diffusion-field retrieval.  This amplification arises because finely-detailed intensity features, in the plane $z=0$, are more sensitive to the microstructure-induced position-dependent blur---when propagating from the plane $z=0$ to the plane $z=\Delta$---compared to intensity distributions that are more slowly spatially varying. Note, the form of the following equations is insensitive to the particular choice of mask.    

\subsubsection{Example: One mask, narrowband illumination, nonconstant sample transmission function} 

This first example begins by considering a  single-mask version of Fig.~\ref{Fig:SimpleImagingWithSourceBlurAndSampleBlurAndMask}.  Assume the mask $\mathcal{M}$ to be constructed so as to induce negligible microstructure-induced blur.  The parallel illumination is taken to be narrowband, with a spatially uniform intensity of unity, immediately upstream of the mask and sample.  With these assumptions in place,  we use Eq.~(\ref{eq:2DVersionOfSimpleBlurWithGeneralisation}) to write the intensity in the plane $z=\Delta > 0$, in the presence of the thin sample, as
\begin{equation}
\label{eq:ScalarDiffusionWithMasks1}
I_{\Delta}^{(s+m)}=TM +\Delta^2\nabla_{\perp}^2(DTM).
\end{equation}
Here, suppressing functional dependence on transverse position $(x,y)$ for clarity, we use the superscript ``$s+m$'' to indicate that both the \underline{s}ample and \underline{m}ask are present, the subscript $\Delta$ to indicate the propagation distance from the thin sample to the detector, $T\equiv T(x,y)$ to denote the intensity transmission function of the sample, and $M\equiv M(x,y)$ to denote the intensity transmission function of the mask $\mathcal{M}$.  If we now use the expression for the diffusion field $D\equiv D(x,y)$ in Eq.~(\ref{eq:PositionDependentScalarDiffusionCoefficient}), we obtain   
\begin{equation}
\label{eq:ScalarDiffusionWithMasks2}
I_{\Delta}^{(s+m)}=TM +\tfrac{1}{2}\Delta^2\theta_0^2\nabla_{\perp}^2(TM)
+ \tfrac{1}{2}\Delta^2\nabla_{\perp}^2(TMF\theta_s^2).
\end{equation}
Also, if the sample is absent while the mask is present, as denoted by the superscript ``$m$'', we may set $T=1$ and $\theta_s=0$ in the above expression, to give 
\begin{equation}
\label{eq:ScalarDiffusionWithMasks3}
I_{\Delta}^{(m)}=M +\tfrac{1}{2}\Delta^2\theta_0^2\nabla_{\perp}^2M.
\end{equation}
Subtract Eq.~(\ref{eq:ScalarDiffusionWithMasks3}) from Eq.~(\ref{eq:ScalarDiffusionWithMasks2}), hence
\begin{align}
\nonumber I_{\Delta}^{(s+m)}-I_{\Delta}^{(m)}=&M(T-1) +\tfrac{1}{2}\Delta^2\theta_0^2\nabla_{\perp}^2[(T-1)M] \\ &+\tfrac{1}{2}\Delta^2\nabla_{\perp}^2(TMF\theta_s^2).
\label{eq:ScalarDiffusionWithMasks4}
\end{align}
The data function
\begin{equation}
\label{eq:ScalarDiffusionWithMasks5}
 \varrho=\frac{2}{\Delta^2}[I_{\Delta}^{(m)}-I_{\Delta}^{(s+m)}+M(T-1)] + \theta_0^2\nabla_{\perp}^2[(T-1)M]
\end{equation}
may be constructed using intensity measurements in the plane $z=
\Delta$ (for $I_{\Delta}^{(m)}$ and $I_{\Delta}^{(s+m)}$) and the plane $z=0$ (for $M$ and $TM$).  Regarding these intensity measurements, note that (i) $I_{\Delta}^{(m)}$ and $M$ require the mask to be present in the absence of the sample, with (ii) $TM$ and $I_{\Delta}^{(s+m)}$ requiring both the mask and the sample to be present.  Upon employing these measurements to construct the data function $\varrho$, Eq.~(\ref{eq:ScalarDiffusionWithMasks4}) becomes
\begin{equation}
\label{eq:ScalarDiffusionWithMasks6}
-\nabla_{\perp}^2(TMF\theta_s^2)=\varrho.
\end{equation}
Hence, using the same notation as Eq.~(\ref{eq:SymbolicSummaryInverseProblem1}), and assuming that $M\equiv M(x,y)$ never vanishes at any point in the field of view, we have the diffusion-field retrieval formula
\begin{equation}
\label{eq:ScalarDiffusionWithMasks7}
    \sqrt{F(x,y)}\,\theta_s(x,y) \!=\! \sqrt{\frac{\hat{\mathcal{L}}^{-1}\varrho(x,y)}{T(x,y) \, M(x,y)}},~\textrm{where}~\hat{\mathcal{L}} \!=\!-\nabla_{\perp}^2.
\end{equation}
While $\hat{\mathcal{L}}^{-1}$ can be computed in any suitable manner, the Fourier sine series solution in Sec.~\ref{sec:FourierSeriesSolutionForPoissonEquation} may be useful when the measured intensity data are obtained on a pixelated grid.  Note, also, that: 
\begin{itemize}
\item In Case \#1---as defined in Eqs.~(\ref{eq:CaseOneFresnelNumberCondition}) and (\ref{eq:TotalBlurAngleCase1})---we may set $F(x,y)=1$ in Eq.~(\ref{eq:ScalarDiffusionWithMasks7}).  In this case, the sample-induced blurring angle $\theta_s(x,y)$ is recovered.
\item In Case \#2---as defined in Eqs.~(\ref{eq:CaseTwoFresnelNumberCondition}), (\ref{eq:F-much-less-than-one-for-case-2}) and (\ref{eq:TotalBlurAngleCase2})---the method is only able to recover $\sqrt{F(x,y)}\,\theta_s(x,y)$.  This recovered quantity plays the role of the effective blur angle\footnote{If $\theta_s(x,y)$ is on the order of unity, but $F$ is much less than unity, the effective blur angle will be sufficiently small to maintain paraxiality.  This observation might be useful for incoherent scattering scenarios, such as neutron scattering from nuclei, where the scatter is isotropic but the scattering fraction $F$ is much less than unity. This raises an important broader point: while the emphasis in the current paper is on the role of unresolved spatially-random microstructure in producing diffuse scatter, there are other mechanisms for generating diffuse scatter, to which the methods of the current paper might be applied. \label{footnote:IncoherentScatterScenarios}} %
\begin{equation}
\label{eq:EffectiveBlurAngle}
\theta_{\textrm{eff}}(x,y)=\sqrt{F(x,y)}\,\theta_s(x,y).
\end{equation}
The individual fields, $\theta_s(x,y)$ and $F(x,y)$, are not recovered by the method.  
\end{itemize}

\subsubsection{Example:  One mask, narrowband illumination, nonabsorbing sample} 

Consider a special case of the previous example, by assuming the sample to have negligible attenuation.  Hence we may set $T=1$ in Eq.~(\ref{eq:ScalarDiffusionWithMasks4}).  The first two terms on the right side disappear, leaving 
\begin{equation}
\label{eq:ScalarDiffusionWithMasks8}
I_{\Delta}^{(s+m)}-I_{\Delta}^{(m)}=\tfrac{1}{2}\Delta^2\nabla_{\perp}^2(MF\theta_s^2) \quad\textrm{if~}T=1.
\end{equation}
We may think of this as a form of continuity equation, in which a flow associated with diffusive current density blurs $I_{\Delta}^{(m)}$ into $I_{\Delta}^{(s+m)}$, with position-dependent smearing that locally conserves the number of registered imaging quanta (e.g., photons, electrons, neutrons etc.).  Under this view,  Eq.~(\ref{eq:ScalarDiffusionWithMasks8}) may be viewed as a diffusion-only version of the refraction-only ``geometric flow'' continuity equation that underpins  Ref.~\cite{PaganinLabrietBrunBerujon2018}.  Comparisons aside, the diffusion-field retrieval formula is (cf.~Eq.~(\ref{eq:EffectiveBlurAngle}))
\begin{align}
\label{eq:ScalarDiffusionWithMasks9}
    &\sqrt{F(x,y)}\,\theta_s(x,y) \\ \nonumber \quad&=\frac{1}{\Delta}\sqrt{\frac{\hat{\mathcal{L}}^{-1}[I_{\Delta}^{(m)}(x,y)-I_{\Delta}^{(s+m)}(x,y)]}{\tfrac{1}{2} M(x,y)}}\quad\textrm{if~}T=1.
\end{align}
Both the transmission function $M(x,y)$ of the single mask $\mathcal{M}$, and the blurred propagated image $I_{\Delta}^{(m)}(x,y)$ of the mask in the absence of the sample, can be measured once and for all.  Further to these once-off measurements, a single additional image is needed in the presence of the sample, namely $I_{\Delta}^{(s+m)}(x,y)$, in order for the diffusion-field retrieval to be performed. This is useful, because it allows us to work with time-dependent data $I_{\Delta}^{(s+m)}(x,y,t)$, with each image frame at each time $t$ able to be analyzed independently, using Eq.~(\ref{eq:ScalarDiffusionWithMasks9}).  This enables, at least in principle, the diffusion-field retrieval of temporally-evolving thin non-absorbing samples.

\subsubsection{Example:  One mask, narrowband polychromatic illumination, nonabsorbing sample}

Our preceding calculations all assume narrowband illumination.  However, given that the notion of a diffusion field does not require a coherent field's capacity to exhibit interference phenomena \cite{Zernike1938}, it is natural to ask whether the method of diffusion-field retrieval can be extended to the case of polychromatic (polyenergetic) illumination.  To examine this point, we revisit the previous example, showing how it may be generalized to the polychromatic case.  Denote the dependence of any quantity on wavelength by a subscript $\lambda$.  The normalized illumination spectrum is  
\begin{equation}
\label{eq:ScalarDiffusionWithMasks10}
S_{\lambda}\equiv S(\lambda), \quad \textrm{where~}\int_{0}^{\infty} S_{\lambda}\,\mathrm{d}\lambda=1\textrm{~and~}S_{\lambda}\ge 0.
\end{equation}
Denote the energy-dependent detector efficiency by
\begin{equation}
\label{eq:ScalarDiffusionWithMasks11}
\mathcal{E}_{\lambda}\equiv \mathcal{E}(\lambda), 
\quad \textrm{where~}
0\le\mathcal{E}_{\lambda}\le 1.
\end{equation}
Denote efficiency-weighted spectrally-averaged quantities with an overline, e.g., the measured propagated polychromatic image in the presence of both mask and sample is
\begin{equation}
\label{eq:ScalarDiffusionWithMasks12}
\overline{I_{{\Delta},\lambda}^{(s+m)}} = \int_{0}^{\infty} \mathcal{E}_{\lambda}I_{\Delta,\lambda}^{(s+m)}(x,y)\,\mathrm{d}\lambda.
\end{equation}
Spectrally averaging Eq.~(\ref{eq:ScalarDiffusionWithMasks8}) gives
\begin{equation}
\label{eq:ScalarDiffusionWithMasks13}
\overline{I_{\Delta,\lambda}^{(s+m)}}
-\overline{I_{\Delta,\lambda}^{(m)}}
=\tfrac{1}{2}\Delta^2\nabla_{\perp}^2\left(\overline{M_{\lambda} F_{\lambda} \theta_{s,\lambda}^2}\right), \quad T_{\lambda}=1.
\end{equation}
However, the formal solution
\begin{equation}
\label{eq:ScalarDiffusionWithMasks14}
\overline{M_{\lambda} F_{\lambda} \theta_{s,\lambda}^2}=(-\nabla_{\perp}^2)^{-1}\left[\,\frac{\overline{I_{\Delta,\lambda}^{(m)}}-\overline{I_{\Delta,\lambda}^{(s+m)}}}{\tfrac{1}{2}\Delta^2}\,\right],  \quad T_{\lambda}=1
\end{equation}
now has the problem that the influence of the mask $M_{\lambda}$ cannot be disentangled from the left side, in general, to yield a diffusion-field recovery that is independent of the mask.  As one way of remedying this, suppose that the spread of illumination energies is sufficiently small, such that for the particular material or materials from which the mask is composed, $M_{\lambda}$ is approximately independent of $\lambda$.  We may then replace $M_{\lambda}$ with $M_{\lambda_0}$ in Eq.~(\ref{eq:ScalarDiffusionWithMasks14}), where $\lambda_0$ is the mean illumination wavelength.   Hence we may now disentangle the influence of the mask, to give a narrowband polychromatic generalization of Eq.~(\ref{eq:ScalarDiffusionWithMasks9}) as
\begin{equation}
\label{eq:ScalarDiffusionWithMasks15}
\overline{ F_{\lambda} \theta_{s,\lambda}^2}=\frac{2}{\Delta^2 M_{\lambda_0}}(-\nabla_{\perp}^2)^{-1}\left[\,\overline{I_{\Delta,\lambda}^{(m)}}-\overline{I_{\Delta,\lambda}^{(s+m)}}\,\right],  \quad T_{\lambda}=1.
\end{equation}

\subsubsection{Example:  Multiple binary masks, broadband polychromatic illumination, nonabsorbing sample}

We now consider broadband polychromatic illumination, of a sample which is nonabsorbing at all wavelengths for which the incident spectrum $S_{\lambda}$ is non-negligible.  For this example, suppose we have an ensemble of $N$ masks $\{\mathcal{M}_j\}$, where $j=1,2,\cdots,N$.  By assumption, each mask has the property that its transmission function is either completely transmitting or completely absorbing, at (i) each transverse position $(x,y)$, and (ii) every wavelength for which $S_{\lambda}$ is non-negligible.  Such a set of masks could comprise, for example, a set of totally opaque sheets, in each of which a different spatially-random series of holes is drilled. The mask ensemble must be such that, for each transverse position in the field of view, at least one of the masks has an opening at that location. A key feature, of this set of masks, is that the associated set of transmission functions $\{M_{j}\}$ will be independent of wavelength, over the range of wavelengths that are present in the illumination.  Moreover, by construction, the only values that any one mask transmission function can take, at any one transverse location, are either zero or unity. With all of these assumptions in place, we may rewrite Eq.~(\ref{eq:ScalarDiffusionWithMasks13}), for the $j$th binary mask, as           
\begin{equation}
\label{eq:ScalarDiffusionWithMasks16}
\overline{I_{\Delta,\lambda,j}^{(s+m)}}
-\overline{I_{\Delta,\lambda,j}^{(m)}}
=\tfrac{1}{2}\Delta^2\nabla_{\perp}^2\left(M_j\overline{ F_{\lambda} \theta_{s,\lambda}^2}\right), \quad T_{\lambda}=1.
\end{equation}
Note that the spectral-average overline on the right side does not include the mask transmission function:~the fact, that each mask's transmission function is either zero or unity, is an energy-independent construction. Next, write down the formal solution for $M_j\overline{ F_{\lambda} \theta_{s,\lambda}^2}$, sum $j$ over the set of $N$ masks, and then isolate $\overline{ F_{\lambda}\theta_{s,\lambda}^2}$.  Hence
\begin{equation}
\label{eq:ScalarDiffusionWithMasks17}
\,\overline{ F_{\lambda} \theta_{s,\lambda}^2}=\frac{\sum_{j=1}^N(-\nabla_{\perp}^2)^{-1}\left[\,{\overline{I_{\Delta,\lambda,j}^{(m)}}-\overline{I_{\Delta,\lambda,j}^{(s+m)}}}\,\right]}{\tfrac{1}{2}\Delta^2\sum_{j=1}^N M_j}, \quad T_{\lambda}=1, 
\end{equation}
where the denominator never vanishes.

\subsection{Second inverse problem: General remarks}

We now turn to the second inverse problem, $\mathcal{I}_2$, as schematically indicated in Eq.~(\ref{eq:SchematicChainOfTwoInverseProblemsForScalarD}).  This deals with the question of taking a retrieved diffusion field $D(x,y)$, and recovering the characteristic length scale $\ell(x,y)$ (see Figs.~\ref{Fig:SimpleImagingWithSourceBlurAndSampleBlur} and \ref{Fig:SimpleImagingWithSourceBlurAndSampleBlurAndMask}) associated with the spatially unresolved random microstructure in the sample \cite{Morgan2022Preprint}. In the absence of any additional knowledge regarding the sample, this second inverse problem is not in general soluble.  Accordingly, below we need to introduce specific models for the spatially-random microstructure, in the context of obtaining $\ell(x,y)$ from the diffusion field that this microstructure imparts to the illuminating beam.  

\subsection{Second inverse problem: Example for Case \#1}\label{sec:SecondInverseProblemExampleForCase1}

In Case \#1---as defined in Eqs.~(\ref{eq:CaseOneFresnelNumberCondition}) and (\ref{eq:TotalBlurAngleCase1})---a geometrical-optics model is assumed.  For wave fields such as visible light or x rays, this corresponds to a ray model in which every ray undergoes random transverse deflections upon traversing the spatially-random fluctuations in scalar refractive index within the volume of the sample.  For matter fields such as neutrons or electrons, this corresponds to a classical point-particle model in the presence of a spatially-random scalar potential, with the trajectory of each particle being akin to the evolving ray described in the previous sentence.  Irrespective of whether we work with radiation or matter models, we are again ignoring the effects of internal degrees of freedom, such as polarization for electromagnetic waves and the spin of electrons or neutrons.

\begin{figure}[ht!]
\centering
\includegraphics[width=0.95\columnwidth]{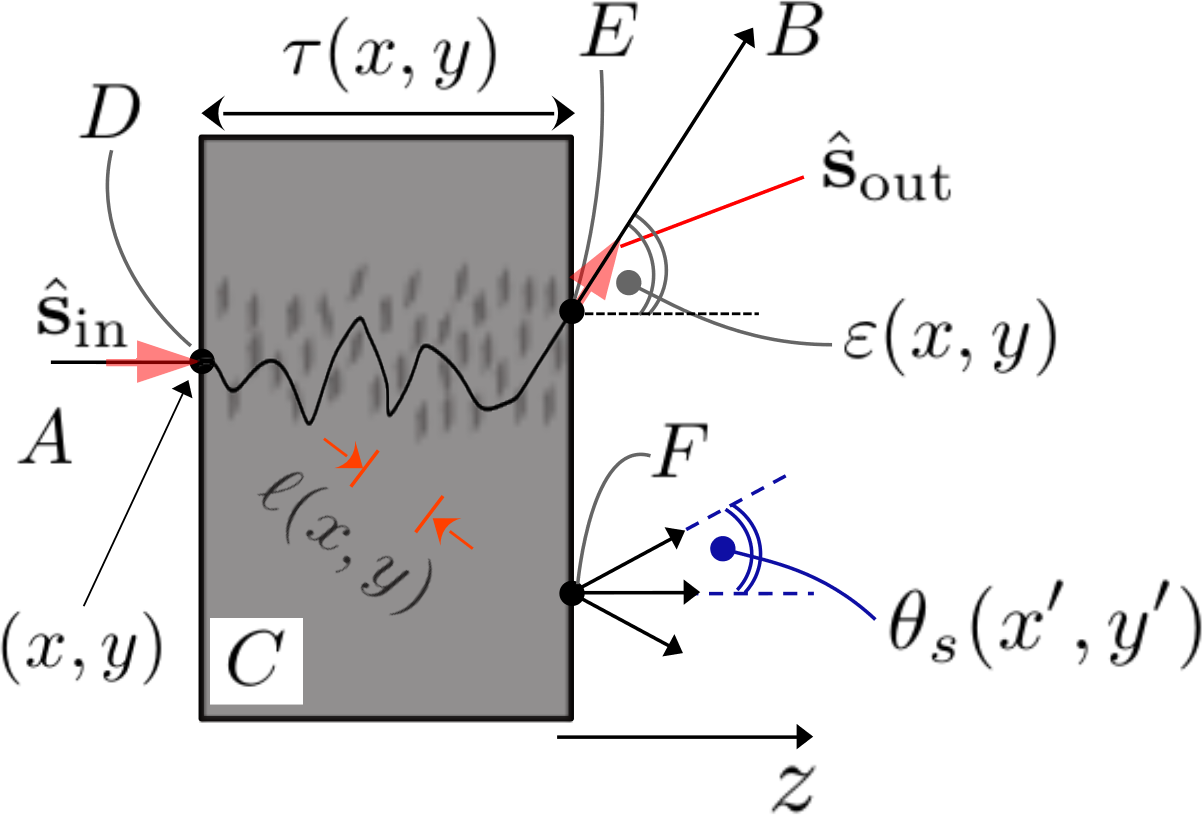}
\caption{Continuous random walk of a paraxial geometrical-optics ray as it traverses a spatially-random medium.  Note that the exit angle $\varepsilon(x,y)$ is exaggerated, since $\vert\varepsilon(x,y)\vert\ll 1$, by assumption.  Only a small part of the illuminated sample is shown, since the thickness $\tau(x,y)$ of the sample can vary with transverse position $(x,y)$.}
\label{Fig:ChernovModel}
\end{figure}

For clarity, we henceforth speak of rays in the context of geometrical optics.  Consider Fig.~\ref{Fig:ChernovModel}, in which the ray path $AB$ corresponds to a trajectory through a medium $C$ with spatially-random fluctuations in refractive index.  The ray direction incident upon the slab at $D$ is given by the unit vector $\hat{\mathbf{s}}_{\textrm{in}}$, with the output ray at $E$ traveling in the direction $\hat{\mathbf{s}}_{\textrm{out}}$.  For the indicated ray $AB$, the deflection angle $\varepsilon(x,y)$ will depend on the transverse coordinate of the point $D$ on the entrance surface of the slab.  At any one point on the exit surface, such as $F$ at the transverse coordinate $(x',y')$, there will be a diffusive cone of rays with root-mean-square (RMS) half-angle
\begin{equation}
    \label{eq:Chernov1}
    \theta_s(x',y')=\sqrt{E([\varepsilon(x',y')]^2)}.
\end{equation}
Here, $E()$ denotes the expectation value of the quantity in round brackets, calculated using an ensemble of realizations of the spatially-random fluctuations in refractive index \cite{Pedersen1976,FrozenPhononModel,MandelWolf,GoodmanSpeckleBook,Nesterets2008,KirklandBook}.  Also, we assume $E(\varepsilon(x,y))$ to vanish, in writing the above expression.  This corresponds to each diffusive cone, as mentioned earlier in this paragraph, having a zero average deflection angle (cf.~Sec.~\ref{Sec:Discussion--FokkerPlanckExtension}).  

For simplicity, assume the spatially-random medium to be non-absorbing, with refractive index distribution\footnote{See e.g.~p.~15 of \citet{ChernovBook1960}. For visible light, $\delta$ is typically positive, but not small compared to unity (see further details above).  For x rays, $\delta$ is typically negative (see e.g.~pp~108-114 of \citet{Paganin2006}).  For neutrons, the Fermi thin-slab formula \cite{FermiThinSlabFormula} implies that $\delta$ can have either sign, since (i) $\delta$ is approximately proportional to the neutron scattering length, and (ii) the neutron scattering length can be positive or negative (see e.g.~p~71 of \citet{SiviaScatteringTheoryBook2011}). Our final result does not depend on the sign of $\delta$.} 
\begin{equation}
    \label{eq:Chernov2}
n(x,y,z)=1+\delta(x,y,z), \quad \vert\delta(x,y,z)\vert\ll 1.
\end{equation}
The above expression is natural for both x rays and neutrons, where the refractive index is very close to unity.  In instances where this is not the case, such as for visible light, the right side should be replaced by $\overline{n}\,[1+\delta(x,y,z)]$, where $\overline{n}$ is the average refractive index.  For further simplicity, assume the refractive index fluctuation $\delta(x,y,z)$ has zero expectation value, when averaged over an ensemble of realizations, at any fixed position $(x,y,z)$ within the volume of the sample. Again, this simplifying assumption is easily dropped, if needed.  

While the ray trajectory in Fig.~\ref{Fig:ChernovModel} is smooth, for the purposes of this paragraph we break it up into straight line segments, that each have a length on the order of the correlation length $
\ell(x,y)$ for the spatially-random medium.  There will be on the order of
\begin{equation}
\label{eq:RoughEstimateForNumberOfAngularDeflections}
\mathcal{N}(x,y)\approx\frac{\tau(x,y)}{\ell(x,y)} 
\end{equation}
angular deflections of the ray as it traverses the medium, where $\tau(x,y)$ is the thickness of the sample along a paraxial ray with transverse coordinate $(x,y)$. Crudely considering each angular deflection to be associated with a prism of refractive index $1+\delta(x,y,z)$ in vacuum, with prism angle $\alpha$ on the order of unity, each angular deflection will be roughly $\delta(x,y,z)$ (in units of radians)\footnote{See e.g.~pp~195-198 of \citet{Paganin2006}.}.  Modeling the sample as a randomly packed ``bag'' of such prism-like elements, the average angular deflection from any one prism will be $\sqrt{E(\delta^2(x,y))}$, with the dependence on $z$ being dropped by assumption.  Adding the $\mathcal{N}(x,y)$ refraction angles on the order of $\sqrt{E(\delta^2(x,y))}$ in quadrature, as befits a random walk in deflection angle as the ray traverses the random medium, we obtain \cite{Beeman1947} (cf.~Ref.~\cite{Khelashvili2006})
\begin{align}
    \nonumber
    \theta_{s}(x,y) &\approx \sqrt{E(\delta^2(x,y))} \sqrt{\mathcal{N}(x,y)}
    \\ &=\sqrt{\frac{E(\delta^2(x,y))\tau(x,y)}{\ell(x,y)}}.
\label{eq:Chernov2.5}
\end{align}
Note that if anomalous rather than normal diffusion holds within the volume of the sample (cf.~Refs.~\cite{kitchen2020emphysema,how2022}), the factor of $\sqrt{\mathcal{N}(x,y)}$ in Eq.~(\ref{eq:Chernov2.5}) would need to be modified to a different positive power of $\mathcal{N}(x,y)$  \cite{EvangelistaLenziBook2018,MetzlerKlafter2000}. Note, also, that our assumption of normal diffusion---of the {\em ray angle} as it traverses the sample---is consistent with the anomalous ballistic diffusion of the exit-surface {\em intensity distribution} as it propagates from the nominally-planar sample exit surface to the detection plane (see Eqs.~(\ref{eq:AnomalousDiffusion1})-(\ref{eq:AnomalousDiffusion2})).

We now supplement this very rough estimate with a more sophisticated analysis.  The correlation function for the sample's spatially-random fluctuations is assumed to have a Gaussian form, associated with a random medium that is both homogeneous and isotropic, in the vicinity of a ray path through $(x,y)$ that is paraxial with respect to the optical axis $z$. We allow the statistics of the medium to change with transverse position $(x,y)$, but only over transverse length scales that are large compared to the correlation length $\ell(x,y)$. Similarly, the thickness $\tau(x,y)$ of the spatially-random sample will in general depend on the transverse location of the paraxial refracted ray. Bearing the above points in mind, we approximate the normalized refractive-index correlation function as\footnote{See e.g.~pp~6 and 11 of \citet{ChernovBook1960}.} 
\begin{align}
\nonumber N(r; x, y) &=\frac{E(\delta(x,y,z)\delta(x+\Delta x,y+\Delta y,z+\Delta z))}{E(\delta^2(x,y,z))}
\\ &\approx
\exp\left[-\frac{r^2}{2 [\ell(x,y)]^2}\right].
\label{eq:Chernov3}
\end{align}
Here, $r$ is the radial distance between (i) any point $(x,y,z)$ on a ray path within the scattering volume, and (ii) a neighboring point $(x+\Delta x,y+\Delta y,z+\Delta z)$ that is also within the scattering volume. The formulation of ray optics via Fermat's principle enables the right side of Eq.~(\ref{eq:Chernov1}) to be evaluated as\footnote{See e.g.~pp~12-17 of \citet{ChernovBook1960}. Note that the correlation length $a$, as defined in that book, is related to our correlation length $\ell$ via $a=\sqrt{2}\,\ell$.  Note, also, that in writing the term $E(\delta^2(x,y))$ in  Eq.~(\ref{eq:Chernov4}), we have assumed the statistical properties of $\delta(x,y,z)$ to be independent of $z$ within the volume of the sample, for fixed transverse location $(x,y)$.}
\begin{equation}
    \label{eq:Chernov4}
    \theta_s(x,y)=\sqrt{E([\varepsilon(x,y)]^2)}=s\sqrt{\frac{  E(\delta^2(x,y)) \tau(x,y)}{\ell(x,y)}}.
\end{equation}
Up to a dimensionless multiplier  $s=(8\pi)^{1/4}$ that is on the order of unity, this agrees with the crude ``bag of random prisms'' model leading to Eq.~(\ref{eq:Chernov2.5}).  It also agrees (again to a multiplicative factor of order unity) with a geometrical-optics model, in which the spatially-random medium is composed of disordered spheres \cite{Nardroff1926,Vineyard1952,Dragsdorf1956,Morgan2022Preprint}. 

To continue, work with the model in Eq.~(\ref{eq:Chernov4}).  Substitute this into Eq.~(\ref{eq:PositionDependentScalarDiffusionCoefficient}), then set $F$ to unity since we are working in the geometrical-optics context of Case \#1. Hence our dimensionless diffusion coefficient is
\begin{equation}
    \label{eq:Chernov5}
    D(x,y) = \tfrac{1}{2}\left[\theta_0^2+\frac{ s^2 E(\delta^2(x,y)) \tau(x,y)}{\ell(x,y)}\right] ~\textrm{(Case~\#1)}.
\end{equation}
This physical model is readily inverted, to give the solution to our second inverse problem $\mathcal{I}_2$ as\footnote{Cf.~Eq.~(4) in \citet{Dragsdorf1956}, which writes an analogous formula to Eq.~(\ref{eq:Chernov6}), based on Vineyard's correction \cite{Vineyard1952} to the random-sphere geometrical-optics result of \citet{Nardroff1926}.}
\begin{equation}
    \label{eq:Chernov6}
    \ell(x,y)=\frac{\tfrac{1}{2}s^2 E(\delta^2(x,y))  \tau(x,y)}{ D(x,y) - \tfrac{1}{2}\theta_0^2} ~\textrm{(Case~\#1)}.
\end{equation}
Both the total local thickness of the sample, $\tau(x,y)$, and the local refractive-index variance, $E(\delta^2(x,y))$, must be known in order to obtain $\ell(x,y)$ from $D(x,y)$.  Alternatively, use Eq.~(\ref{eq:RoughEstimateForNumberOfAngularDeflections}) to rewrite Eq.~(\ref{eq:Chernov6}) as
\begin{equation}
    \label{eq:Chernov7}
    \mathcal{N}(x,y)\approx\frac{\tau(x,y)}{\ell(x,y)}=\frac{ D(x,y) - \tfrac{1}{2}\theta_0^2}{\tfrac{1}{2}s^2 E(\delta^2(x,y))  } ~\textrm{(Case~\#1)}.
\end{equation}
This is clarifying, since it tells us that $D(x,y)$ is proportional to $\mathcal{N}(x,y)$. We can use the measurement of $D(x,y)$ to infer the number of interfaces (or scatterers, or correlation lengths) $\mathcal{N}(x,y)$ that each ray passes through at each transverse location $(x,y)$, provided that $E(\delta^2(x,y))$ is known.  If $E(\delta^2(x,y))$ is not known, we can instead determine $\mathcal{N}(x,y) E(\delta^2(x,y))$.  It is also worth pointing out that, for a compact object that is well contained within the field of view, $D(x,y)-\tfrac{1}{2}\theta_0^2$ can be readily calculated by subtracting the mean value of the reconstructed function $D(x,y)$, that corresponds to regions of the field of view where the sample is known to be absent.\footnote{Cf.~the sentence following Eq.~(\ref{eq:DiffusionCoefficientOutsideTheSample}), below.}  

\subsection{Second inverse problem: Hard-sphere example for Case \#2}\label{SecondInverseProblemExampleForCase2}

In Case \#2---as defined in Eqs.~(\ref{eq:CaseTwoFresnelNumberCondition}), (\ref{eq:F-much-less-than-one-for-case-2}) and (\ref{eq:TotalBlurAngleCase2}))---the characteristic length scale $\ell$ is sufficiently small, such that a diffractive (rather than a refractive) model is required.  We need a particular diffractive model, in order to proceed with our second inverse problem $\mathcal{I}_2$, which seeks to relate a measured diffusion field to the properties of the sample that resulted in such a diffusion field. 

\begin{figure}[ht!]
\centering
\includegraphics[width=0.9\columnwidth]{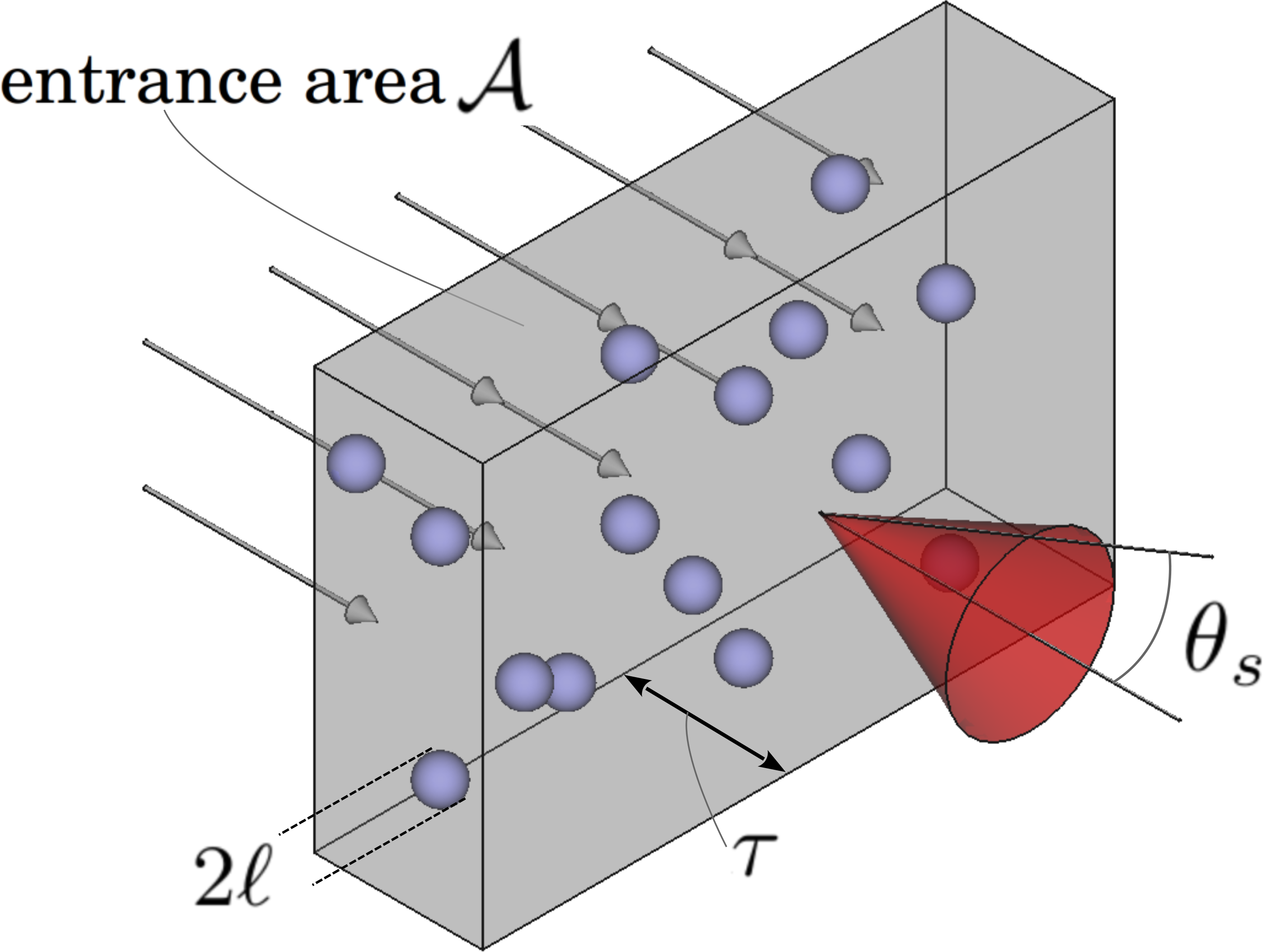}
\caption{Small-angle scattering of incident plane-wave radiation, by a vacuum slab of thickness $\tau$ that is dilutely filled with randomly distributed spheres of radius $\ell$.  One small-angle-scatter cone at the exit surface of the sample is shown, but its coaxial cone of unscattered radiation is omitted. }
\label{Fig:SimpleSASmodel}
\end{figure}

Figure~\ref{Fig:SimpleSASmodel} shows a simple diffractive model, which captures some of the basic physics underpinning the theory of small-angle scattering (e.g.~in both the x-ray and neutron domains).  Here, a small portion of our thin sample is considered to be a slab of thickness $\tau$ that is randomly filled in a dilute manner, with rotationally symmetric particles of radius $\ell$.  Let $\sigma_t$ denote the total scattering cross section for each particle, with this quantity having units of area.  If there are $N$ particles in a small patch of area $\mathcal{A}$, which is illuminated by normally-incident plane waves, the total cross section $N \sigma_t$ of the $N$ illuminated particles corresponds to a fraction
\begin{equation}
\label{eq:SecondInverseProblemForDiffractionCase1}
    F=\frac{N \sigma_t}{\mathcal{A}}\ll 1
\end{equation}
of the beam being scattered.  We require $F$ to be much less than unity, by the previously mentioned conditions that the scatterers be dilute and the sample be thin. Identify $F$ with the fraction of the beam that is converted to diffuse scatter, and then write the illuminated volume $\mathcal{V}$ as $\mathcal{V}=\mathcal{A}\tau$. Hence 
\begin{equation}
\label{eq:SecondInverseProblemForDiffractionCase2}
    F=\frac{N \sigma_t \tau}{\mathcal{V}}=\varrho_n \,  \sigma_t \tau \ll 1,
\end{equation}
where $\varrho_n$ is the number of scatterers per unit volume.

For simplicity, let our dilute scattering particles be hard spheres, in the sense that the potential for each particle is infinite inside a volume of radius $\ell$, and zero outside.\footnote{See e.g.~pp~43-44, 48 and 59 of \citet{GuinierFournetBook} in the context of small-angle x-ray scattering, or pp~393-395 of \citet{Messiah} in the context of quantum-mechanical potential scattering. Note, also, that if Eq.~(\ref{eq:SecondInverseProblemForDiffractionCase3}) is replaced by $\sigma_t=4\pi b^2$ (see e.g.~p.~19 of \citet{FeiginSverginBook}), where $b$ is the neutron nuclear scattering length, the work of the present section may be applied to neutron  scattering from amorphous materials. \label{footnote:neutron-scattering-length}}  In the high-energy limit of Eq.~(\ref{eq:Ell1MuchBiggerThanTheWavelength}), the total scattering cross section $\sigma_t$ for each sphere is twice the geometrical-optics result of $\pi \ell^2$, and so (Ref.~\cite{Messiah}, p.~394)
\begin{equation}
\label{eq:SecondInverseProblemForDiffractionCase3}
     \sigma_t =  2\pi\ell^2 \quad\textrm{(hard~sphere)}.
\end{equation}
This allows us to recast Eq.~(\ref{eq:SecondInverseProblemForDiffractionCase2}) as
\begin{equation}
\label{eq:SecondInverseProblemForDiffractionCase4}
    F= 2\pi\varrho_n \,  \ell^2 \tau\ll 1.
\end{equation}

Put the above expression to one side for a moment. The optical form of the uncertainty principle \cite{MansuripurBook} implies that the small-angle-scatter cone, in Fig.~\ref{Fig:SimpleSASmodel}, will have a half-angle that obeys \cite{Beeman1947,PaganinMorgan2019} 
\begin{equation}
\label{eq:SecondInverseProblemForDiffractionCase5}
    \theta_s \approx \frac{1}{\ell \, k},
\end{equation}
where $k$ is the mean wavenumber of the illumination. The same result follows from the Guinier approximation of small-angle scattering \cite{GuinierFournetBook,SiviaScatteringTheoryBook2011}, with $\ell$ being identified with the radius of gyration of the scattering particle.\footnote{Equation (\ref{eq:SecondInverseProblemForDiffractionCase5}) also follows from applying a small-angle approximation, to the intensity distribution that is diffusely scattered by a Gaussian-correlated isotropic homogeneous distribution of refractive index having a correlation length of $\ell$, under the first Born approximation. See e.g.~p.~121 of \citet{ThinWolfBook} for an optical-physics example of this statement, and p.~52 of \citet{ChernovBook1960} for an acoustical-physics example.  Even though these models differ from that in this paper, the same uncertainty principle holds, on account of its significant degree of generality. Cf.~Eq.~(\ref{eq:TensorDiffusionFieldRetrieval120}).}     

To proceed further, let $\varrho_n, \ell,$ and $\tau$ vary with respect to transverse position $(x,y)$, over length scales that are large compared to  $\ell$. We may then substitute Eqs.~(\ref{eq:SecondInverseProblemForDiffractionCase4}) and (\ref{eq:SecondInverseProblemForDiffractionCase5}) into  Eq.~(\ref{eq:PositionDependentScalarDiffusionCoefficient}), to give
\begin{equation}
\label{eq:SecondInverseProblemForDiffractionCase6}
    D(x,y) = \tfrac{1}{2}\theta_0^2+\pi k^{-2}\varrho_n(x,y) \tau(x,y).
\end{equation}
Note the cancellation of the correlation length $\ell$ that has occurred.\footnote{In the refractive-sphere model considered in Sec.~\ref{sec:Refractive-sphere model}, this cancellation does not occur. }  This prompts us to reformulate our second inverse problem, in the context of the current model only, towards using a measurement of $D(x,y)$ to recover the sample-based quantity $\varrho_n(x,y)$, together with quantities derivable from this number density.

Subject to the particular Case \#2 model under which it was derived, Eq.~(\ref{eq:SecondInverseProblemForDiffractionCase6}) has a very simple physical interpretation. Up to an additive constant that represents the blur associated with a finite source size, we see that the position-dependent diffusion field $D(x,y)$ is proportional to the product $\varrho_n(x,y) \tau(x,y)$.  Since this product is a number density multiplied by a thickness along the optical axis, we see that $\varrho_n(x,y) \tau(x,y)$ is the {\em number of scatterers per unit area, when viewed along the optical axis}.  This relates the diffusion field---which pertains to both the properties of the sample and the properties of the illumination that interrogates the sample---to a quantity that depends only upon the sample that is being probed.  

We now mention a tomographic generalization of Eq.~(\ref{eq:SecondInverseProblemForDiffractionCase6}).  If we now allow the number density $\varrho_n$ to vary in all three dimensions within the sample, over length scales that are large compared to the correlation length of the unresolved microstructure, then (cf.~Refs.~\cite{Khelashvili2006,Wang2009,Bech2010})
\begin{equation}
\label{eq:SecondInverseProblemForDiffractionCase7}
    \frac{k^2}{\pi}\left[D(x,y) - \tfrac{1}{2}\theta_0^2\right] =  \int_{z=-T_s}^{z=0}\varrho_n(x,y,z) \,\mathrm{d}z.
\end{equation}
Here, the lower and upper limits on the integral correspond to the nominally-planar entrance and exit surfaces of the sample, respectively (see Fig.~\ref{Fig:SimpleImagingWithSourceBlurAndSampleBlur}).  Since the right side of the above equation is equal to the projection of the sample's scattering-sphere number density along the optical axis, the methods of computed tomography \cite{kak1988principles,Natterer} may now be employed.  In particular, the sample can be rotated to a number of equally-spaced angles $\Theta$ about an axis that (i) passes through the sample and (ii) is perpendicular to the optical axis.  After the scaling given by the left side of Eq.~(88), the resulting set of diffusion fields $D(x,y;\Theta)$ can then be input into any standard tomographic reconstruction process, such as filtered backprojection, to yield a three-dimensional map of the number density $\varrho_n(x,y,z)$ of the scatterers.  

\subsection{Second inverse problem: Refractive-sphere example for Case \#2}\label{sec:Refractive-sphere model}

Here we revisit the analysis of Sec.~\ref{SecondInverseProblemExampleForCase2}, by replacing the hard spheres of Fig.~\ref{Fig:SimpleSASmodel} with refractive spheres.  Following the notation of Eq.~(\ref{eq:Chernov2}), let the uniform refractive index of the spheres be $1+\delta$, with these spheres of radius $\ell$ being randomly distributed, as shown in Fig.~\ref{Fig:SimpleSASmodel}.  

\begin{figure}[ht!]
\centering
\includegraphics[width=0.75\columnwidth]{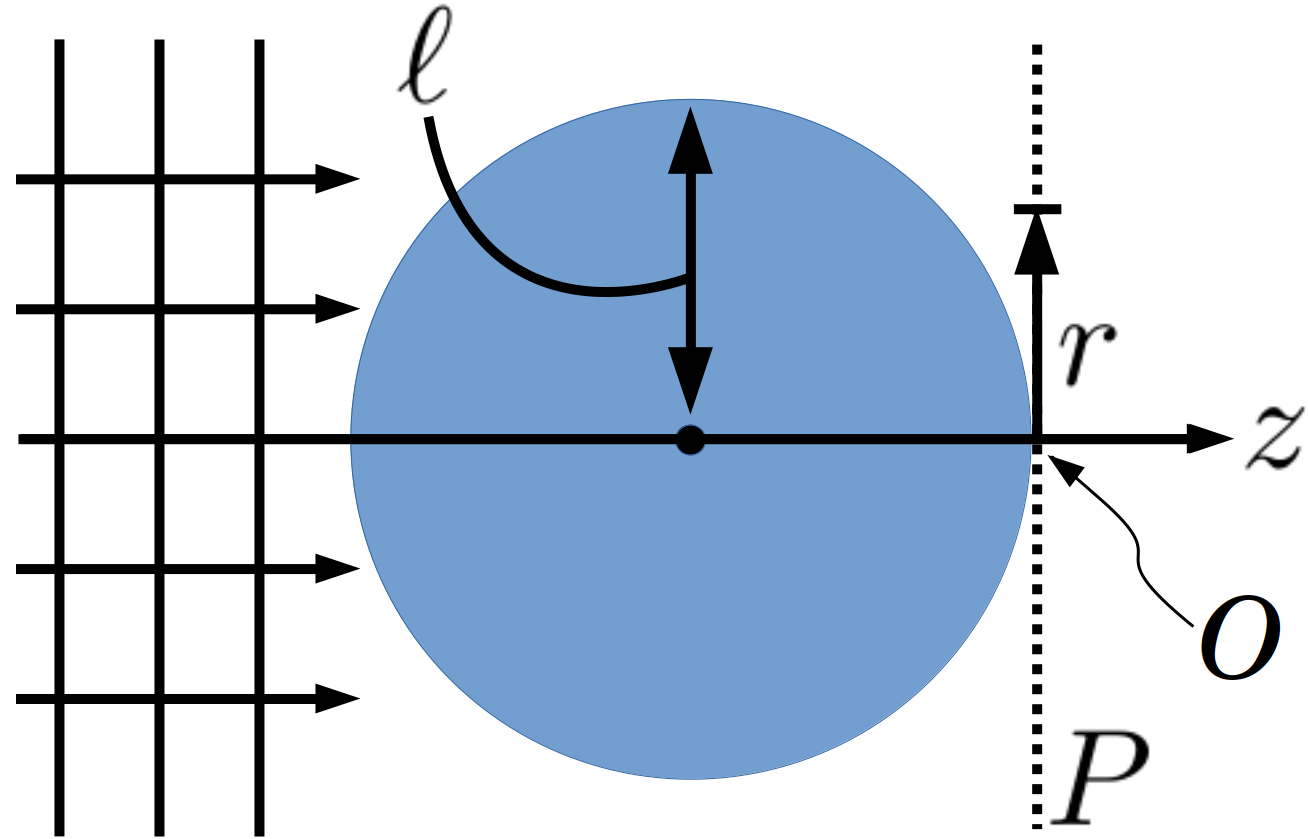}
\caption{One refractive sphere is illuminated by narrowband scalar plane waves, under the projection approximation.}
\label{Fig:SingleRefractiveSphere}
\end{figure}

To estimate the total scattering cross section $\sigma_t$, for any one  refractive sphere, assume it to be sufficiently weakly refracting that the projection approximation \cite{Paganin2006} holds.  As shown in  Fig.~\ref{Fig:SingleRefractiveSphere}, if a single sphere is illuminated by narrowband complex scalar plane waves, then relative to propagation through vacuum the complex wavefield $\psi_P(r)$ over a nominally-planar exit surface $P$ will be
\begin{align}
\nonumber
\psi_P(r) &= \begin{cases}
  \exp(2ik\delta\sqrt{\ell^2-r^2}), &r\le \ell,  \\
  1, &r > \ell,
\end{cases}
\\ &\approx \begin{cases}
  1+2ik\delta\sqrt{\ell^2-r^2}, ~\,&r\le \ell,  \\
  1, &r > \ell.
\end{cases}
\label{eq:AppendixA=1}
\end{align}
Here, $r$ denotes radial distance in cylindrical polar coordinates, relative to the $z$ axis in Fig.~\ref{Fig:SingleRefractiveSphere}, and $k$ is the average wavenumber of the illumination. Note, also, that we assume the weak-object approximation, in passing from the upper to the lower part of the above expression.

We estimate the scattered field $\psi_s(r)$, within the geometrical-optics shadow $r \le \ell$ of the sphere over the exit surface $P$, to be
\begin{equation}
\label{eq:AppendixA=2}
\psi_s(r)\approx 2ik\delta\sqrt{\ell^2-r^2}, \quad r \le \ell.    
\end{equation}
The unscattered field $\psi_0(r)$ may be approximated by unity, over the same region.  The scattering cross section $\sigma_t$ can then be obtained via
\begin{equation}
\label{eq:AppendixA=3}
    \frac{\sigma_t}{\pi\ell^2}
    \approx
    \frac{\iint_{r\le\ell}\vert\psi_s(r)\vert^2\,r\,\mathrm{d}r\,\mathrm{d}\theta}{\iint_{r\le\ell}\vert\psi_0(r)\vert^2\,r\,\mathrm{d}r\,\mathrm{d}\theta},
\end{equation}
where $(r,\theta)$ denotes plane polar coordinates over the tangent plane $P$, relative to the origin $O$ that is indicated in Fig.~\ref{Fig:SingleRefractiveSphere}.  Evaluating the resulting integrals gives
\begin{equation}
\label{eq:TotalCrossSectionRefractiveSphere}
    \sigma_t\approx 2\pi k^2 \delta^2 \ell^4 \quad\textrm{(refractive~sphere)}.
\end{equation}
A detailed classical-electrodynamics analysis, analogous to the partial-waves formalism of scattering theory, gives exactly the same result in the high-energy limit implied by Eq.~(\ref{eq:Ell1MuchBiggerThanTheWavelength}) (Ref.~\cite{JacksonElectrodynamicsBook}, p.~510). 

The form for the scattering fraction $F$, as given in Eq.~(\ref{eq:SecondInverseProblemForDiffractionCase2}), may be combined with Eq.~(\ref{eq:TotalCrossSectionRefractiveSphere}) to give 
\begin{equation}
\label{eq:AppendixA=4}
F \approx 2\pi k^2 \varrho_n \, \delta^2 \ell^4  \tau   \ll 1.
\end{equation}
Here, we again use $\varrho_n$ to denote the number of scattering spheres per unit volume, with $\tau$ being the thickness of the scattering slab (cf.~Fig.~\ref{Fig:SimpleSASmodel}).   

In the diffractive regime employed in Sec.~\ref{SecondInverseProblemExampleForCase2}, we may substitute Eqs.~(\ref{eq:AppendixA=4}) and (\ref{eq:SecondInverseProblemForDiffractionCase5})  into Eq.~(\ref{eq:PositionDependentScalarDiffusionCoefficient}), to give the dimensionless diffusion coefficient
\begin{equation}
\label{eq:AppendixA=5}
D(x,y) = \tfrac{1}{2}\theta_0^2+\pi \varrho_n(x,y) \, [\ell(x,y)]^2  \tau(x,y) \delta^2.
\end{equation}
Observe that the correlation length $\ell$ has not canceled away, contrary to the case in Eq.~(\ref{eq:SecondInverseProblemForDiffractionCase6}) (which used a hard-sphere model rather than a refractive-sphere model).  Also, recall our earlier comment, that $\varrho_n(x,y)\tau(x,y)$ is the average number of scatterers per unit area, when viewed along the direction of the optical axis.  Since this product is now multiplied by $[\ell(x,y)]^2$---which is proportional to the surface area $4\pi\ell^2$ of each scattering sphere, in addition to being proportional to the surface area $2\pi\ell^2$ of two circles of radius $\ell$---we may interpret $2\pi \varrho_n(x,y) \, [\ell(x,y)]^2  \tau(x,y)$ in Eq.~(\ref{eq:AppendixA=5}) as {\em the average total area of the spheres, per unit area, when projected along the direction of the optical axis}. Stated differently, this dimensionless quantity is equal to the number of spherical interfaces that are present, when the sample is viewed along a line of sight that is parallel to the optical axis.    

In the context of two-dimensional diffusion-field retrieval, the inverse problem $\mathcal{I}_2$---namely the question of relating $D(x,y)$ to sample microstructure---may be addressed by algebraically isolating $\varrho_n(x,y) \, [\ell(x,y)]^2$ in the preceding expression, if the slab thickness $\tau(x,y)$ is known.  If the slab thickness is not known, we can instead algebraically isolate $2\pi \varrho_n(x,y) \, [\ell(x,y)]^2  \tau(x,y)$, which may be interpreted as described in the previous paragraph. Finally, if $\varrho_n$ and $\ell$ are now allowed to vary with $z$, but $\delta$ is taken to be fixed, the tomographic variant of Eq.~(\ref{eq:AppendixA=5}) is (cf.~Eq.~(\ref{eq:SecondInverseProblemForDiffractionCase7}) and Refs.~\cite{Khelashvili2006,Wang2009,Bech2010}) 
\begin{equation}
\label{eq:AppendixA=6}
\frac{1}{\pi\delta^2}\left[D(x,y) - \tfrac{1}{2}\theta_0^2\right] =  \int_{z=-T_s}^{z=0}\varrho_n(x,y,z) [\ell(x,y,z)]^2 \,\mathrm{d}z.
\end{equation}
Rotating the sample through equally-spaced angles $\Theta$, about an axis through the sample that is perpendicular to $z$, allows the resulting diffusion fields $D(x,y;\Theta)-\tfrac{1}{2}\theta_0^2$ to be input into computed tomography \cite{Natterer}.  After multiplying by $4/\delta^2$, this yields a three-dimensional map of $4\pi\varrho_n(x,y,z)[\ell(x,y,z)]^2$, namely the {\em surface-area-to-volume ratio of the sample that is randomly filled with refractive spheres whose size may vary with position}.  The same result holds for the inverse structure, namely a homogeneous sample of refractive index $1 \pm \delta$, randomly filled with spherical cavities or voids of radius $\ell(x,y,z)$.

\section{Tensorial generalization of diffusion-field retrieval}\label{sec:IV}

In the preceding discussions, the local small-angle-scatter blurring cone is assumed to be rotationally symmetric.  This is rotationally-isotropic diffusion, in the sense that there is no preferred transverse direction for the microstructure-induced blur.  While rotationally-isotropic position-dependent blur naturally occurs when there is no preferred direction for the spatially-random unresolved microstructure, this becomes inapplicable if there is some directional character to the microstructure.  This leads to a tensorial generalization of diffusion-field retrieval.  The  small-angle-scatter distribution is now considered to have fan-like rather than cone-like character, corresponding to the transverse slices of these distributions being locally elliptical rather than locally circular.  We restrict consideration to what is called ``Case \#2'' earlier in the paper, namely the case where diffraction rather than refraction leads to the diffuse scatter.

\subsection{The forward problem}

Consider a transparent sample with small-magnitude refractive-index fluctuation defined in Eq.~(\ref{eq:Chernov2}).  Assume this fluctuation to be a zero-mean stochastic function, with normalized two-point correlation function 
\begin{equation}
\label{eq:TensorDiffusionFieldRetrieval1}    
C(\mathbf{r},\mathbf{r}+\Delta\mathbf{r})
\equiv\frac{E\left(\delta(\mathbf{r})\delta(\mathbf{r}+\Delta\mathbf{r})\right)}{E\left([\delta(\mathbf{r})]^2\right)}
\end{equation}
having the anisotropic Gaussian form (cf.~Eq.~(\ref{eq:Chernov3}) and pp 402-403 of Ref.~[56])
\begin{equation}
\label{eq:TensorDiffusionFieldRetrieval2}    
C(\mathbf{r},\mathbf{r}+\Delta\mathbf{r})
\approx \exp\!\left[-\tfrac{1}{2}
\{\mathcal{R}^{T}_{\mathbf{\Phi}(\mathbf{r})}(\Delta\mathbf{r})\}^T\mathbf{P}(\mathbf{r})\{\mathcal{R}^{T}_{\mathbf{\Phi}(\mathbf{r})}(\Delta\mathbf{r})\}
\right].
\end{equation}
Here, $E()$ again denotes statistical averaging over an ensemble of realizations of the refractive-index distribution \cite{Pedersen1976,FrozenPhononModel,MandelWolf,Nesterets2008,GoodmanSpeckleBook}, $\mathbf{r}$ and $\mathbf{r}+\Delta\mathbf{r}$ denote any pair of positions within the volume of the sample, $\Delta\mathbf{r}$ is considered to be a three-component column vector 
\begin{equation}
    \Delta\mathbf{r}=
    \begin{pmatrix}\Delta x\\ \Delta y\\ \Delta z\end{pmatrix} \equiv (\Delta x, \Delta y, \Delta z)^T,
\end{equation}
a superscript $T$ denotes matrix transposition, $\mathcal{R}_{\mathbf{\Phi}(\mathbf{r})}$ is a $3\times 3$ rotation-matrix field corresponding to the vector correlation-ellipse orientation field\footnote{For an example of Euler-angle fields in the context of materials-science texture analysis, see p.~42 of \citet{BungeBook}, together with primary references cited therein. See, also, p.~158 of \citet{SiviaScatteringTheoryBook2011}.} 
\begin{equation}
\label{eq:ThreeEulerAngles}
\mathbf{\Phi}(\mathbf{r}) =\left(\Phi_1(\mathbf{r}),\Phi_2(\mathbf{r}),\Phi_3(\mathbf{r})\right)^T  
\end{equation}
of Euler angles $\Phi_1,\Phi_2,\Phi_3$ in any of the numerous available conventions,\footnote{For Euler-angle conventions, see e.g.~pp~147 and 606-610 in \citet{GoldsteinBook}.} and $\mathbf{P}(\mathbf{r})$ is a diagonal $3\times 3$ matrix 
\begin{eqnarray}
\nonumber
\mathbf{P}(\mathbf{r})=
\begin{pmatrix}
[\ell_1(\mathbf{r})]^{-2} & 0 & 0\\
0 & [\ell_2(\mathbf{r})]^{-2} & 0\\
0 & 0 & [\ell_3(\mathbf{r})]^{-2}
\end{pmatrix}, \quad\quad
\\ \ell_1(\mathbf{r}) \le \ell_2(\mathbf{r}) \le \ell_3(\mathbf{r}),~
\label{eq:TensorDiffusionFieldRetrieval3}
\end{eqnarray}
containing the principal correlation-ellipse characteristic lengths $\ell_1(\mathbf{r})$, $\ell_2(\mathbf{r})$, and $\ell_3(\mathbf{r})$.  The form in Eq.~(\ref{eq:TensorDiffusionFieldRetrieval2}) is intuitive, being equivalent to the statements that (i) the random microstructure has three correlation lengths that may be viewed as aligned with the three principal axes of a position-dependent correlation ellipsoid, (ii) this correlation ellipsoid has arbitrary orientation, specified by a suitable rotation matrix obtained using Euler angles, and (iii) the random microstructure obeys Gaussian statistics. Under this model, the statistical properties of the sample are modeled by seven numbers at each point $\mathbf{r}$:
\begin{itemize}
    \item the three correlation lengths $\ell_1(\mathbf{r})$, $\ell_2(\mathbf{r})$, and $\ell_3(\mathbf{r})$, used to assemble the diagonal-matrix field $\mathbf{P}(\mathbf{r})$,
    \item the three angles in the Euler-angle vector field $\mathbf{\Phi}(\mathbf{r})$, used to specify the orientation of the correlation ellipsoid at each point in the sample, and
    \item the refractive-index variance $E([\delta(\mathbf{r})]^2)$, which normalizes the correlation function $C(\mathbf{r},\mathbf{r}+\Delta\mathbf{r})$.
\end{itemize}
The six numbers, listed in the first two dot points, may be combined into the six independent components of the symmetric second-rank $3 \times 3$ tensor field
\begin{equation}
\label{eq:TensorDiffusionFieldRetrieval4}    
\mathbf{\Sigma}(\mathbf{r})=
\mathcal{R}_{\mathbf{\Phi}(\mathbf{r})}
\mathbf{P}(\mathbf{r})\,
[\mathcal{R}_{\mathbf{\Phi}(\mathbf{r})}]^T. 
\end{equation}
Hence Eq.~(\ref{eq:TensorDiffusionFieldRetrieval2}) may be written in the compact form
\begin{equation}
\label{eq:TensorDiffusionFieldRetrieval5}   
C(\mathbf{r},\mathbf{r}+\Delta\mathbf{r})
\approx \exp\left[-\tfrac{1}{2}
(\Delta \mathbf{r})^T \mathbf{\Sigma}(\mathbf{r}) \,\Delta \mathbf{r}
\right].
\end{equation}

\begin{figure}[ht!]
\centering
\includegraphics[width=1.0\columnwidth]{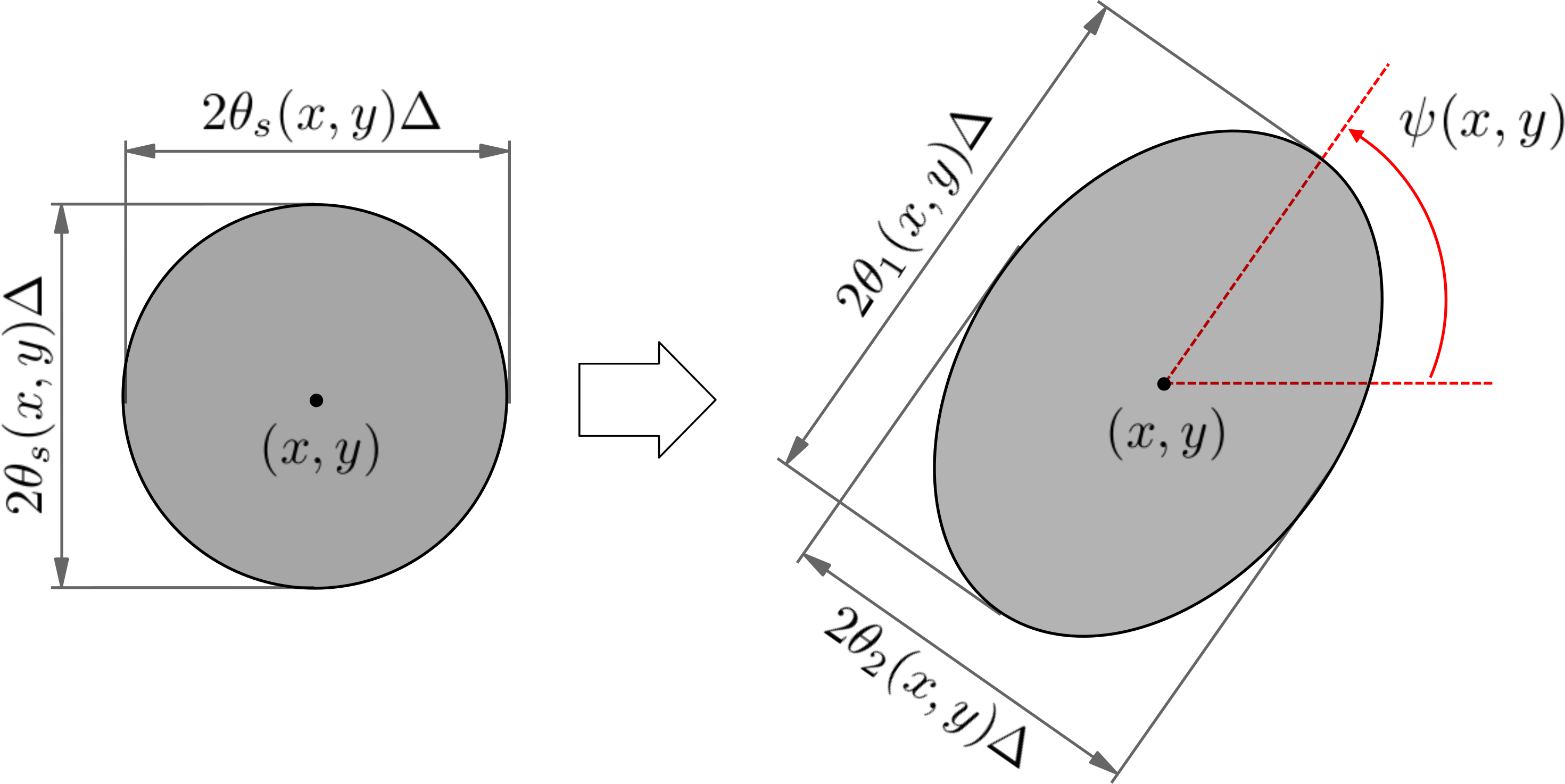}
\caption{For a sample with locally-isotropic random microstructure, illuminated with a pencil beam at the transverse location $(x,y)$, the resulting cone of small-angle diffuse scatter will have the circular transverse section shown to the left. Allowing the random microstructure to be anisotropic, as described by Eq.~(\ref{eq:TensorDiffusionFieldRetrieval5}), causes the scatter distribution to have the elliptical transverse section shown to the right.}
\label{Fig:EllipticalSASfan}
\end{figure}

Suppose that a sample having locally-anisotropic spatially-random microstructure \cite{jensen2010a,jensen2010b,Yashiro2011,Wieczorek2016}, as modeled above, is illuminated via the experiment in Fig.~\ref{Fig:SimpleImagingWithSourceBlurAndSampleBlur}.  Rather than the local small-angle-scatter distribution being a cone with transverse slice looking as shown on the left of Fig.~\ref{Fig:EllipticalSASfan}, we instead have the locally-elliptical fan \cite{jensen2010b} with transverse slice shown on the right of Fig.~\ref{Fig:EllipticalSASfan}.  The sample-induced cone half-angle $\theta_s(x,y)$ thereby generalizes to (i) two half-angles $\theta_1(x,y)$ and $\theta_2(x,y) \le \theta_1(x,y)$, respectively associated with the semi-major and semi-minor axes of the elliptical transverse slice of the small-angle-scatter distribution, together with (ii) the ellipse inclination angle $\psi(x,y)$.    

If the effects of finite source size may be neglected, by considering $\theta_0$ to be negligible, the isotropic-diffusion effective angle in Eq.~(\ref{eq:EffectiveBlurAngle}) generalizes to the pair of angles
\begin{align}
\nonumber 
\theta_j^{\textrm{(eff)}}(x,y) &=\sqrt{F(x,y)}\,\theta_j(x,y), \\ &j=1,2, \quad \theta_1^{\textrm{(eff)}}\ge\theta_2^{\textrm{(eff)}}.
\label{eq:TensorDiffusionFieldRetrieval6}   
\end{align}
Here, in the same notation that was employed earlier in the paper, $F(x,y) \ll 1$ denotes the fraction of the incident illumination that is converted to diffuse scatter, at the transverse position $(x,y)$.  Thus, at each point in the propagated image that is registered over the plane $z=\Delta$, there will be local blurring over an elliptical region $E_a$ that has semi-major and semi-minor axes of $\theta_1^{\textrm{(eff)}}(x,y)\,\Delta$ and $\theta_2^{\textrm{(eff)}}(x,y) \,\Delta$, respectively, with this blurring ellipse being rotated by $\psi(x,y)$ with respect to the positive $x$ axis.  We emphasize that the effective blurring ellipse $E_a$ is much smaller than the small-angle-scatter ellipse $E_b$ that is sketched in the right part of Fig.~\ref{Fig:EllipticalSASfan}.  This difference arises because the unscattered component of the transmitted radiation or matter waves, at each transverse position $(x,y)$,  implies the local blurring of intensity to have a smaller diameter than the small-angle scatter fan associated with that transverse position.\footnote{Cf.~our statements regarding ``two coaxial diffusive cones'', in the text below Eq.~(\ref{eq:F-much-less-than-one-for-case-2}).}  Importantly, however, the ellipses $E_a$ and $E_b$ are similar to one another, since (i) they have the same eccentricity
\begin{align}
\nonumber 
e(x,y) 
&=\sqrt{1-\left[\frac{\theta_2^{\textrm{(eff)}}(x,y)}{\theta_1^{\textrm{(eff)}}(x,y)}\right]^2}
\\ &=\sqrt{1-\left[\frac{\theta_2(x,y)}{\theta_1(x,y)}\right]^2}
\label{eq:TensorDiffusionFieldRetrieval7}
\end{align}
for any nonzero diffusion-fraction field $F(x,y)$, as well as (ii) having the same angular orientation $\psi(x,y)$. Thus, both the effective-blur eccentricity field $e(x,y)$ and the effective diffusion-ellipse orientation field $\psi(x,y)$ are invariant with respect to $F(x,y)$ \cite{MISTdirectional}.\footnote{This statement is predicated on the previously-stated assumption that the influence of $\theta_0$ may be neglected.  If $\theta_0$ is not negligible, then the eccentricity of $E_a$ and $E_b$ will no longer be equal, although their orientation angles $\psi(x,y)$ will still be equal. \label{footnote:e-and-psi-are-independent-of-F}}  We shall make use of this invariance, later in the present section. 

Consider the transverse gradient $\nabla_{\perp}$ to be the two-component column-vector operator $(\partial/\partial x,\partial/\partial y)^T$, with the transverse divergence $\nabla_{\perp}\cdot$ being the row-vector operator $(\partial/\partial x,\partial/\partial y)$. Using the ``$\widetilde{D}$'' representation in Sec.~\ref{sec:AlternativeFormulationsForDiffusionField},  Eq.~(\ref{eq:2DVersionOfSimpleBlurWithGeneralisation==ALTERNATIVE-FORM}) has the anisotropic-diffusion generalization\footnote{Cf.~e.g.~pp~4-6 of \citet{CrankBook}.}
\begin{align}
\nonumber I(x,y,z=&\Delta \ge 0)=I(x,y,z=0)
\\ &+\Delta^2\nabla_{\perp}\cdot[\widetilde{\mathbf{D}}(x,y) \nabla_{\perp}I(x,y,z=0)].
\label{eq:TensorDiffusionFieldRetrieval8}
\end{align}
Here, analogous to Eq.~(\ref{eq:TensorDiffusionFieldRetrieval4}), we introduce the symmetric $2\times 2$ second-rank anisotropic-diffusion tensor field
\begin{align}
\label{eq:TensorDiffusionFieldRetrieval9}    
\widetilde{\mathbf{D}}&(x,y)= 
\\ \nonumber &\mathcal{R}_{\psi(x,y)} 
\begin{pmatrix}
\tfrac{1}{2}[\theta_1^{\textrm{(eff)}}(x,y)]^2 & 0\\
0 & \tfrac{1}{2}[\theta_2^{\textrm{(eff)}}(x,y)]^2
\end{pmatrix}
[\mathcal{R}_{\psi(x,y)}]^T,
\end{align}
where the matrix field 
\begin{equation}
\label{eq:TensorDiffusionFieldRetrieval10}    
\mathcal{R}_{\psi(x,y)}=
\begin{pmatrix}
\cos\psi(x,y) & -\sin\psi(x,y)\\
\sin\psi(x,y) & \cos\psi(x,y)
\end{pmatrix}
\end{equation}
gives a local rotation of each effective-diffusion ellipse by the angle $\psi(x,y)$. If we expand out Eq.~(\ref{eq:TensorDiffusionFieldRetrieval9}), we may write the components of our dimensionless diffusion tensor as 
\begin{equation}
\label{eq:TensorDiffusionFieldRetrieval11}    
\widetilde{\mathbf{D}}(x,y)= 
\begin{pmatrix}
\widetilde{D}_{xx}(x,y) & \widetilde{D}_{xy}(x,y)\\
\widetilde{D}_{xy}(x,y) & \widetilde{D}_{yy}(x,y)
\end{pmatrix},
\end{equation}
where
\begin{align}
\nonumber 
\widetilde{D}_{xx}(x,y) &= \tfrac{1}{2}\Big\{[\theta_1^{\textrm{(eff)}}(x,y)]^2\cos^2[\psi(x,y)] \\ \nonumber &\quad\quad\quad+[\theta_2^{\textrm{(eff)}}(x,y)]^2\sin^2[\psi(x,y)]\Big\}, \\  \label{eq:TensorDiffusionFieldRetrieval12}    
\widetilde{D}_{yy}(x,y)&= \tfrac{1}{2}\Big\{
[\theta_1^{\textrm{(eff)}}(x,y)]^2\sin^2[\psi(x,y)]
\\ \nonumber &\quad\quad\quad+ [\theta_2^{\textrm{(eff)}}(x,y)]^2\cos^2[\psi(x,y)]
\Big\}, \\
\nonumber 
\widetilde{D}_{xy}(x,y)&=\tfrac{1}{4}\sin[2\psi(x,y)]\Big\{[\theta_1^{\textrm{(eff)}}(x,y)]^2 \\ \nonumber &\quad\quad\quad\quad\quad\quad\quad\quad\quad -[\theta_2^{\textrm{(eff)}}(x,y)]^2\Big\}.
\end{align}

\begin{figure}[ht!]
\centering
\includegraphics[width=0.9\columnwidth]{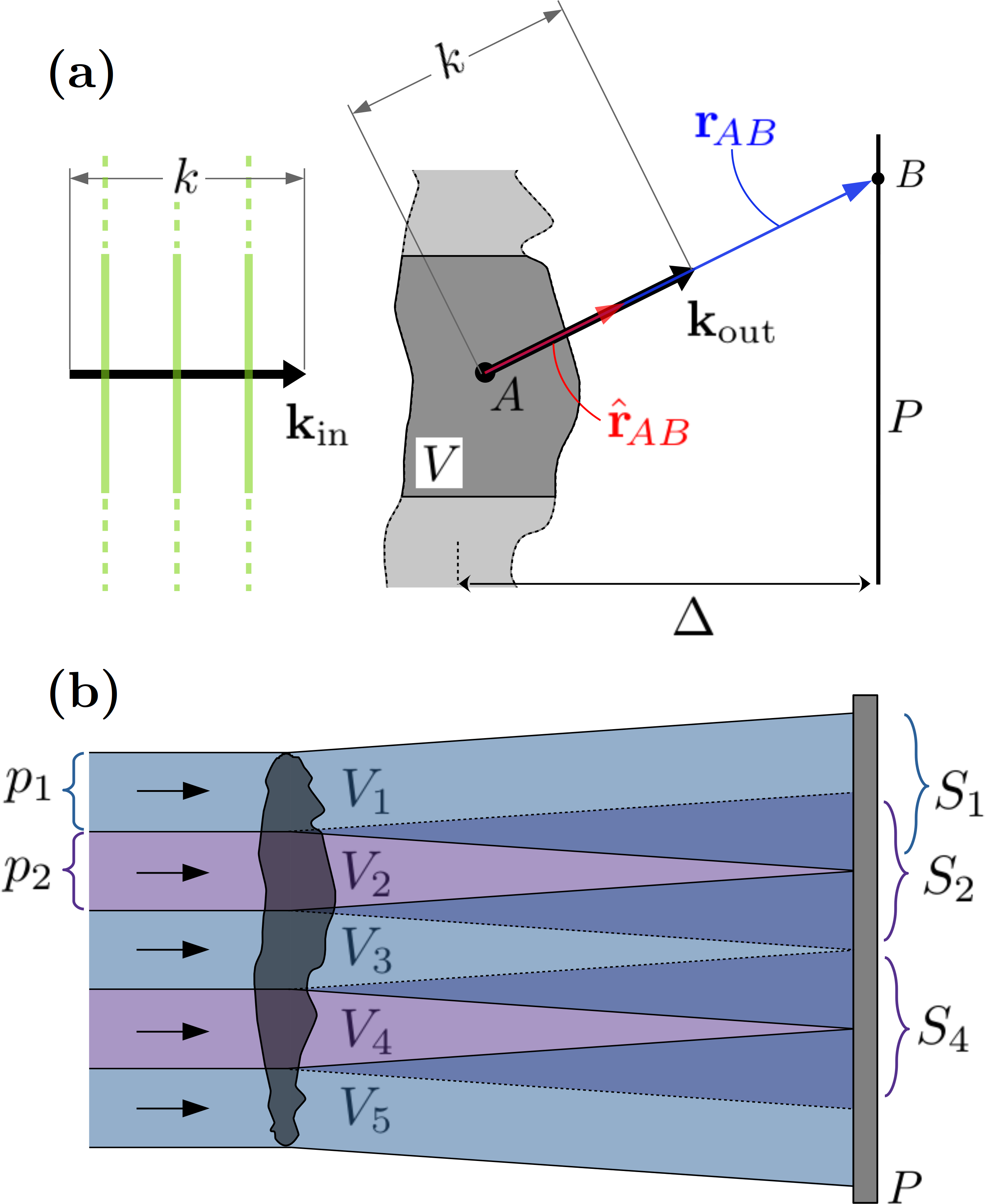}
\caption{(a) Notation for incident and scattered wavevectors, associated with small-angle scattering from anisotropic microstructure in a small volume $V$.  Only one plane-wave component, out of the distribution of diffuse small-angle scatter from $V$, is shown.  By assumption, in propagating to the detection plane $P$, the effective diffuse-scatter fan (not shown) only spreads a transverse extent on the order of the width of $V$.  (b) Hence there is only a moderate degree of overlap between effective diffuse-scatter fans $S_1,S_2,\cdots$ from respective contiguous scattering volumes $V_1, V_2,\cdots$.  Each contiguous scattering volume is illuminated by a separate ``pencil'', such as pencil $p_1$ for volume $V_1$, and pencil $p_2$ for volume $V_2$.}
\label{Fig:MomentumTransfer}
\end{figure}

To complete our description of the forward problem associated with the tensorial generalization of diffusion-field retrieval, we need to show how  $\widetilde{\mathbf{D}}(x,y)$ may be calculated from  $C(\mathbf{r},\mathbf{r}+\Delta\mathbf{r})$ and $E([\delta(\mathbf{r})]^2)$. For simplicity, assume that for any fixed transverse location $(x,y)$, the statistical properties of the sample are independent of the $z$ position within the sample. We also assume the first Born approximation to hold.  Assume a small component of the illuminated anisotropic microstructure to correspond to Fig.~\ref{Fig:MomentumTransfer}(a), where for simplicity we have placed the component at the origin of coordinates $A$. 

In the language of the theory of small-angle scatter \cite{GuinierFournetBook,SiviaScatteringTheoryBook2011,GlatterKratky1982}, let $\mathbf{k}_{\textrm{in}}$ be the wavevector of an incident $z$-directed plane wave,  $\mathbf{k}_{\textrm{out}}$ be the wavevector of a scattered plane wave downstream of the sample, and 
\begin{equation}
\label{eq:TensorDiffusionFieldRetrieval13}    
\mathbf{Q}=(Q_x,Q_y,Q_z)^T=\mathbf{k}_{\textrm{out}}-\mathbf{k}_{\textrm{in}}=k \, \hat{\mathbf{r}}_{\!AB}-\mathbf{k}_{\textrm{in}}
\end{equation}
be the associated ``momentum transfer''.  This momentum transfer corresponds to the scattered wave traveling in the direction of the unit vector $\hat{\mathbf{r}}_{\!AB}$, which points from the center $A$ of the scattering volume $V$, to the detector location $B$. Since the scattering is elastic, by assumption, both $\mathbf{k}_{\textrm{in}}$ and $\mathbf{k}_{\textrm{out}}$ have a magnitude equal to the average wavenumber $k$ of the narrowband illumination.  

In small-angle scattering, the intensity $I_{\textrm{scattered}}(\mathbf{Q})$ of the diffusely scattered beam is proportional to the Fourier transform of the autocorrelation function of the spatially-random scattering potential.\footnote{In making this statement, the first Born approximation is assumed.  Moreover, the measurement plane is assumed to be in the far field of the spatially-random scattering volume. See e.g.~p.~77 of \citet{GuinierFournetBook}, pp~20-21 of \citet{GlatterKratky1982}, or p.~127 of \citet{SiviaScatteringTheoryBook2011}. }  Using the correlation function in Eq.~(\ref{eq:TensorDiffusionFieldRetrieval5}), we may therefore write 
%
\begin{align}
\nonumber
I_{\textrm{scattered}}(\mathbf{Q}) \propto \iiint_{V} &\exp\left[-\tfrac{1}{2}
(\Delta \mathbf{r})^T \mathbf{\Sigma}(\mathbf{r}_A) \,\Delta \mathbf{r}
\right]  \\ &\times \exp(-i\mathbf{Q}\cdot \Delta\mathbf{r}) \,\mathrm{d}(\Delta\mathbf{r}).
\label{eq:TensorDiffusionFieldRetrieval14}    
\end{align}
Here, the integral is taken over the small volume $V$---namely, one of the volumes $V_j$, where $j=1,2,\cdots$---that is illuminated by a small ``pencil'' within the full illuminating plane wave (see Fig.~\ref{Fig:MomentumTransfer}(b)). This pencil is (i) small in transverse extent compared to both the illuminating plane wave and the transverse length scale over which the statistical properties of the sample vary appreciably, but (ii) large in transverse extent compared to all correlation lengths of the illuminated region.  Hence  $\mathbf{\Sigma}(\mathbf{r})\approx\mathbf{\Sigma}(\mathbf{r}_A)$ at any point in $V$, where the point $A$ in the middle of $V$ is taken to have a position vector $\mathbf{r}_A$. Further, in Eq.~(\ref{eq:TensorDiffusionFieldRetrieval14}), $\Delta\mathbf{r}$ is a position vector relative to an origin at $A$. Also, in writing Eq.~(\ref{eq:TensorDiffusionFieldRetrieval14}), the omitted proportionality factor depends on transverse location $(x,y)$.  This proportionality factor is itself proportional to $F(x,y)$. We do not calculate $F(x,y)$, since we only consider the $F$-independent quantities $e(x,y)$ and $\psi(x,y)$ \cite{MISTdirectional}.

Using the standard result that the Fourier transform of a Gaussian having standard deviation $\sigma$ is
\begin{equation}
\label{eq:TensorDiffusionFieldRetrieval15}
\int_{-\infty}^{\infty}\!\!\exp\!\left(-\frac{x^2}{2\sigma^2}\right)\exp(-ikx)\,\mathrm{d}x=\sqrt{2\pi}\sigma \exp(-\tfrac{1}{2}k^2\sigma^2),
\end{equation}
the integral in Eq.~(\ref{eq:TensorDiffusionFieldRetrieval14}) becomes (cf.~Ref.~\cite{Khelashvili2006})
\begin{align}
\nonumber &I_{\textrm{scattered}}(\mathbf{Q}) \propto \exp\left\{-\tfrac{1}{2}
 [\mathcal{R}^T_{\mathbf{\Phi}(\mathbf{r}_A)}\mathbf{Q}]^T \mathbf{P}^{-1}(\mathbf{r}_A)\,[\mathcal{R}^T_{\mathbf{\Phi}(\mathbf{r}_A)}
\mathbf{Q}]
\right\}
\\ &\quad = \exp\left\{-\tfrac{1}{2}
 \mathbf{Q}^T \mathcal{R}_{\mathbf{\Phi}(\mathbf{r}_A)}\mathbf{P}^{-1}(\mathbf{r}_A)\,\mathcal{R}^T_{\mathbf{\Phi}(\mathbf{r}_A)}
\mathbf{Q}
\right\}. 
\label{eq:TensorDiffusionFieldRetrieval16}
\end{align}
Here, $\mathbf{P}^{-1}(\mathbf{r})$ is the inverse of the matrix in Eq.~(\ref{eq:TensorDiffusionFieldRetrieval3}), so 
\begin{eqnarray}
\nonumber
\mathbf{P}^{-1}(\mathbf{r}_A)=
\begin{pmatrix}
[\ell_1(\mathbf{r}_A)]^2 & 0 & 0\\
0 & [\ell_2(\mathbf{r}_A)]^2 & 0\\
0 & 0 & [\ell_3(\mathbf{r}_A)]^2
\end{pmatrix}, 
\\ \ell_1(\mathbf{r}_A) \le \ell_2(\mathbf{r}_A) \le \ell_3(\mathbf{r}_A).
\label{eq:TensorDiffusionFieldRetrieval17}
\end{eqnarray}

As a simple illustrative example, consider the  isotropic-scatter special case where
\begin{equation}
\label{eq:TensorDiffusionFieldRetrieval18}    \ell_1(\mathbf{r}_A) = \ell_2(\mathbf{r}_A) = \ell_3(\mathbf{r}_A) \equiv \ell.
\end{equation} 
This implies $\mathbf{P}^{-1}(\mathbf{r}_A)$ is proportional to the $3 \times 3$ unit matrix $\mathbf{I}_3$.  If we also make the paraxial approximation 
\begin{equation}
\label{eq:TensorDiffusionFieldRetrieval19}    \mathbf{Q} \approx (Q_x,Q_y,0)^T \equiv \mathbf{Q}_{\perp},
\end{equation} 
then Eq.~(\ref{eq:TensorDiffusionFieldRetrieval16}) becomes
\begin{align}
\nonumber
I_{\textrm{scattered}}(\mathbf{Q}_{\perp}) &\propto \exp\left\{-\tfrac{1}{2}
 [\mathcal{R}^T_{\mathbf{\Phi}(\mathbf{r}_A)}\mathbf{Q}_{\perp}]^T\! \,\ell^2\mathbf{I}_3\,[\mathcal{R}^T_{\mathbf{\Phi}(\mathbf{r}_A)}
\mathbf{Q}_{\perp}]
\right\}
\\ \nonumber &=\exp\left[-\tfrac{1}{2}\ell^2
 \mathbf{Q}_{\perp}^T\mathcal{R}_{\mathbf{\Phi}(\mathbf{r}_A)} \mathcal{R}^T_{\mathbf{\Phi}(\mathbf{r}_A)}
\mathbf{Q}_{\perp}
\right]
\\ \nonumber &=\exp\left(-\tfrac{1}{2}\ell^2
 \mathbf{Q}_{\perp}^T
\mathbf{Q}_{\perp}
\right)
\\ \nonumber &= \exp\left[-\tfrac{1}{2}\ell^2
(Q_x^2+Q_y^2)
\right]
\\ &\approx \exp\left(-\tfrac{1}{2}\ell^2
k^2\theta^2
\right). 
\label{eq:TensorDiffusionFieldRetrieval120}
\end{align}
In the final line of the above expression \cite{ShullRoess1947a}, we  introduce the scattering angle
\begin{equation}
\label{eq:TensorDiffusionFieldRetrieval21}
\theta=\tan^{-1}\left(\frac{\sqrt{Q_x^2+Q_y^2}}{k}\right)\approx \frac{\sqrt{Q_x^2+Q_y^2}}{k}, \quad \vert\theta\vert\ll 1.
\end{equation} 
The angular width $1/(\ell k)$ of the diffuse-scatter fan, implied by the final line of Eq.~(\ref{eq:TensorDiffusionFieldRetrieval120}), reproduces the optical uncertainty principle in Eq.~(\ref{eq:SecondInverseProblemForDiffractionCase5}).

We are primarily interested in the fully anisotropic case of Eq.~(\ref{eq:TensorDiffusionFieldRetrieval16}).  The paraxial approximation in Eq.~(\ref{eq:TensorDiffusionFieldRetrieval19}) implies that the exponent in Eq.~(\ref{eq:TensorDiffusionFieldRetrieval16}) will be a known linear combination of $Q_x^2$, $Q_y^2$ and $Q_x Q_y$. Defining the weighting coefficients $\frak{A},\frak{B},\frak{C}$ via
\begin{equation}
\label{eq:TensorDiffusionFieldRetrieval22}
\frak{A}Q_x^2+\frak{B}Q_y^2+\frak{C}Q_xQ_y \equiv 
 \mathbf{Q}^T_{\perp} \mathcal{R}_{\mathbf{\Phi}(\mathbf{r}_A)}\mathbf{P}^{-1}(\mathbf{r}_A)\,\mathcal{R}^T_{\mathbf{\Phi}(\mathbf{r}_A)}
\mathbf{Q}_{\perp},
\end{equation} 
Eq.~(\ref{eq:TensorDiffusionFieldRetrieval16}) may be written in the paraxial-scatter approximation as 
\begin{align}
\nonumber &I_{\textrm{scattered}}(\mathbf{Q}_{\perp}) \propto \exp\left[-\tfrac{1}{2} \left(\frak{A}Q_x^2+\frak{B}Q_y^2+\frak{C}Q_xQ_y\right)\right] 
\\ &\quad\quad=\exp\left[-\tfrac{1}{2}(Q_x,Q_y)
\begin{pmatrix}
\frak{A} & \tfrac{1}{2}\frak{C} \\ 
\tfrac{1}{2}\frak{C} & \frak{B}
\end{pmatrix}
\begin{pmatrix}
Q_x \\
Q_y
\end{pmatrix}
\right]. \quad
\label{eq:TensorDiffusionFieldRetrieval23}
\end{align}

Since it is real and symmetric, the $2 \times 2$ matrix in the previous expression may be diagonalized by an orthogonal matrix.\footnote{See e.g.~p.~320 of \citet{AntonRorresBook}.} Now, an arbitrary $2\times 2$ orthogonal real matrix will be either (i) a standard $2\times 2$ rotation matrix, or (ii) a standard $2\times 2$ rotation matrix multiplied by a matrix that causes a reflection through a line passing through the origin. The second case can be ignored, here, since the elliptical level surfaces of our small-angle-scatter fan all have the symmetry that a reflection is equivalent to a suitable rotation.  Therefore there exists an orthogonal $2\times 2$ rotation matrix $\mathcal{R}_{\psi}$ (see Eq.~(\ref{eq:TensorDiffusionFieldRetrieval10})) such that
\begin{equation}
\label{eq:TensorDiffusionFieldRetrieval24}
\begin{pmatrix}
\frak{A} & \tfrac{1}{2}\frak{C} \\ 
\tfrac{1}{2}\frak{C} & \frak{B}
\end{pmatrix} = \mathcal{R}_{\psi} 
\begin{pmatrix}
a & 0 \\ 
0 & b
\end{pmatrix}
\mathcal{R}^T_{\psi},
\end{equation}
where $a,b\ge a$ are real positive numbers having units of squared length, and $\psi$ is a rotation angle.  Since it describes the orientation of an ellipse, as shown in  Fig.~\ref{Fig:EllipticalSASfan}, $\psi$ is only defined modulo $\pi$ radians.  The position-dependent ellipse-orientation field $\psi(x,y)$ is a ``director field'', which may be loosely thought of as a field of arrowless vectors \cite{PismenBook}.    

To continue, we need to  convert from Fourier-space coordinates $(Q_x,Q_y)$ to real-space coordinates $(x + \delta x,y + \delta y)$, where $(x,y)$ is the transverse location on the illuminated sample that results in a given fan of small-angle scatter, and $(\delta x, \delta y)$ specifies relative transverse coordinates used to describe the distribution of scatter in that particular fan.  The required conversion is given by similar triangles as
\begin{equation}
\label{eq:TensorDiffusionFieldRetrieval25}
\frac{Q_x}{k}=\frac{\delta x}{\Delta} \quad\textrm{and}\quad \frac{Q_y}{k}=\frac{\delta y}{\Delta}. 
\end{equation}
When this change of coordinates is combined with Eq.~(\ref{eq:TensorDiffusionFieldRetrieval24}), Eq.~(\ref{eq:TensorDiffusionFieldRetrieval23}) becomes (cf.~Ref.~\cite{jensen2010b})
\begin{align}
\nonumber I_{\textrm{scattered}}&(x,y;\delta x, \delta y) \propto \exp\Bigg[-\frac{k^2}{2\Delta^2}(\delta x,\delta y)
\mathcal{R}_{\psi(x,y)}
\\ &\begin{pmatrix}
a(x,y) & 0 \\ 
0 & b(x,y)
\end{pmatrix}
\mathcal{R}^T_{\psi(x,y)}
\begin{pmatrix}
\delta x \\
\delta y
\end{pmatrix}
\Bigg]. \quad
\label{eq:TensorDiffusionFieldRetrieval26}
\end{align}
Comparison with the right part of Fig.~\ref{Fig:EllipticalSASfan} shows that
\begin{itemize}
  \item $\psi(x,y)$ (as obtained in Eq.~(\ref{eq:TensorDiffusionFieldRetrieval24})) is the inclination angle of the locally-elliptical diffuse-blur fan, and
  \item the principal-axis blur-fan half-widths are given in units of length by $\sqrt{\Delta^2/[k^2 a(x,y)]}$ and $\sqrt{\Delta^2/[k^2 b(x,y)]}$, with corresponding half-widths in angular units of
  \begin{equation}
  \begin{cases}
  \theta_1(x,y)=1/[{k\sqrt{a(x,y)}}],  \\
  \theta_2(x,y)=1/[{k\sqrt{b(x,y)}}],
  \end{cases}
  \label{eq:TensorDiffusionFieldRetrieval27}
  \end{equation}
  where $\theta_1(x,y)\ge \theta_2(x,y)$ since $b(x,y) \ge a(x,y)$.
  \end{itemize}
  
Having obtained $\psi(x,y)$, $\theta_1(x,y)$, and $\theta_2(x,y)$, we may construct the eccentricity field $e(x,y)$, using the lower line of Eq.~(\ref{eq:TensorDiffusionFieldRetrieval7}).  Thus both the eccentricity and angular orientation of the small-angle-scatter fans may be calculated, as a function of transverse position, even though we have not computed the scatter fraction $F(x,y)$ (cf.~the sentence to which footnote~\ref{footnote:e-and-psi-are-independent-of-F} is attached).

If we wish to calculate the diffusion-field tensor $\widetilde{\mathbf{D}}(x,y)$ using Eq.~(\ref{eq:TensorDiffusionFieldRetrieval9}), the effective blur angles in Eq.~(\ref{eq:TensorDiffusionFieldRetrieval6}) should be computed.  Hence, an expression for the scatter fraction $F(x,y)$ should be derived. To keep the paper to a reasonable length, we will not calculate this quantity here.\footnote{\label{footnote:WaysToCalculateF}(i) A simple way to estimate $F(x,y)$, for a statistically anisotropic microstructure, would be to adapt the method of Sec.~\ref{sec:Refractive-sphere model} to an arbitrarily-oriented ellipsoid. This leads to an ellipsoidal-scatterer generalization of Eq.~(\ref{eq:TotalCrossSectionRefractiveSphere}). Our position-dependent Euler-angle field (Eq.~(\ref{eq:ThreeEulerAngles})) can then be replaced by a position-dependent orientation distribution function (Ref.~\cite{SiviaScatteringTheoryBook2011}, pp~158-159), namely a statistical distribution of Euler-angle triplets at each position within the sample.  Statistical averaging, using this distribution function, allows $F(x,y)$ to be calculated in the high-energy limit given by Eq.~(\ref{eq:Ell1MuchBiggerThanTheWavelength}). (ii) Another way to calculate $F(x,y)$ could begin by using the first Born approximation to derive the proportionality constant in Eq.~(\ref{eq:TensorDiffusionFieldRetrieval14}), e.g.~by taking the squared modulus of Eq.~(2.76) on p.~87 of Ref.~\cite{Paganin2006}, and then statistically averaging over an ensemble of realizations of the sample microstructure.  Here, the contribution due to the unscattered beam should be retained, since having an expression for both the scattered and unscattered beam intensities enables $F(x,y)$ to be calculated directly.  Equations (\ref{eq:TensorDiffusionFieldRetrieval2}) and (\ref{eq:TensorDiffusionFieldRetrieval19}) could be employed, in such a calculation.}  Rather, suppose the scatter fraction to have been calculated.  For the model considered in this section, which is based on the sample-microstructure correlation function in Eq.~(\ref{eq:TensorDiffusionFieldRetrieval2}) together with the far-field form of the first Born approximation, we may write the scatter fraction as $F(x,y;\ell_1,\ell_2,\ell_3,\Phi_1,\Phi_2,\Phi_3,E(\delta^2))$ or $F(x,y;\mathbf{P},\mathbf{\Phi},E(\delta^2))$.  This fraction can then be used to convert the diffuse-scatter fan angles $\theta_1$ and $\theta_2$ into their corresponding effective angles $\theta_1^{\textrm{(eff)}}$ and $\theta_2^{\textrm{(eff)}}$, using Eq.~(\ref{eq:TensorDiffusionFieldRetrieval6}). Since $\theta_1^{\textrm{(eff)}}$, $\theta_2^{\textrm{(eff)}}$, and $\psi$ are all now known, we may construct the diffusion field $\widetilde{\mathbf{D}}(x,y)$ using Eq.~(\ref{eq:TensorDiffusionFieldRetrieval9}).  This diffusion field may then be used in  Eq.~(\ref{eq:TensorDiffusionFieldRetrieval8}).

\begin{figure}[ht!]
\centering
\includegraphics[width=0.9\columnwidth]{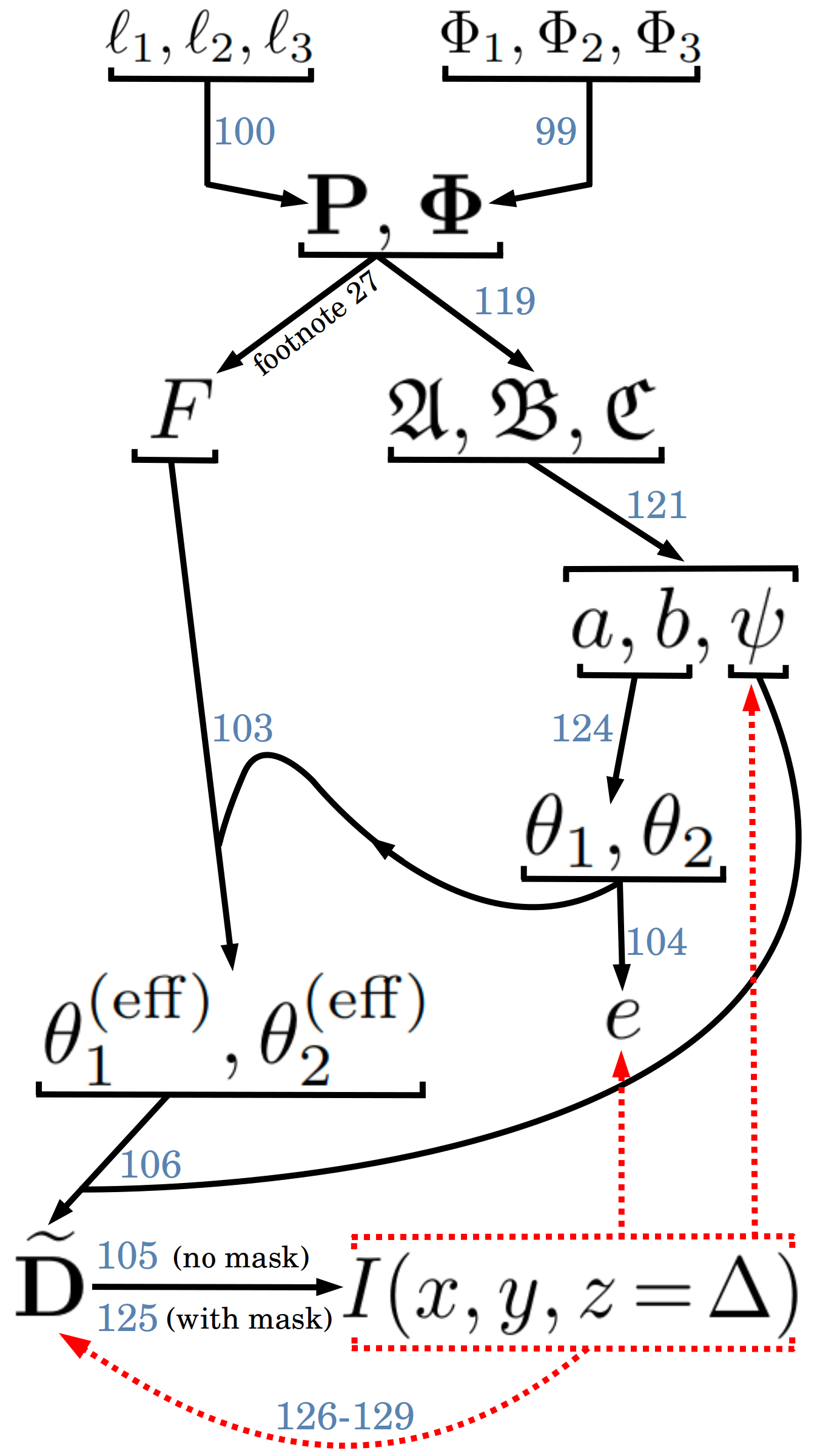}
\caption{Flowchart for the forward problem (solid-line arrows)
and a corresponding inverse problem (dotted-line arrows) in the tensorial generalization of paraxial diffusion-field retrieval. Relevant equation numbers are as indicated. }
\label{Fig:Flowchart}
\end{figure}

This completes our description of the forward problem.  The process is summarized as follows:
\begin{enumerate}
    \item Model a thin weakly-diffracting three-dimensional sample, whose statistical properties are approximately independent of distance $z$ along the optical axis, using these sample-related quantities:
        \begin{itemize}
        \item the diagonal matrix $\mathbf{P}(x,y)$ in Eq.~(\ref{eq:TensorDiffusionFieldRetrieval3}), which specifies the principal-axis half-lengths $\ell_1(x,y),\ell_2(x,y),\ell_3(x,y)$ of the correlation ellipsoid at each transverse location $(x,y)$, and
        \item the vector $\mathbf{\Phi}(x,y)$ of three Euler angles, using any suitable convention, which specifies the orientation of the correlation ellipsoid at each transverse location (see Eq.~(\ref{eq:ThreeEulerAngles})).  
    \end{itemize}
    \item The two sample-related quantities, as specified in the previous step, may be used to calculate the right side of Eq.~(\ref{eq:TensorDiffusionFieldRetrieval22}).  The coefficients of $Q_x^2$, $Q_y^2$ and $Q_xQ_y$ will then be equal to $\frak{A}(x,y)$, $\frak{B}(x,y)$, and $\frak{C}(x,y)$, respectively.  
    \item Use $\frak{A}(x,y)$, $\frak{B}(x,y)$, and $\frak{C}(x,y)$ to construct the $2 \times 2$ matrix field on the left side of Eq.~(\ref{eq:TensorDiffusionFieldRetrieval24}). Diagonalization of this matrix field then gives the quantities $a(x,y)$, $b(x,y)$, and $\psi(x,y)$, that appear on the right side of Eq.~(\ref{eq:TensorDiffusionFieldRetrieval24}).  The angle $\psi(x,y)$ is the inclination of the diffuse-scatter fan, as shown in the right part of Fig.~\ref{Fig:EllipticalSASfan}.
    \item The blur angles $\theta_1(x,y)$ and $\theta_2(x,y)$ may now be calculated, using Eg.~(\ref{eq:TensorDiffusionFieldRetrieval27}).
    \item The eccentricity field $e(x,y)$ may be calculated, using the lower line of Eq.~(\ref{eq:TensorDiffusionFieldRetrieval7}).
    \item  The scattering fraction $F(x,y;\mathbf{P},\mathbf{\Phi},E(\delta^2))$ should then be calculated (see footnote \ref{footnote:WaysToCalculateF}).
    \item  The effective scatter angles $\theta_1^{\textrm{(eff)}}(x,y)$ and $\theta_2^{\textrm{(eff)}}(x,y)$ may then be calculated, using Eq.~(\ref{eq:TensorDiffusionFieldRetrieval6}).
    \item The dimensionless diffusion-field tensor, $\widetilde{\mathbf{D}}(x,y)$, may then be assembled using Eq.~(\ref{eq:TensorDiffusionFieldRetrieval9}).
    \item Equation~(\ref{eq:TensorDiffusionFieldRetrieval8}) may then be used to evolve a specified exit-surface intensity distribution $I(x,y,z=0)$, to give the propagated intensity distribution $I(x,y,z=\Delta\ge 0)$ (see~Fig.~\ref{Fig:SimpleImagingWithSourceBlurAndSampleBlur}, for the special case where $\theta_0$ is negligible).
\end{enumerate}
For a visual representation of this summary, see Fig.~\ref{Fig:Flowchart}. 

\bigskip

We close this section by noting that, if a mask is placed immediately before the sample (see Fig.~\ref{Fig:SimpleImagingWithSourceBlurAndSampleBlurAndMask}), we need only change the final step in the above chain of logic.  In this case, we may generalize our transition from Eq.~(\ref{eq:2DVersionOfSimpleBlurWithGeneralisation}) to Eq.~(\ref{eq:ScalarDiffusionWithMasks1}), in replacing Eq.~(\ref{eq:TensorDiffusionFieldRetrieval8}) by
\begin{align}
\label{eq:TensorDiffusionFieldRetrieval28}
I(x,&y,z=\Delta \ge 0)=T(x,y)M(x,y)
\\ \nonumber &+\Delta^2 \nabla_{\perp}\cdot\{\widetilde{\mathbf{D}}(x,y) \nabla_{\perp}[T(x,y)M(x,y)]\}.
\end{align}
Here, $T(x,y)$ is the sample transmission function, $M(x,y)$ is the mask transmission function, and uniform unit-intensity illumination of the sample is assumed. 

\subsection{The inverse problem}

To limit the length of this paper, we give only one example of tensorial diffusion-field retrieval.  Adapting the terminology used by \citet{jensen2010a,jensen2010b} in a different but closely related context, we speak of our approach as ``directional diffusion-field retrieval'' (DDFR).  The particular DDFR problem we consider here is a mask-based variant which follows the dotted-line arrows in Fig.~{\ref{Fig:Flowchart}}, using intensity measurements to infer both (i) the small-angle-scatter ellipse orientation field $\psi(x,y)$ and (ii) the corresponding eccentricity field $e(x,y)$.  We reiterate a point made earlier, that these quantities are both independent of $F(x,y)$ \cite{MISTdirectional}, when the effects of $\theta_0$ are negligible (cf.~footnote \ref{footnote:e-and-psi-are-independent-of-F}). 

For the mask-based experiment in Fig.~\ref{Fig:SimpleImagingWithSourceBlurAndSampleBlurAndMask}, let there be $j_{\textrm{max}} \ge 7$ different mask transmission functions $M_j(x,y)$.  These different mask transmission functions may be obtained, for example, by transversely displacing a single spatially-random screen.  Expanding the final line of Eq.~(\ref{eq:TensorDiffusionFieldRetrieval28}), we obtain
\begin{align}
\nonumber \Delta^{-2} I_{j,\Delta} =& \Delta^{-2} T M_j
+  [\partial_x\widetilde{D}_{xx}] \partial_x(TM_j) 
\\  \nonumber &+ [\partial_x\widetilde{D}_{xy}] \partial_y(TM_j)
+ [\partial_y\widetilde{D}_{yy}] \partial_y(TM_j)
\\  \nonumber &+ [\partial_y\widetilde{D}_{xy}] \partial_x(TM_j)
+ [\widetilde{D}_{xx}] \partial_x^2 (TM_j)
\\  \nonumber &+ [\widetilde{D}_{yy}] \partial_y^2 (TM_j)
+ 2 [\widetilde{D}_{xy}] \partial_{x}\partial_{y} (TM_j), 
\\ &\quad\quad\quad\quad\quad j=1,2,\cdots,j_{\textrm{max}}\ge 7,
\label{eq:TensorDiffusionFieldRetrieval29}
\end{align}
for the propagated intensity 
\begin{equation}
I_{j,\Delta} \equiv I_j(x,y,z=\Delta \ge 0) 
\end{equation}
associated with the $j$th mask.  In writing Eq.~(\ref{eq:TensorDiffusionFieldRetrieval29}), we omit transverse coordinates $(x,y)$, and introduce the abbreviations  $\partial_x\equiv\partial/\partial x$, $\partial_y\equiv\partial/\partial y$. 

If the exit-surface intensity distributions $TM_j$ are measured for each mask, together with the propagated intensity distributions $I_{j,\Delta}$, then for each pixel location $(x,y)$, Eq.~(\ref{eq:TensorDiffusionFieldRetrieval29}) gives a system of $j_{\textrm{max}} \ge 7$ linear equations for the seven quantities in square brackets. It will typically be advantageous to use more than the minimum number of $j_{\textrm{max}}=7$ mask positions, to improve noise robustness by working with an overdetermined system of linear equations.  This system may be solved, for example, using the form of linear-least-squares matrix inversion that is based on QR decomposition \cite{Press,MISTdirectional}.  

Having solved the linear equations in Eq.~(\ref{eq:TensorDiffusionFieldRetrieval29}) for the seven terms in square brackets, we could simply keep the reconstructed $\widetilde{D}_{xx}$, $\widetilde{D}_{yy}$, and $\widetilde{D}_{xy}$, while discarding the other terms in square brackets.  However, it may be preferable to aggregate all seven reconstructions in the following manner. (i) Given the estimates $\widetilde{D}_{xx}^{\textrm{(est)}}$ and $\partial_x\widetilde{D}_{xx}^{\textrm{(est)}}$ for both $\widetilde{D}_{xx}$ and $\partial_x\widetilde{D}_{xx}$, respectively, form the linear combination $\widetilde{D}_{xx}^{\textrm{(est)}} + \aleph\, \partial_x\widetilde{D}_{xx}^{\textrm{(est)}}$, where $\aleph$ is any real number.  Let $\mathcal{F}$ denote Fourier transformation with respect to $x$ and $y$, with $(k_x,k_y)$ being corresponding Fourier-space coordinates. Use any convention for which the Fourier representation of $\partial/\partial x$ is multiplication by $ik_x$.\footnote{See e.g.~p.~395 of \citet{Paganin2006}.}  Form the aggregated estimate $\widetilde{D}_{xx}^{\textrm{(agg)}}$ for $\widetilde{D}_{xx}$, 
\begin{equation}
\widetilde{D}_{xx}^{\textrm{(agg)}}=\mathcal{F}^{-1}\left\{\frac{\mathcal{F}[\widetilde{D}_{xx}^{\textrm{(est)}} + \aleph\, \partial_x\widetilde{D}_{xx}^{\textrm{(est)}}]}{1 + i \aleph k_x}\right\}.    
\end{equation}
The real parameter $\aleph$ can be chosen to maximize signal-to-noise ratio.  (ii) An aggregated estimate $\widetilde{D}_{yy}^{\textrm{(agg)}}$ for $\widetilde{D}_{yy}$ is obtained by replacing all instances of $x$ with $y$, in the previous formula. (iii) To aggregate the three remaining estimates  $\widetilde{D}_{xy}^{\textrm{(est)}}$, $\partial_x\widetilde{D}_{xy}^{\textrm{(est)}}$, and $\partial_y\widetilde{D}_{xy}^{\textrm{(est)}}$, we use    
\begin{align}
\widetilde{D}_{xy}^{\textrm{(agg)}}
&=\!\mathcal{F}^{-1}\left(\frac{\mathcal{F}\{\widetilde{D}_{xy}^{\textrm{(est)}} + \aleph [\partial_x\widetilde{D}_{xy}^{\textrm{(est)}} + \partial_y\widetilde{D}_{xy}^{\textrm{(est)}}]\}}{1 + i \aleph  (k_x+k_y)}\right).    
\end{align}

The aggregated estimates $\widetilde{D}_{xx}^{\textrm{(agg)}}$, $\widetilde{D}_{yy}^{\textrm{(agg)}}$, and $\widetilde{D}_{xy}^{\textrm{(agg)}}$ may then be used to assemble the symmetric matrix $\widetilde{\mathbf{D}}$, using Eq.~(\ref{eq:TensorDiffusionFieldRetrieval11}). Diagonalization, as indicated by Eq.~(\ref{eq:TensorDiffusionFieldRetrieval9}), then yields $\psi(x,y)$, $\theta_1^{\textrm{(eff)}}$, and $\theta_2^{\textrm{(eff)}}$.  Finally, $e(x,y)$ can be obtained using the upper line of Eq.~(\ref{eq:TensorDiffusionFieldRetrieval7}).   

For brevity, we only studied one DDFR inverse problem.  A much broader range of inverse problems may be considered, by generalizing Eq.~(\ref{eq:SchematicChainOfTwoInverseProblemsForScalarD}) to 
\begin{align}
 \nonumber I &\stackrel{{\mathcal{I}}_1} {\longrightarrow} \left(\left\{\psi,e,\theta_1^{\textrm{(eff)}},\theta_2^{\textrm{(eff)}},\cdots\right\} \longrightarrow \left\{\widetilde{\mathbf{D}},F\right\}\right) \\ &\stackrel{{\mathcal{I}}_2}{\longrightarrow} \ell_1,\ell_2,\ell_3,\Phi_1,\Phi_2,\Phi_3.
\label{eq:SchematicChainOfTwoInverseProblemsForTensorD}
\end{align}
This schematically indicates that the first inverse problem $\mathcal{I}_1$ is associated with recovering quantities such as $\psi,e,\theta_1^{\textrm{(eff)}},\theta_2^{\textrm{(eff)}},\cdots$ that pertain to the diffusion field over a planar exit surface of an illuminated sample, with all of these recovered quantities being ultimately derivable from $\widetilde{\mathbf{D}}$ and $F$.  Weaker forms of $\mathcal{I}_1$, such as the particular example studied above, consist in recovering fewer diffusion-field quantities than would be needed to reconstruct both $\widetilde{\mathbf{D}}$ and $F$.  A stronger form of $\mathcal{I}_1$, which is beyond the scope of the present paper, seeks to recover both $\widetilde{\mathbf{D}}$ and $F$. Regarding the DDFR inverse problem $\mathcal{I}_2$, which is also beyond our scope, it would appear to be essential to have an analytical model for $F(\ell_1,\ell_2,\ell_3,\Phi_1,\Phi_2,\Phi_3,E(\delta^2))$. Such an analytical model, together with the formalism of tensor tomography \cite{Gullberg1999,Malecki_2014,bayer2014reconstruction,liebi2015nanostructure,schaff2015,Vogel2015,Sharma2016,Wieczorek2016,Schaff2017,liebi2018,TensorTomo2022}, are key ingredients in studying the DDFR inverse problem $\mathcal{I}_2$.

\section{Discussion}\label{sec:V}

We now discuss several points of interest.  Section~\ref{Sec:Discussion--ParallelSAS} develops an earlier comment, that diffusion-field retrieval may be viewed as a parallelized form of small-angle scattering.  Section~\ref{Sec:Discussion--RelationToDarkField} explains how diffusion-field retrieval is a form of dark-field imaging. A connection with computational imaging is drawn in Sec.~\ref{Sec:Discussion--ComputationalImaging}.  Conceptual overlaps with speckle tracking and single-grid imaging are given in Sec.~\ref{Sec:Discussion--RelationToSingleMaskImaging}, with particular reference to existing literature using x-ray or neutron beams.  Section~\ref{Sec:Discussion--NRU} argues that a form of noise-resolution tradeoff implies that shot noise in intensity measurements gives an additional additive offset to the dimensionless diffusion field. Resolution limits for paraxial diffusion-field retrieval are examined in Sec.~\ref{Sec:Discussion--ResolutionLimits}. Topological diffusion-field defects are considered in Sec.~\ref{Sec:Discussion--TopologicalDefects}.  Section~\ref{Sec:Discussion--FokkerPlanckExtension} points to a Fokker-Planck extension of diffusion-field retrieval, with connections being drawn to several recent papers on this topic.          

\subsection{Relation to small-angle scattering}\label{Sec:Discussion--ParallelSAS}

Paraxial diffusion-field retrieval may be viewed as a parallelized form of small-angle scattering (SAS), in the sense of the latter term that is employed in small-angle x-ray scattering (SAXS) and small-angle neutron scattering (SANS) \cite{SiviaScatteringTheoryBook2011}. This connection is suggested by Fig.~\ref{Fig:MomentumTransfer}(b), which conceptually divides an incident narrowband plane wave into a set of parallel pencil beams $p_1,p_2,\cdots$, that simultaneously create a set of moderately-overlapping small-angle-scatter fans $S_1,S_2,\cdots$, respectively. 

This parallelization of SAS comes at the cost of a very strong simplifying assumption.  In particular, our method utilizes only a minute fraction of the information that would be present, in a full SAS curve.  Full SAS curves can be measured at each transverse location, using  (i) sequential illumination by a true pencil beam 
such as an illuminated pinhole \cite{Fratzl1997} or focused probe \cite{schaff2015}, or (ii) simultaneous illumination by an array of beamlets \cite{GalvanJosaOPUS2022}. Conceptually-related work employs arrays of circular gratings \cite{CircularGratings1,CircularGratings2,CircularGratings3,CircularGratings4} for parallelized SAS measurements. By way of comparison, for the scalar-diffusion case, the only SAS-related quantities that appear in Eq.~(\ref{eq:PositionDependentScalarDiffusionCoefficient}) are
\begin{itemize}
\item the angular width $\theta_s(x,y)$ of the SAS cone in the Guinier approximation, where this width is related to the radius of gyration of the scattering particles;
\item 
the fraction $F(x,y)$ of the illumination converted to small-angle scatter (this information is often unavailable in conventional SAS measurements, but is available e.g.~in absolute SAXS \cite{SAXSabsolute} and absolute SANS \cite{SANSabsolute} measurements). 
\end{itemize}
Similar remarks apply to the tensor-diffusion case.  Here, the determinant (``det'') of Eq.~(\ref{eq:TensorDiffusionFieldRetrieval9}) implies that\footnote{In Eq.~(\ref{eq:TensorDiffusionFieldRetrieval9=DETERMINANT-FORM}), use has been made of the facts that (i) the determinant of a product of square matrices is equal to the product of the respective determinants, (ii) the determinant of a square matrix is equal to the determinant of the transpose of that matrix, (iii) the rotation matrix $\mathcal{R}$ has unit determinant, and (iv) the area of an ellipse with semi-major axis $r_1$ and semi-minor axis $r_2$ is $\pi r_1r_2$. Also, Eq.~(\ref{eq:TensorDiffusionFieldRetrieval6}) has been used in the middle line.}
\begin{align}
\nonumber2\sqrt{\det [\widetilde{\mathbf{D}}(x,y)]} &=
\theta_1^{\textrm{(eff)}}(x,y) 
\, \theta_2^{\textrm{(eff)}}(x,y) 
\\ \nonumber &= F(x,y) \, \theta_1(x,y) \, \theta_2(x,y)
\\ &= (1/\pi) \, F(x,y) \, \mathrm{d}\Omega(x,y).
\label{eq:TensorDiffusionFieldRetrieval9=DETERMINANT-FORM}    
\end{align}
This is proportional to (i) the solid angle $\mathrm{d}\Omega(x,y) \ll 1$ subtended by the central peak of the local SAS fan in the Guiner approximation, and (ii) the fraction $F$ of the incident beam converted to the diffuse-scatter channel due to unresolved random microstructure in the sample.

There is another key point of difference, between paraxial diffusion-field retrieval and SAS.  The former intrinsically incorporates the unscattered beam, as there is no spatial separation between scattered and unscattered beams, in intensity measurements used to construct the data functions (e.g.~Eqs.~(\ref{eq:DataFunction}) or (\ref{eq:ScalarDiffusionWithMasks5})) employed in diffusion-field retrieval.  Conversely, SAS experiments typically discard the unscattered beam using a beam stop, with the unscattered beam being spatially well separated from the scattered beam, for all but the smallest momentum transfers. 

In light of our previous mentions of the Guinier approximation, we now discuss how the process of diffusion-field retrieval is affected by situations where this approximation becomes invalid.  One example, where such a breakdown occurs, is long-tailed SAXS and SANS associated with random-fractal or self-affine microstructure \cite{BaleSchmidt1984,Wong1985,Teixeira1988, Sinha1988,Nesterets2008,Yashiro2010}.  While long-tailed scattering distributions may be well approximated by a Guinier-approximation Gaussian for the smallest momentum transfers, when broad non-Gaussian tails at higher momentum transfers have a non-negligible influence on intensity data employed for paraxial diffusion-field retrieval, these tails can no longer be neglected.  The formalism developed earlier in the present paper then becomes inapplicable.  In this case, a higher-order form of the diffusion equation might be employed, akin to the passage from the Fokker-Planck to the Kramers-Moyal equation \cite{MorganPaganin2019,PaganinMorgan2019} (cf.~p.~66 of Ref.~\cite{Risken1989}). The moments of the SAXS or SANS distributions \cite{DeblurByDefocus,Modregger2012,modregger2017,modregger2018,PaganinMorgan2019}, as a function of transverse location, enable calculation of the corresponding higher-order diffusion terms.  These additional diffusion terms involve progressively higher-order derivatives with respect to transverse position, augmenting those that appear in Eqs.~(25) and (\ref{eq:TensorDiffusionFieldRetrieval8}). Note, also, that the role of anomalous diffusion will need to be accounted for, in an analogous manner to that given earlier in the present paper. Finally, if the tails of the diffuse-scatter fans are too broad, it is likely that the previously-mentioned Kramers-Moyal-type extension will itself break down.  

\subsection{Relation to dark-field imaging}\label{Sec:Discussion--RelationToDarkField}

Once the uniform background due to source-size blur is removed, diffusion-field retrieval is a form of dark-field imaging \cite{gage1920}, in the sense that only photons (or neutrons, electrons etc.) that are scattered by the sample will contribute to the recovered diffusion field.  Note, in this context, that the diffusion field $D(x,y)$ (for the scalar-diffusion case) and the eccentricity field $e(x,y)$ (for the tensorial case) are both non-negative. Hence images of these quantities will appear as bright regions within certain areas of the sample, on a dark background corresponding to sample-free regions in the field of view.  Accordingly, we follow \citet{Beltran2022} in a different but closely related context, by speaking of the work in the present paper as ``diffusive dark-field'' imaging.  Use of the term ``dark field'', to describe any method for recovering such a diffusive signal, follows the tradition established by much work in both x-ray optics and neutron optics, using single grids \cite{wen2008,  how2022}, single random gratings (i.e. speckle-generators) \cite{berujon2012b, zanette2014, zdora2017}, grating interferometers \cite{Pfeiffer2008df, Strobl2008, Bech2010}, analyzer-crystal optics \cite{ando2002, Pagot2003, rigon2003}, and edge illumination \cite{Olivo2001,olivo2007coded,Olivo2021,NeutronEdgeIllumination}.  

The previously-cited papers typically speak of the diffusive dark-field signal in terms of the visibility reduction $\mathcal{V}_S/\mathcal{V}_R$ of mask illumination patterns that arises due to unresolved spatially-random microstructure in the sample. Here, $\mathcal{V}_R$ is the local visibility of the reference pattern in the absence of the sample, and $\mathcal{V}_S$ is the local visibility in the presence of the sample. The language of the preceding two sentences is rather different from that employed in the present paper, which works instead in terms of a dimensionless diffusion coefficient $D$.  Adapting Eq.~(10) in \citet{MorganPaganin2019} to the dimensionless form for $D$, the reduction in visibility is 
\begin{equation}
\frac{\mathcal{V}_S}{\mathcal{V}_R} = \exp\left({-\frac{4 \pi^2 D \Delta^2}{p^2}}\right), 
\end{equation}
where $p$ is the period of the reference pattern (note the equation above incorporates the factor of $2\pi$ not explicitly included when $p$ is used in \citet{MorganPaganin2019}). Since these dark-field imaging approaches use a reference pattern image (or a reference scan), the effect of source-size blur is separated, meaning the resulting extracted images do not show a diffusive signal arising from the finite source size, effectively $\theta_0 = 0$.

\subsection{Relation to computational imaging}\label{Sec:Discussion--ComputationalImaging}

Gabor's famous paper on inline holography conceives of imaging as a two-step process, namely recording of image data followed by a reconstruction step that is based on those data \cite{Gabor1948}.  Beyond the particular context of inline holography, we may consider this two-step process as ``computational imaging'', wherein the computer forms an intrinsic part of the imaging system.  This is not merely image processing, but rather the use of virtual optical elements (software optics, software lenses) based on the physics of the imaging process, together with the associated inverse problem of image reconstruction \cite{paganin2004}. Application of our inverse-problem algorithms to measured intensity data may therefore be spoken of as ``computational diffusive dark-field imaging''.  Rather than hardware optical elements being used to generate a dark-field image, such as the use of a knife edge in Schlieren imaging \cite{HechtOpticsBook}, the filtration is instead performed in the second (i.e., reconstruction stage) of a two-step process. This enables computational diffusive dark-field images to be obtained using bright-field data \cite{PaganinMorgan2019,Gureyev2020,Aminzadeh2022,Leatham2021}.

\subsection{Relation to speckle tracking and single-grid imaging}\label{Sec:Discussion--RelationToSingleMaskImaging}

The x-ray speckle-tracking literature \cite{berujon2012b, zanette2014, wang2015, zdora2017}, together with both the x-ray \cite{Pfeiffer2008df, Bech2010, Yashiro2011} and neutron \cite{Strobl2008} grating-based-imaging literature \cite{Strobl2014}, measures a position-dependent diffusive dark-field signal.  As mentioned in Sec.~\ref{Sec:Discussion--RelationToDarkField}, these works are couched in terms of visibility reduction---of speckles and fringes, respectively---that is due to unresolved microstructure in an imaged sample. While not explicitly formulated from a diffusion perspective, these works may be considered as studying a more complicated diffusion-retrieval problem than that considered in our paper, since small-angle-scatter diffusion effects and sample attenuation are there augmented by the influence of sample refraction.  Nevertheless, the resulting measurement of position-dependent visibility reduction
\begin{equation}
\nu(x,y)\equiv\mathcal{V}_S(x,y)/\mathcal{V}_R(x,y) 
\end{equation}
may be related to a corresponding diffusion coefficient $D(x,y)$, in an algebraically simple manner (see Eq.~(8) of Ref.~\cite{MorganPaganin2019}, together with Ref.~\cite{Morgan2022Preprint} and Sec.~\ref{Sec:Discussion--RelationToDarkField} above).  By thereby linking $D(x,y)$ to $\nu(x,y)$, as indicated by extending Eq.~(\ref{eq:SchematicChainOfTwoInverseProblemsForScalarD}) to
\begin{align}
\nonumber \label{eq:ExtendedSchematicChainOfTwoInverseProblemsForScalarD}
 I \stackrel{{\mathcal{I}}_1}{\longrightarrow}\,\,&D \stackrel{{\mathcal{I}}_2}{\longrightarrow} \ell 
 \\ &\,\big\updownarrow
 \\ &\,\,\nonumber \nu\equiv\mathcal{V}_S/\mathcal{V}_R \quad\quad\quad ,
\end{align}
techniques employed in the previously cited references may incorporate aspects of our $\mathcal{I}_2$ inverse problem, to relate visibility reduction $\nu$ (and its associated diffusion coefficient $D$) to sample-related properties such as $\ell$.  

A similar remark applies to using the  $\mathcal{I}_2$ methods of the present paper, for the Fokker--Planck implicit\footnote{This approach to x-ray speckle tracking is termed ``implicit'', since it does not explicitly track the deformation (e.g.~refraction-induced transverse deflection and microstructure-induced diffusion) of speckles in the structured illuminating intensity distribution, which results due to passage of those speckles through a sample that is being imaged.} approach to x-ray speckle tracking \cite{PaganinMorgan2019,MIST,MISTdirectional,alloo2022dark}, since the latter approach is also couched in terms of a diffusion field (cf.~Sec.~\ref{Sec:Discussion--FokkerPlanckExtension}).  Thus, for example, suppose implicit x-ray speckle tracking is employed, to recover the dimensionless diffusion coefficient $D(x,y; \Theta)$ for each projection of a sample, obtained using a tomographic data set.  This data set corresponds to the sample being  rotated through a number of angular orientations $\Theta$, with respect to a rotation axis that is perpendicular to the optical axis.  Suppose, also, that the refractive-sphere model of Sec.~\ref{sec:Refractive-sphere model} may be employed, to model the unresolved microstructure that may be present within the sample.  The reconstructed tomogram, obtained when the recovered diffusion fields $D(x,y; \Theta)$ are input into a standard tomographic reconstruction, will yield a tomogram that is proportional to the surface-area-to-volume ratio of the unresolved microstructure in the sample.  The associated proportionality constant is given in Sec.~\ref{sec:Refractive-sphere model}. 

In light of the preceding general comments and indicative example, the methods of the present paper may be viewed as a form of ``speckle tracking without refraction'', when spatially-random masks are employed in Fig.~\ref{Fig:SimpleImagingWithSourceBlurAndSampleBlurAndMask}.  An analogous term, ``single-grid imaging without refraction'', applies to single-grid diffusion-only dark-field imaging.  

\subsection{Noise-dependent contribution to diffusion field}\label{Sec:Discussion--NRU}

\begin{figure}[ht!]
\centering
\includegraphics[width=0.85\columnwidth]{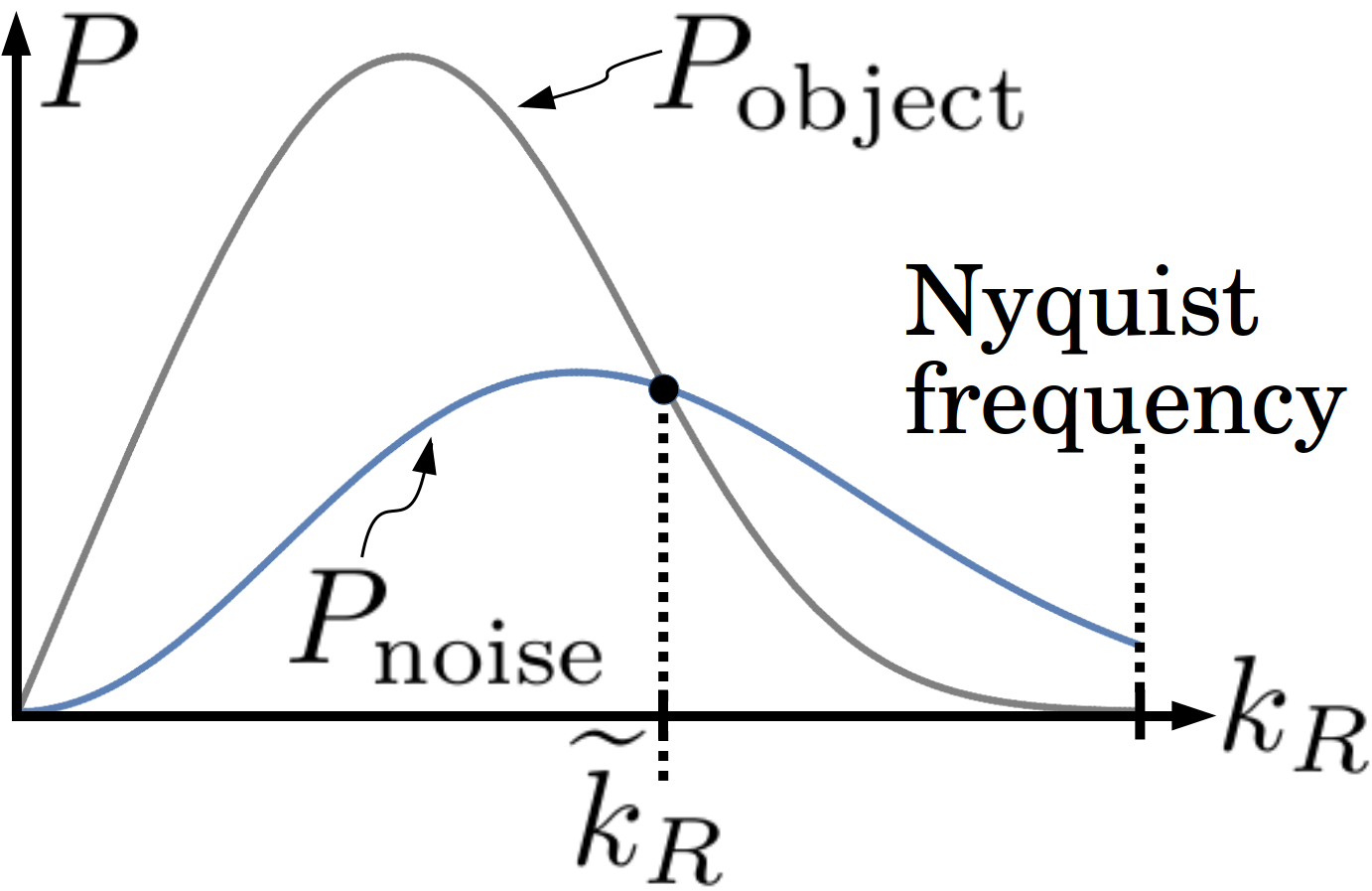}
\caption{The rotationally averaged power spectra $P$, associated with (i) the image of a sample in the absence of noise, $P_{\textrm{object}}$, and (ii) the contribution of noise to the image, $P_{\textrm{noise}}$.  These power spectra  decay at large radial spatial frequencies $k_R$, according to Eqs.~(\ref{eq:PowerLawDecayOfObjectInAbsenceOfNoise}) and (\ref{eq:PowerLawDecayOfNoiseInAbsenceOfObject}), respectively. }
\label{Fig:NRU}
\end{figure}

Consider the intensity power spectrum
\begin{equation}
\label{eq:IntensityPowerSpectrumDefinition}
P_{\textrm{object}}(k_x,k_y)=\vert\mathcal{F}[I_{\textrm{object}}(x,y)]\vert^2   
\end{equation}
for an idealized noise-free intensity image $I_{\textrm{object}}(x,y)$, which is sampled on a pixelated grid. Rotationally average $P_{\textrm{object}}(k_x,k_y)$ to give $P_{\textrm{object}}(k_R)$, where
\begin{equation}
\label{eq:RadialSpatialFrequencyDefinition}
k_R=\sqrt{k_x^2+k_y^2}    
\end{equation}
denotes radial spatial frequency.  For simplicity, suppose that $P_{\textrm{object}}(k_R)$ decays according to the power law \cite{Ruderman1994}
\begin{equation}
\label{eq:PowerLawDecayOfObjectInAbsenceOfNoise}
P_{\textrm{object}}(k_R)\sim \frac{A_{\textrm{object}}}{(k_R)^{\gamma}}, \quad A_{\textrm{object}}>0, \quad \gamma > 0, 
\end{equation}
for large radial spatial frequencies (namely, spatial frequencies that are a non-negligible fraction of the Nyquist frequency \cite{Press} associated with the pixelated grid).  Here, $A_{\textrm{object}}$ is a measure of the total radiant exposure in $I_{\textrm{object}}(x,y)$, and the exponent $\gamma$ governs the decay rate of the noise-free power spectrum $P_{\textrm{object}}(k_R)$.  Similarly, let $P_{\textrm{noise}}(k_R)$ denote the rotationally averaged intensity power spectrum of the noise that will be present in a given measured image of the sample.  Assume the power-law noise model \cite{PowerLawNoise1,PowerLawNoise2,PowerLawNoise3}
\begin{equation}
\label{eq:PowerLawDecayOfNoiseInAbsenceOfObject}
P_{\textrm{noise}}(k_R)\sim \frac{A_{\textrm{noise}}}{(k_R)^{\beta}}, \quad A_{\textrm{noise}}>0, \quad 0<\beta < \gamma, 
\end{equation}
where $A_{\textrm{noise}}$ is a measure of the total power contained in the noise, and the positive exponent $\beta$ is smaller than $\gamma$ because we assume the noise to decay more slowly than the signal at high radial spatial frequency. As sketched in Fig.~\ref{Fig:NRU}, let 
\begin{equation}
\label{eq:CrossoverSpatialFrequencyForNRU}
\widetilde{k}_R=\left(\frac{A_{\textrm{object}}}{A_{\textrm{noise}}}\right)^{\!\!1/(\gamma-\beta)} 
\end{equation}
denote the radial spatial frequency where the power spectra $P_{\textrm{object}}(k_R)$ and $P_{\textrm{noise}}(k_R)$ cross.  This corresponds to the maximum radial spatial frequency for which the signal power $P_{\textrm{object}}(k_R)$ is larger than the noise power $P_{\textrm{noise}}(k_R)$.  Hence we have a noise-induced spatial resolution
\begin{equation}
\label{eq:NoiseDependentResolution}
\ell_{\textrm{noise}}=\frac{2\pi}{\widetilde{k}_R}=2\pi\left(\frac{A_{\textrm{noise}}}{A_{\textrm{object}}}\right)^{\!\!1/(\gamma-\beta)}. 
\end{equation}
As the relative noise level improves, namely as the measure of the noise-to-signal ratio given by    $A_{\textrm{noise}}/A_{\textrm{object}}$ becomes progressively smaller, $\ell_{\textrm{noise}}$ decreases.\footnote{Once  $\ell_{\textrm{noise}}$ becomes smaller than the pixel size, the effects of noise-induced resolution reduction become ignorable.  This corresponds to $\widetilde{k}_R$ exceeding the Nyquist limit for the pixelated imaging device.}  This noise-dependent resolution, which is an example of the well known tradeoff between noise and spatial resolution 
\cite{NRU=1997,NRU==1998,NRU==2001,NRU=A,NRU=B,NRU=C,NRU=D,NRU=E}, may be thought of as the washing out of high-frequency detail by high-frequency noise. In the present context, Eq.~(\ref{eq:NoiseDependentResolution}) implies that the source-size angular blur can be generalized to include the additional effects of noise-induced resolution reduction.  This can be achieved by augmenting the term $\theta_0$, which is associated with the source-size contribution to the blur-cone angle in expressions such as Eqs.~(\ref{eq:BlurConeAngleSimpleForm}) and (\ref{eq:PositionDependentScalarDiffusionCoefficient}), with a noise-induced blur angle $\theta_{\textrm{noise}}$.  Adding these angles in quadrature,\footnote{Note, the effects of detector point-spread-function (PSF) blur can be included in a similar manner, by adding $\theta_0$ (in quadrature) to the effective angular blur implied by a specified PSF.} 
\begin{equation}
\label{eq:AddingBlurAnglesInQuadrature}
\theta_0^2\longrightarrow \theta_0^2 + \theta_{\textrm{noise}}^2,
\end{equation}
with the noise contribution being given under the paraxial approximation as
\begin{equation}
\theta_{\textrm{noise}}=\tan^{-1}\!\left(\frac{\ell_{\textrm{noise}}}{\Delta}\right)\approx\frac{\ell_{\textrm{noise}}}{\Delta}=\frac{2\pi}{\Delta} \left(\frac{A_{\textrm{noise}}}{A_{\textrm{object}}}\right)^{\!\!1/(\gamma-\beta)}\!\!\!.     
\end{equation}
Hence, for example, Eq.~(\ref{eq:PositionDependentScalarDiffusionCoefficient}) generalizes to
\begin{equation}
\label{eq:PositionDependentScalarDiffusionCoefficientWithNRU}
    D(x,y) = \tfrac{1}{2}\left\{\theta_0^2+\theta_{\textrm{noise}}^2+F(x,y)[\theta_s(x,y)]^2\right\}.
\end{equation}
Thus $\theta_0$ and $\theta_{\textrm{noise}}$ contribute an additive constant
\begin{equation}
\label{eq:DiffusionCoefficientOutsideTheSample}
    D_0=\tfrac{1}{2}(\theta_0^2+\theta_{\textrm{noise}}^2)
\end{equation}
to $D(x,y)$. For a compact object well contained within the field of view, we can subtract $D_0$ from the  retrieved diffusion field $D(x,y)$, with $D_0$ estimated using regions of the field of view that are outside the object.  The noisier the intensity data, the larger $D_0$ will be.  Subtracting $D_0$ from $D(x,y)$ yields a diffusion field that is entirely due to the compact sample.  This will be a dark-field image in the sense that it will appear on a black background, since only scatter contributes to $D(x,y)-D_0$ (cf.~Sec.~\ref{Sec:Discussion--RelationToDarkField}). 

\subsection{Resolution limit for paraxial diffusion-field retrieval}\label{Sec:Discussion--ResolutionLimits}

The expression for the noise-dependent spatial resolution, in Eq.~(\ref{eq:NoiseDependentResolution}), leads to the closely-related question of the ultimate resolution limit for paraxial diffusion-field retrieval.  This implies the two distinct questions of (i) the spatial resolution with which $D(x,y)$ may be recovered, and (ii) the smallest unresolved characteristic length $\ell(x,y)$ that can be retrieved using our method's indirect approach.  We do not know how to give a general answer to these interesting questions, and suggest that they would form an interesting topic for future investigation. The following suggestions might provide some guidance:
\begin{itemize}
\item The achievable spatial resolution for $D(x,y)$ is bounded from below by $L$, defined as the larger of the following quantities: (i) the pixel size of the position-sensitive detector employed to register the measured intensity data, and (ii) the width of the imaging-system point-spread function. In practice, we conjecture that the typical spatial resolution for $D(x,y)$ would be on the order of $2 L$, when employing the methods for paraxial diffusion-field retrieval that are developed in the present paper, under empirically-optimized experimental scenarios.    
\item The achievable resolution for $D(x,y)$ and $\ell(x,y)$ will depend on the noise level in the data, in general.  Accordingly, it is natural to conjecture that the resolution of either quantity will be bounded from below by a monotonically increasing function of the noise-to-signal ratio that is present in the measured intensity data.  Furthermore, for dose-sensitive samples that are damaged by the irradiation process, there will be a minimum permissible noise-to-signal ratio.
\item The achievable resolution for $\ell(x,y)$ will depend on the scattering model pertinent to a particular class of unresolved spatially-random microstructure.  For example, in a refractive ``large microstructure'' model such as that presented in Sec.~\ref{sec:SecondInverseProblemExampleForCase1}, $\ell(x,y)$ cannot be smaller than a specified limiting value, in order for the underpinning geometrical-optics assumption to remain valid.  Similarly, in a diffractive ``small microstructure'' regime which employs a scalar-radiation form of the first Born approximation for the case of electromagnetic-wave illumination (cf.~Eq.~(\ref{eq:TensorDiffusionFieldRetrieval14})), $\ell(x,y)$ cannot be smaller than a specified limiting value, since if $\ell(x,y)$ is too small then the polarization degree of freedom cannot be ignored (see e.g.~pp~65-70 of Ref.~\cite{Paganin2006}).  
\end{itemize}

\subsection{Topological diffusion-field defects}\label{Sec:Discussion--TopologicalDefects}

In tensor diffusion-field retrieval, the small-angle-scatter ellipse-orientation field $\psi(x,y)$ plays a key role.  As stated earlier, $\psi(x,y)$ is a director field since $\psi(x,y)$ is only defined modulo $\pi$ radians. Topological defects \cite{VilenkinShellard1994,PismenBook,SethnaBook} in this director field may exist, e.g.~those associated with non-zero values for the integer  
\begin{equation}
    m(x,y)=\frac{1}{\pi}\oint_{\Gamma} \mathrm{d}\psi(x,y). 
\end{equation}
Here, $\Gamma$ is an infinitesimally small clockwise-traversed simple closed contour containing $(x,y)$, with $\psi(x,y)$ being (by assumption) both continuous and differentiable at each point on $\Gamma$. By way of context, we may point out that topological defects have been studied in many physical systems, such as screw-type and edge-type defects in the phase of coherent visible-light \cite{NyeBerry1974,Nye1999} and electron-optical fields \cite{Bliokh2017}, polarization singularities in electromagnetic fields \cite{DennisProgOpt2009}, several forms of topological defect in otherwise-perfect crystal lattices \cite{KittelBook}, and topological defects in liquid-crystal films \cite{PismenBook}.  This last-mentioned example includes director-field topological defects such as disclinations and ring singularities.  In general, $\psi(x,y)$ may admit such topological defects, and in our view it would be interesting to investigate this possibility in future work.  Similarly, it would be interesting to investigate topological defects in the correlation-ellipse field \cite{FreundOnEllipseFieldDefects}, associated with  $\mathbf{\Sigma}(\mathbf{r})$ in Eqs.~(\ref{eq:TensorDiffusionFieldRetrieval4}) and (\ref{eq:TensorDiffusionFieldRetrieval5}).

\subsection{Fokker--Planck merging of diffusion-field retrieval with the transport-of-intensity equation}\label{Sec:Discussion--FokkerPlanckExtension}

If the illuminating source is sufficiently spatially coherent that the influence of sample refraction cannot be ignored, both diffusive and coherent transport will determine the propagation of intensity from $z=0$ to $z=\Delta$ in Figs.~\ref{Fig:SimpleImagingWithSourceBlurAndSampleBlur} and \ref{Fig:SimpleImagingWithSourceBlurAndSampleBlurAndMask}.  In particular, refraction by the sample implies that we can no longer assume the line $PU$ in Fig.~\ref{Fig:SimpleImagingWithSourceBlurAndSampleBlur} to be parallel to the optical axis $z$, since refraction by the sample means that the centroid $U$ of the diffuse-scatter cone will be transversely displaced in the plane $C$, in this diagram. 
The Fokker--Planck equation \cite{Risken1989} provides a natural formalism, here, in the form provided by the Fokker--Planck extension \cite{MorganPaganin2019,PaganinMorgan2019,PaganinPelliccia2020} to the transport-of-intensity equation \cite{Teague1983,TIE==LongReviewArticle2020} of paraxial optics. 

The Fokker--Planck extension to Eq.~(\ref{eq:2DVersionOfSimpleBlurWithGeneralisation}) is \cite{MorganPaganin2019,PaganinMorgan2019}
\begin{align}
\nonumber I(&x,y,z=\Delta \ge 0) =I(x,y,z=0) 
\\  \nonumber &-\frac{\Delta}{k} \nabla_{\perp}\cdot \{[1-F(x,y)]I(x,y,z=0) \nabla_{\perp}\phi(x,y)\}
\\ &+\Delta^2\nabla_{\perp}^2[D(x,y) I(x,y,z=0)],
\label{eq:2DVersionOfSimpleBlurWithGeneralisation==FokkerPlanckForm}
\end{align}
where $\phi(x,y)$ is the phase of the coherent component of the field over the exit surface of the sample, and $D(x,y)$ is given by Eq.~(\ref{eq:PositionDependentScalarDiffusionCoefficient}).  Note that $\phi(x,y)$ encodes the refractive properties of the sample, with the two components of $k^{-1}\nabla_{\perp}\phi(x,y)$ corresponding to paraxial deflection angles in each of the two transverse directions. Note, also, that for partially coherent illumination, the phase $\phi(x,y)$ may be defined using the approach of \citet{paganin1998}. Similarly to the preceding equation, the Fokker--Planck extension to Eq.~(\ref{eq:TensorDiffusionFieldRetrieval8}) is (cf.~Refs.~\cite{MorganPaganin2019,PaganinMorgan2019})
\begin{align}
\nonumber I(&x,y,z=\Delta \ge 0)=I(x,y,z=0)
\\  \nonumber &-\frac{\Delta}{k} \nabla_{\perp}\cdot \{[1-F(x,y)] I(x,y,z=0) \nabla_{\perp}\phi(x,y)\}
\\  &+\Delta^2\nabla_{\perp}\cdot[\widetilde{\mathbf{D}}(x,y) \nabla_{\perp}I(x,y,z=0)], 
\label{eq:TensorDiffusionFieldRetrieval8==FokkerPlanckForm}
\end{align}
where $\widetilde{\mathbf{D}}(x,y)$ is given by Eqs.~(\ref{eq:TensorDiffusionFieldRetrieval6}), (\ref{eq:TensorDiffusionFieldRetrieval11}) and (\ref{eq:TensorDiffusionFieldRetrieval12}). 

Three special cases are worth mentioning:
\begin{itemize}
\item If $F\!\ll\!1$, then $1-F(x,y)$ may be replaced with unity, in the middle lines of Eqs.~(\ref{eq:2DVersionOfSimpleBlurWithGeneralisation==FokkerPlanckForm}) and (\ref{eq:TensorDiffusionFieldRetrieval8==FokkerPlanckForm}).
\item If $F(x,y)$ and $\theta_0$ are negligible, Eqs.~(\ref{eq:2DVersionOfSimpleBlurWithGeneralisation==FokkerPlanckForm}) and (\ref{eq:TensorDiffusionFieldRetrieval8==FokkerPlanckForm}) reduce to a forward-finite-difference form of the transport-of-intensity equation \cite{Teague1983}.
\item If $F(x,y)=1$, then Eqs.~(\ref{eq:2DVersionOfSimpleBlurWithGeneralisation==FokkerPlanckForm}) and (\ref{eq:TensorDiffusionFieldRetrieval8==FokkerPlanckForm}) reduce to Eqs.~(\ref{eq:2DVersionOfSimpleBlurWithGeneralisation}) and (\ref{eq:TensorDiffusionFieldRetrieval8}), respectively.
\end{itemize}

For imaging systems that can access both positively and negatively defocused images,  the data function
\begin{equation}
    q(x,y)=\frac{I(x,y,\Delta)+I(x,y,-\Delta)-2I(x,y,0)}{2\Delta^2}
\end{equation}
respectively reduces Eqs.~(\ref{eq:2DVersionOfSimpleBlurWithGeneralisation==FokkerPlanckForm}) and (\ref{eq:TensorDiffusionFieldRetrieval8==FokkerPlanckForm}) to
\begin{align}
    \nabla_{\perp}^2[D(x,y) I(x,y,z=0)] &=q(x,y),
    \\
\nabla_{\perp}\cdot[\widetilde{\mathbf{D}}(x,y) \nabla_{\perp}I(x,y,z=0)] &=q(x,y).   
\end{align}
The above expressions are amenable to the methods developed in the present paper, even when the middle lines of Eqs.~(\ref{eq:2DVersionOfSimpleBlurWithGeneralisation==FokkerPlanckForm}) and (\ref{eq:TensorDiffusionFieldRetrieval8==FokkerPlanckForm}) cannot be ignored, provided that both $I(x,y,z=0)$ and $q(x,y)$ have been measured. This and related points, regarding Fokker--Planck and associated extensions to diffusion-field retrieval, will be further explored in a future companion manuscript.

\section{Conclusion}\label{sec:VI}

We studied diffusion-field retrieval, for low-coherence paraxial imaging systems in which the output intensity is a blurred form of the input intensity. This seeks to (i) recover a position-dependent diffusion coefficient associated with a propagating or defocused intensity distribution, downstream of an illuminated sample, and then (ii) relate the retrieved diffusion field to the spatially-random microstructure in that sample.  The cases of isotropic and anisotropic random microstructure were considered, with the former leading to a scalar diffusion field and the latter leading to a tensorial diffusion field. Two classes of object were considered, corresponding to the influence of the sample microstructure being well described by refractive or diffractive models, respectively.  Two classes of imaging system were treated, corresponding to no beam-shaping mask, or the presence of single beam-shaping mask upstream of the illuminated sample.  The method may be viewed as a parallel form of small-angle scattering in the Guinier regime.  It may also be spoken of as computational diffusive dark-field imaging, based on a simplified form of the Fokker-Planck equation of paraxial optics, in which the effects of diffuse-flow intensity transport dominate over those of coherent flow.  While we focused on the mechanism of diffuse scatter associated with unresolved spatially-random microstructure, this is not the only means of generating such scatter. Hence the methods of the present paper might also be applicable to certain incoherent scattering scenarios, such as x-ray Compton scatter and neutron nuclear scattering.


\section*{Acknowledgments}

K.\,S.\,M.~acknowledges funding via the Australian Research Council (ARC) Future Fellowship FT180100374.  K.\,S.\,M. and D.\,M.\,P. acknowledge funding from ARC Discovery Project DP230101327. All authors acknowledge useful discussions with Samantha Alloo, Mario Beltran, Matthieu Boone, Stef Claeys, Michelle Croughan, Christian Dwyer, Ying Ying How, Andrew Kingston, Kieran Larkin, Thomas Leatham, Konstantin Pavlov, Tim Petersen, and Imants Svalbe.

\end{document}